\documentclass[12pt]{article}
\usepackage{amsmath,amsthm}
\usepackage{eufrak}
\usepackage[pdftex]{graphicx}
\newcommand{\ar}{\renewcommand{\arraystretch}{1}} % 1.0 % 0.6
\DeclareMathAlphabet{\bb}{U}{msb}{m}{n} \gdef\C{\bb C} \gdef\dZ{\bb
Z}   \gdef\dS{\bb S} \gdef\R{\bb R}
\gdef\K{\bb K} \gdef\BH{\bb H} \gdef\F{\bb F} 

 \DeclareMathOperator{\spin}{{\bf
Spin}} \DeclareMathOperator{\pin}{{\bf Pin}}

\DeclareMathOperator{\Sym}{Sym}

 \DeclareMathOperator{\SL}{SL}
\DeclareMathOperator{\SO}{SO}\DeclareMathOperator{\SU}{SU}
 \DeclareMathOperator{\GO}{O}\DeclareMathOperator{\GU}{U}

\newcommand{\cA}{\mathcal{A}}

\newcommand{\bsA}{{\bf\sf A}}
\newcommand{\sA}{{\sf A}}
\newcommand{\sB}{{\sf B}}
\newcommand{\sI}{{\sf I}}

\newcommand{\sC}{{\sf C}}

\newcommand{\sH}{{\sf H}}
\newcommand{\sT}{{\sf T}}
\newcommand{\sS}{{\sf S}}

\newcommand{\sQ}{{\sf Q}}
\newcommand{\sX}{{\sf X}}
\newcommand{\sY}{{\sf Y}}

\newcommand{\bsH}{{\boldsymbol{\sf H}}}

\newcommand{\bsU}{{\boldsymbol{\sf U}}}
\newcommand{\bsX}{{\boldsymbol{\sf X}}}
\newcommand{\bsY}{{\boldsymbol{\sf Y}}}
\newcommand{\bsE}{{\boldsymbol{\sf E}}}
\newcommand{\bsC}{{\boldsymbol{\sf C}}}
\newcommand{\bsB}{{\boldsymbol{\sf B}}}

\newcommand{\fA}{\mathfrak{A}}

\newcommand{\fK}{\mathfrak{K}}

\newcommand{\fW}{\mathfrak{W}}

\newcommand{\fa}{\mathfrak{a}}

\newcommand{\fg}{\mathfrak{g}}

\newcommand{\cl}{C\kern -0.2em \ell}

\newcommand{\ld}{\left[}
\newcommand{\rd}{\right]}
\newcommand{\lf}{\left\{}
\newcommand{\rf}{\right\}}

\newtheorem{thm}{Theorem}
\newtheorem{defn}{Definition}
\begin{document}
\title{Group Theory and Mass Quantization}
\author{V.~V. Varlamov\thanks{Siberian State Industrial University,
Kirova 42, Novokuznetsk 654007, Russia, e-mail:
varlamov@sibsiu.ru}}
\date{}
\maketitle
\begin{abstract}
The mass spectrum problem (the 14th Ginzburg’s problem) is analyzed in terms of the conventional reductional and alternative holistic frameworks. From the holistic viewpoint, substance (the same as energy) is the primary concept and particles are secondary and emergent. An axiomatic system is proposed where the basic notion of the spectrum of matter is defined, and the particle spectrum emerges as a result of the energy/mass quantization. A new mass formula is derived in terms of the quantum numbers related to the Lorentz group Casimir eigenvalues. It is shown that the masses of states in the lepton and hadron sectors of the spectrum of matter are proportional to the electron rest mass (quantum of mass) with an accuracy of $0,41\%$ on average.
\end{abstract}
\begin{flushright}
\begin{minipage}{16pc}{{\it
In the beginning there was symmetry
}

Werner Heisenberg \cite{Heisen}
}\end{minipage}
\end{flushright}
\section{Introduction\label{sec1}}
Despite of more than half-a-century investigations, no satisfactory description of the mass spectrum of particles has been obtained. This issue has got No.14 in the Ginzburg's list of the thirty most important problems facing contemporary physics \cite{Ginz}:

Ginzburg's 14th problem: {\it Mass spectrum of particles}.

It is well known that the quark model based on $\SU(3)$ flavor symmetry gives only an incomplete description of the mass spectrum of particles. We will give a brief review of models in Section 2. This unsatisfactory situation is well recognized now. Summarizing the achievements of particle physics during the second half of the 20th century S. Weinberg wrote: ``The 70s were a period when everything was flowing beautifully, then it jammed up again and we haven't really been making much progress since then'' \cite{Wein}. This jam has not been cleared yet.

What is the reason for the current no-go position of particle physics? Although the answer may seem provocative, we suggest that the main drawback is the atomic hypothesis, which treats physical substance as an agglomerate of indivisible fundamental elements. The atomistic air is present even in the CERN mission formulation: “At CERN, we probe the fundamental structure of the particles that make up everything around us” (https://home.cern/about).

The ideas that everything is built of particles, the whole is the sum of its parts, and the global properties of a system $S$ are entirely determined by the states and interactions of its parts, make up the core of {\it reductionism} and {\it separability principle}. In a reductional model, a system is necessarily treated as a mechanical sum of its parts. Although such mechanical approach is illegal at the quantum level, it is still applied both at the atomic (the Bohr planetary model) and even subatomic (the quark model) scales.

The fact that separability (which is the main principle of reductionism) is only partly applicable in quantum mechanics is well known. If some parts of a system $S$ are entangled (a nonseparable state of $S$), then none of the global properties of $S$ are determined by the properties of its parts. On the other hand, no part of $S$ can be in a pure state, i.e., none of  subsystems $S_1$, $S_2$, $\ldots$, $S_N$  of $S$ can exist independently. In no case the parts of an entangled quantum system can be treated as autonomous. In other words, the parts of a system in nonseparable state dissolve themselves in the whole, thus depriving it of a structure in ordinary sense. This does not mean that the system is completely amorphous, but only that its structure is not borrowed from the alien “repertoire of classical physics” (as Heisenberg called it) but naturally stems from the mathematics of quantum theory, such as quantum state vectors, symmetry groups, Hilbert space, tensor products of Hilbert and $\K$-Hilbert spaces, etc. This study is an attempt to find such a structure.

The proposed theory is based on the Heisenberg’s principles and von Neumann’s concept of quantum formalism. This rigorous construction enables computation of the particle mass spectrum, which can be readily verified to the values provided by Particle Data Group (PDG).

This paper is organized as follows. Section 2 briefly reviews the most applicable mass formulas with special attention paid to the Balmer-like ones, which will play an important role in the following. Section 3 introduces ideological grounds of our model: the general holistic view and various notions of particle (Section 3.1). Section 3.2 describes the Heisenberg’s theory as a rare example of holistic approach to the quantum formalism. In Section 4, the spectrum of matter theory is elaborated from scratch: starting from Axiomatics (Section 4.1), specifying the symmetries (Section 4.2) and the Hilbert space structure (Section 4.3), and finally arriving at the mass formula (Section 4.4). This formula was used to compute more than three hundred states comprising the mass spectrum of matter. The results are listed in Section 5 along with the values provided by PDG. It is seen that they coincide to within $0,41\%$ accuracy. In conclusion we stress once again that in holistic terms, all of the matter should be taken as a single quantum system and particles can be conceived as certain states of this system. The basic element in this machinery is described by spinor, which has nothing to do with spinning but rather expresses the fundamental doubling, which creates other symmetries.

\section{Mass Spectrum: Past and Present\label{sec2}}
\subsection{Mass spectrum and quark model: matching unsuitable?\label{sec2_1}}

The celebrated Gell-Mann-Okubo mass formula describes only the two types of splitting within $\SU(3)$ theory, namely, the hypercharge and charge splitting of a supermultiplet into isotopic multiplets. Similar to the Zeeman effect in atomic spectra, each charge multiplet splits into a number of mass levels via $\SU(3)/\SU(2)$ reduction. The mass spectrum resulting for each $\SU(3)$ supermultiplet is given by the Gell-Mann--Okubo formula \cite{Gel61,OR64}
\begin{equation}\label{GMOQ}
m^2=m^2_0+\alpha+\beta Y+\gamma\left[I(I+1)-\frac{1}{4}Y^2\right]
+\alpha^\prime-\beta^\prime
Q+\gamma^\prime\left[U(U+1)-\frac{1}{4}Q^2\right],
\end{equation}
where $Q$ and $Y$ are the particle charge and hypercharge, respectively; $I$, $U$ are isotopic spins corresponding to different choices of subalgebras $\mathfrak{su}(2)$ within algebra $\mathfrak{su}(3)$; and factors $\alpha$, $\alpha^\prime$, $\beta$, $\beta^\prime$, $\gamma$, $\gamma^\prime$ satisfy the ratio
\[
\frac{\alpha^\prime}{\alpha}=\frac{\beta^\prime}{\beta}=\frac{\gamma^\prime}{\gamma}=\theta,\quad
|\theta|\ll 1.
\]
In case $\delta m^2/m^2_0\ll 1$, the quadratic mass formula (\ref{GMOQ}) can be replaced by a linear one
\begin{equation}\label{GMOL}
m=m_0+\alpha+\beta Y+\gamma\left[I(I+1)-\frac{1}{4}Y^2\right]
+\alpha^\prime-\beta^\prime
Q+\gamma^\prime\left[U(U+1)-\frac{1}{4}Q^2\right].
\end{equation}
In the approach based on the $\SU(6)$ ``spin-flavor'' symmetry, the mass spectrum is described by the Beg-Singh formula \cite{BS,RF70},
\begin{multline}\label{BS}
m^2=m^2_0+\mu_1C_2(3)+\mu_2\cdot 2J(J+1)+\mu_3Y+
\mu_4\left[2S(s+1)+\frac{1}{4}Y^2-C_2(4)\right]+\\
+\mu_5\left[-\frac{1}{2}Y^2+2T(T+1)\right]+
\mu_6\left[2N(N+1)-2S(S+1)\right],
\end{multline}
which includes two $\SU(6)$ Casimir operators, as well as the spin, isospin, hypercharge, and the so-called strange and non-strange spin numbers. As in the case of $\SU(3)$ symmetry, the $\SU(6)$ group action produces a Zeeman-like effect, i.e. splits quantum states within the $\SU(6)$ hypermultiplets into 56-baryon and 35-meson multiplets via the $\SU(6)/\SU(3)$- and $\SU(6)/\SU(4)$-reductions, respectively.

According to the Particle Data Group (PDG) concept, classification of high-spin quantum states is based on the $\SU(6)\times\GO(3)$ spin-flavor-orbital symmetry. The adopted quantum numbers are those of the $\SU(6)\times\GO(3)$ group: $I(J^P)$ for baryons and $I(J^{PC})$ for mesons, with spin $J$, isotopic spin $I$, parity $P$, and charge parity $C$.

There is a crucially important circumstance about the spin definition in the quark $\SU(6)\times\GO(3)$ model. That is, the spin of a $q\bar{q}$-meson or a $qqq$-baryon is taken equal to the total angular momentum of the corresponding $q\bar{q}$- or $qqq$- quark system: $\mathbf{J}=\mathbf{L}+\mathbf{S}$, where $\mathbf{L}$ is the orbital momentum of the quark system and $\mathbf{S}$ is the total spin of the quarks. This is a mechanical definition of spin. However, spin is not a mechanical concept.

V. A. Fock noted \cite[c.~111]{Fock71} that, though the spin degree of freedom has formal relation to the angular momentum, an electron cannot be seen as a rotating ball; mechanical assimilations are decidedly out of place here since spin is not a mechanical concept. Recall that the concept of spin was introduced in 1925 by W. Pauli, who surmised a two-valued nature of a radiating electron in attempt to explain the doublet structure of the alkali metal spectrum (anomalous Zeeman effect) and the Larmor theorem violation \cite{Pauli}. Later, this classically nondescribable two-valuedness of electron was adopted under the name of spin (see, e.g., Van der Waerden \cite{Waer}).

In our opinion, the use of a formal mechanical model to define a non-classical concept of spin is the main weak point of the quark model; it is also the cause of problems in baryon spectroscopy and of the so-called proton spin puzzle.

Let us see how the mechanical interpretation of the hadron spin in the quark model leads to a problem with the baryon spectra. In terms of the $\SU(6)\times\GO(3)$  model, the system of three quarks $q_1,q_2,q_3$ can be reduced to a system of two separate 3d oscillations of Jacobi parameters $\boldsymbol{\rho}$ and $\boldsymbol{\lambda}$ (at a frequency $\omega=\sqrt{3K/m}$ each), where $\boldsymbol{\rho}$ refers to the $(q_1,q_2)$ diquark oscillations and $\boldsymbol{\lambda}$ describes the $q_3$ motion in the ($q_1,q_2$) diquark rest frame. The baryon mass has the form
\begin{equation}\label{orbital}
M=3m+\text{const}+\omega(N+3),
\end{equation}
where $N=2k_\rho+2k_\lambda+l_\rho+l_\lambda$ is the band number, $m$ is the quark mass, and $k,l=0,1,2,\ldots$ for both $\boldsymbol{\rho}$ and $\boldsymbol{\lambda}$ oscillators. However, formula (\ref{orbital}) lacks experimental support in the range of excited baryons. The multiplets in the first three bands are as follows:
\begin{eqnarray}
N=0&:&\quad(56,0^+),\nonumber\\
N=1&:&\quad(70,1^-),\nonumber\\
N=2&:&\quad(56^\prime,0^+),\;(70,0^+),\;(56,2^+),\;(70,2^+),\;(20,1^+).\nonumber
\end{eqnarray}
First number in the round brackets is the $\SU(6)$-multiplicity and second is the total orbital angular momentum $\mathbf{L}$ of the $q_1q_2q_3$-system with the parity $P$ in the superscript. Experimental data are in good correspondence with values predicted for $N=0$ (basic state $l_\rho=l_\lambda=0$) and $N=1$ but not for $N=2$, in which case more than half of the expected excited states are actually absent. No baryonic state is observed within the (20,$1^+$)-multiplet. This fact is known as the “missing resonance” problem, which has still no solution within the quark framework. The situation is so serious that the authors of \cite{CR20} write: “If no new baryons are found, both QCD and the quark model will have made incorrect predictions, and it would be necessary to correct the misconceptions that led to these predictions. Current understanding of QCD would have to be modified and the dynamics within the quark model would have to be changed”.

Note that neither the Gell—Mann--Okubo formulas (\ref{GMOQ})-(\ref{GMOL}) nor the Beg—Singh formula (\ref{BS}) correctly specifies the $m_0$ value because it depends on a chosen $\SU(3)$ supermultiplet (or an $\SU(6)$ hypermultiplet). At the same time, plenty of data observed on barionic octets (see, e.g., pdg.lbl.gov) cannot be found among the values given above. As a rule, the accuracy of mass prediction by the $\SU(3)$-, $\SU(6)$- and $\SU(6)\times\GO(3)$-models is at a level of 4\%--6\% only.

As is known, the Standard Model (SM) has 18 parameters:
\[
m_e,\;m_u,\;m_d;\quad m_\mu,\;m_s,\;m_c;\quad m_\tau,\;m_b,\;m_t;
\]
\[
\theta_u,\;\theta_d,\;\theta,\;\delta;
\]
\[
M_w,\;M_h;
\]
\[
\alpha,\;\alpha_s,\alpha_W.
\]
Thirteen of them are related to fermion masses: three lepton masses $(m_e,m_\mu,m_\tau)$, six quark masses $(m_u,m_d,m_s,m_c,m_b,m_t)$, and four mixing angles $(\theta_u,\theta_d,\theta,\delta)$. However, the number of SM parameters is not rigidly fixed. One reason why the SM might need extension is neutrino oscillations taking which into account adds three neutrino masses and at least four parameters of the PMNS (Pontecorvo–Maki–Nakagawa–Sakata) neutrino mixing matrix, CKM quark mixing matrices, and possibly two more parameters in case neutrino appears to be a Majorana particle. All these parameters are derived from experiment data, they cannot be predicted within the SM framework and are thus taken as fundamental. The current list of the SM elementary particles includes three sorts of them: quarks, leptons, and gauge bosons. This motley grass of elementaries suggests that SM can be upgraded to somewhat a more perfect construction.

\subsection{Universal length and Balmer-like formulas\label{sec2_2}}

Heisenberg considered the difficulties of particle physics to be of deep principal nature related to physics fundamentals, much like the difficulties encountered by the electromagnetic theory before the discovery of relativity or the inconsistencies in describing atomic phenomena before the advent of the quantum theory. In 1938 he wrote \cite{Heisen38} that divergences are the reason why physicists cannot formulate a consistent quantum particle theory that would explain much of the data observed in nuclear physics and cosmic rays. As a remedy, Heisenberg proposed to introduce a universal constant of length dimension due account of which should eliminate the divergences \cite{Heisen38}. According to the data of nuclear experiments, this constant is comparable to the classical radius of electron
\[
r_e=\frac{1}{4\pi\epsilon_0}\frac{e^2}{m_e c^2}=2,8179403227(19)\times 10^{-15}\text{m},
\]
This value $r_e\sim r_0=2,81\cdot 10^{-13}$ defines the fundamental scale to within an order of magnitude.

This value of the universal length Heisenberg substantiated by a number of arguments: (1) a lot of particles including neutrons, protons, and heavy electrons have a mass of about $\frac{\hbar}{r_0 c}$, (2) $r_0$ is a typical range of nuclear forces, and (3) the explosion-like processes occur when the particle collision energy exceeds $\frac{\hbar c}{r_0}$  (in the center-of-mass frame). Heisenberg assumed \cite{Heisen38} that the universal length limits the applicability of the quantum theory, just as the Planck constant $\hbar$ and the speed of light $c$ limit the applicability of the classical physics.

The existence of fundamental scale $l_0\sim 10^{-13}$ cm was the basic hypothesis underlying a nonlinear quantum field theory developed by Heisenberg in the mid-1950s \cite{Heisen3}. His main idea was to find a nonlinear generalization of the Dirac equation in case where mass is represented as a field. Within this approach, all the matter (rather, the pramatter) is described by a spinor field function solving a nonlinear wave equation in which the mass term is taken nonlinear in the field function and includes scale $l_0$ as a factor:
\[
i\sigma^\nu\frac{\partial\chi(x)}{\partial x^\nu}+l^2_0\sigma^\nu:\chi(x)\left(\chi^\ast(x)\sigma_\nu\chi(x)\right):=0.
\]
Here, field function $\chi(x)$ is defined as a two-component Weyl spinor operator with respect to the Lorenz transform and as a two-component spinor in the isospin space, $\sigma^\nu$ are the Pauli matrices. Heisenberg presumed that the excited states of pramatter field $\chi(x)$ correspond to observable particle states constituting the spectrum of matter.

By Heisenberg’s thought, the value of $l_0$ also characterizes the quantum mass scale. Choosing mass unit in the form
\[
\frac{\hbar}{cl_0}=\frac{70\;\text{MeV}}{c^2}
\]
one gets rather accurate mass values for the following particles \cite{Sakharov}:
\[
\begin{array}{lcl}
\mu-\text{meson}\;=\;3/2&\quad&\Lambda-\text{hyperon}\;=\;16\\
\pi-\text{meson}\;=\;2&\quad&\Sigma-\text{hyperon}\;=\;17\\
K-\text{meson}\;=\;7&\quad&\Xi-\text{hyperon}\;=\;19\\
\eta-\text{meson}\;=\;8&\quad&\text{electron}\;=\;1/137\\
\text{proton and neutron}\;=\;13,5&\quad&\text{photon, neutrino and graviton}\;=\;0.
\end{array}
\]

{\it Balmer-like formulas}
The Heisenberg’s conjecture that the mass scale is related to fundamental length $l_0$ (made in 1938) was picked up by Yoichiro Nambu \cite{Num52} who studied experimental mass spectra of particles and paid attention to the notable Balmer-like patterns:
\begin{equation}\label{Numbu}
m_N=\left(N/2\right)137\cdot m_e,
\end{equation}
where $N$ is a positive number and $m_e$ is the electron mass. After the work of Nambu, a lot of similar data have been reported by other authors \cite{MG70,MG07,Pal07,SM08,MS08,Gro,Chiatti}.

Nambu formula (\ref{Numbu}) can be recast in terms of the fine-structure constant
\begin{equation}\label{Alpha}
m=\frac{N}{2\alpha}m_e,
\end{equation}
leading to the so-called $\alpha$-quantization of the particle mass.

In 2003, Sidharth \cite{Sid1} proposed the following empiric formula:
\begin{equation}\label{Sidharth}
\text{mass}=137\cdot m\left(n+\frac{1}{2}\right),
\end{equation}
where $m$ and $n$ are positive integers. It describes the entire mass spectrum of particles (known at the time of 2003) with an accuracy of 3\%. One can immediately notice that the Sidharth’s formula closely resembles the formula for quantum harmonic oscillations $E_n=\hbar\omega\left(n+\frac{1}{2}\right)$, $n=0,1,2,\ldots$. Sidharth tried to find a relation between numbers $m$ and $n$ in his formula and the quantum number of harmonic oscillator. However, the meaning of these numbers, as well as the meaning of number $N$ in the other Balmer-like formulas (e.g., (\ref{Numbu}), (\ref{Alpha})), is still unclear today.

In 1979, Azim Orhan Barut \cite{Bar79} proposed a mass formula for leptons in the form
\begin{equation}\label{Barut}
m(N)=m_e\left(1+\frac{3}{2}\alpha^{-1}\sum^{n=N}_{n=0}n^4\right),
\end{equation}
where $\alpha\approx 1/137$ is the fine-structure constant and number $N$ is taken equal to 0, 1, or 2 for the electron, muon, or $\tau$-lepton masses, respectively.

One more empirical mass formula relating the masses of charged leptons was proposed by Koide in 1983 \cite{Koi83}
\[
m_e+m_\mu+m_\tau=\frac{2}{3}\left(\sqrt{m_e}+\sqrt{m_\mu}+\sqrt{m_\tau}\right)^2.
\]
Curiously, the $\tau$-lepton mass predicted by Koide (1777 MeV) slightly differed from the experimental value adopted at the time of 1983 (1784 MeV) but almost perfectly fits the present data (According to PDG \cite{PDG}, the current world average value of the $\tau$-lepton mass is 1776.86 $\pm$ 0.12 MeV). For more on the Koide formula interpretations and applications, see \cite{Gsp05,Foot,Esp}.

The considerations above show that there is still no explanation for the Balmer-like patterns observed in the mass spectra. In what follows we offer a rigorous group-theoretical substantiation of these spectra and a clear interpretation of the $N$ factor in Namby formula (\ref{Numbu}).

\section{A Holistic View\label{sec3}}
In recent time, reductional ideology continuously gives ground to its antithesis, which is holistic approach. It becomes evident that quantum phenomena are essentially holistic and quantum systems are naturally nonseparable. This key fact is confirmed by numerous experimental tests of the Bell's inequalities reported by Friedman-Clauser, Aspe, Greenberg-Horn-Zailinger and others authors. The 2022 Nobel Prize in Physic was awarded to Alain Aspe, John F. Clauser, and Anton Zeilinger  “for experiments with entangled photons, establishing the violation of Bell inequalities and pioneering quantum information science”. Quantum mechanics is nonlocal and, hence, needs no reference to space-time. Since particle physics acts completely in the field of quantum phenomena, one does not need space-time to construct a particle model and should replace the traditional localization in space-time by a purely holistic framework. Looking at the historical retrospective of physics, can we find there a model which describes particles in a holistic way? The first and perhaps the only one attempt of this kind was the nonlinear spinor theory of matter proposed by Heisenberg \cite{Heisen3,Ivan}. We will consider the pro and contras of this theory in more detail in Section 3.2. Before that we will define a particle holistically and show that this holistic concept naturally embraces most of the known definitions.

\subsection{What is a particle?\label{sec3_1}}
This key question was posed by Heisenberg \cite{HeisenEP}, Schrödinger \cite{Shred50}, Markov \cite{Mark}, and other pioneers of quantum mechanics, but none of them found an irrefragable answer. Up to now, we have no commonly accepted notion of this quantum object (or better to say quantum phenomenon) but only a wide range of descriptions, sometimes rather ambiguous. An incomplete list of ways to define quantum particle is given by Wolchover \cite{Wolhover}:\\
(1) a result of the wave function collapse;\\
(2) a quantum of field excitation;\\
(3) a group irreducible representation;\\
(4) a string vibration;\\
(5) a deformation of the information ocean.

The first concept (which is also the earliest one) originates from the idea of Dirac and von Neumann about the reduction of quantum superposition, also known as the wave function collapse. As the quantum superposition of states they conceived an infinite set of probable states; then, particle is the tip of this iceberg, i.e., the state that has been actually observed. The second point will be discussed in more detail below. The third interpretation goes back to Wigner \cite{Wig39}, who described particles by irreducible representations of the Poincare group, emphasizing the main role of symmetry. Heisenberg noted that although particles are the embodiments and the simplest expressions of symmetries, they are only their consequence \cite{Heisen}. His phrase in the epigraph to this paper gets much closer to the truth than the Democritus’s ``In the beginning there was a particle''. The same idea of symmetry with an air of topology can be related to point 4. Finally, the fifth concept refers to the It-From-Bit Ur-hypothesis of von Weizsäcker \cite{Wa55,Wa92} and the space-time code hypothesis of Finkelstein \cite{Fin}. According to these ideas, the world at its primary level is composed of information bits, and all physical objects are constructs of the information fields.

Holistic approach also faces the problem of defining a particle. In contrast to the reductional framework where particles are principal elements and parts make up a whole, in holistic Universe parts are only subordinates under the whole. A part has no autonomous status but represents some state of the whole, a certain mode of its being. This view appeals to the famous theorems on substance formulated by Spinoza, the founder of European pantheism \cite{Spinoza}:\\
Theorem 1: Substance by nature is the first of its states.\\
Theorem 8: Every substance is necessarily infinite.\\
Theorem 13: Substance absolutely infinite is indivisible.\\
These are the central pillars of holism: the whole is greater than its parts, the substance (energy) is infinite, the whole (substance) is indivisible.

Thus, in the holistic approach, particle is not an individual object but rather a state of substance, which is considered as the whole. Tracing the evolution of our views on particles during the 20th century, M.A. Markov wrote: “Given a concept of field, physicists define a particle as an atom of this field, with only a replacement of "atom" for "quantum". Particle of a certain kind is the simplest element of the respective field, or simply a quantum of this field” \cite{Mark}. Here, particle is considered as a secondary concept derived from the field, which is the primary one. This definition falls into the second place on Woolhower’s list (particle as a field excitation). It is also in line with the emergent nature of particle and even with the Spinoza's Theorem 1. But the notions of field and substance are not the same. Field is the main attribute of the short-range-action concept, where the spacetime continuum is necessarily needed to carry interactions. Strictly speaking, the notion of quantum is not rigidly related to the notion of field; moreover, the discreteness of energy given by the Planck’s law contradicts the continuity of spacetime. This contradiction is absent in holistic approach, where particle is not a quantum of field, but a quantum of energy.

Observe that none of the five formulations mentioned in the previous section claims to be complete but captures (conceivably, except for the fourth one) some essential feature of a quantum object. Below we show that each of the first three points of the Woolhower’s list is a special case of the holistic description of particle as a mode of substance.
\begin{itemize}
    \item The notion of particle as a result of {\it wave function collapse} leads one to the Heisenberg-Fock concept \cite{Heisenberg,Fock} where (with reference to Aristotle) the being has two different ontological statuses: $\delta\upsilon\nu\alpha\mu\sigma$ (dunamis), which is the being in potency, and $\varepsilon\nu\tau\varepsilon\lambda\varepsilon\chi\varepsilon\iota\alpha$ (entelechy), which is the being in reality, the manifested world. A quantum object (aka a quantum of energy) arises in the mode of reality from the collapse of a quantum superposition, which exists in potency as a nonseparable state of the substance taken as a whole \cite{Var15a}.
    \item The representation of a particle as a {\it perturbation of a quantized field} grasps the emergent nature of states; but in holistic terms the states of a field are replaced by the states of the substance (energy).
    \item Finally, the Wigner's interpretation of particle as {\it an irreducible representation of a group} is naturally inherent to the holistic description by virtue of the canonical state-representation correspondence $\omega\leftrightarrow\pi_\omega$ of the Gelfand-Naimark-Segal (GNS) construction.
\end{itemize}

\subsection{Heisenberg’s theory of spinor matter\label{sec3_2}}

Heisenberg was one of those who came very close to the holistic understanding of quantum objects. In \cite{Heisen52}, he noted that {\it particle} is just a way for a {\it unified field} to be realized in a certain state: “... the future theory of elementary particles will have to start with a unified field, referring simply to {\it matter} or {\it energy}, not to any specific type of particle; for this unified field some kind of exchange relations and field equations may be given, which lead to the existence of continuous and discrete eigenvalues. The discrete eigenvalues describe {\it particles} which may be called elementary or compound particles according to convenience without sharp distinction between both definitions”. Here, the original {\it unified field} is conceived as energy/matter and its states defined by discrete eigenvalues (energy levels) are attributed to {\it particles}, which are thus viewed as secondary with respect to the matter, i.e., as the modes of its being. In fact, a strong echo of the Spinoza's Theorem 1 is heard in the Heisenberg’s proposal \cite{Heisen3} to conceive elementary particles as various forms in which fundamental substance, be that matter or energy, can exist.

However, it was not easy to define rigorously an equation for pramatter $\chi(x)$: first, Heisenberg believed that it should be formulated on the spacetime continuous background; in addition, a Hilbert space with an indefinite metric was required to introduce the nonlinear term which generates the spectrum of field $\chi(x)$ excitations. Elementary particles were considered as these excitations. Even at that time, at the end of 1950s, this spectrum-generating framework appeared to be too narrow to describe the increasingly growing number of the observed particles. Possibly, Bohr had something like that in mind when he doubted whether the Heisenberg’s theory was “crazy enough to be true”.
%{Footnote: This happened in 1958 at Columbia University when Pauli had been giving a lecture on Heisenberg's nonlinear spinor theory of matter. It is known that in 1957 Pauli joined Heisenberg in studying the group properties of his equation. In \cite{HeisenUTF} Heisenberg noted that Pauli had a firm hope that this equation, being uniquely simple and highly symmetrical, should be the right starting point from which a unified field theory of particles should be developed}.

Heisenberg relied on the field concept and defined his nonlinear equation for the field of matter on the spacetime continuum, which he took as a fundamental background. In our opinion, this was the main cause of a deficient spectrum generation mechanism in the spinor theory of matter.
%The feeling of some impedance is expressed by L. de Broglie in his celebrated phrase that trying to squeeze microscopic phenomena into the spacetime framework is like trying to squeeze a diamond into a frame that does not suit it \cite{deBroglie}.
Since that time, the fundamental status of spacetime has been increasingly doubted and recently has transformed into a well-supported belief that continuum is an emergent construction (twistor program of R.Penrose \cite{Penrose}, binary systems of complex relations by Yu.S. Vladimirov \cite{Vladimirov}, Zurek's theory of decoherence \cite{Zurek}, hierarchy of spacetime emergency levels by D. Oriti \cite{Oriti}.

%\s more, it should break the continuum framework and go beyond it.

\section{Theory of the Spectrum of Matter\label{sec4}}
Despite of some difficulties, the Heisenberg's nonlinear spinor theory of matter has a very important merit: it explicitly introduces a holistic description into the quantum world. The Democritus’s {\it In the beginning there was a particle} is replaced by {\it In the beginning there was symmetry (energy)}, or in other words, {\it the whole precedes its parts}. In this hierarchy, particles are not fundamental building blocks of the universe but emergent, derivative states (modes) of the matter. Accordingly, the spectrum of particles loses its fundamental meaning and gives place to the {\it spectrum of matter} (or, the spectrum of substance).

\subsection{Axiomatics\label{sec4_1}}
Following Heisenberg, we assume that the major observable at the fundamental microscopic level is energy, which is quantized in accordance with the Planck's law. Note that the continuum concept is out of the play from the very beginning. The fundamental level is understood as one characterized by the elementary length $l_0$ scale. Particle is interpreted as a quantum of energy and described by an algebraic quantization procedure known as the Gelfand-Naimark-Segal (GNS)-construction, where the energy operator $H$ is considered as a $C^\star$-algebra operator. We use the following system of definitions (axioms) \cite{Var18}:

\begin{enumerate}%[leftmargin=*,labelsep=6.9mm,topsep=3pt]

\item[\textbf{A.I}] \textrm{(Energy and fundamental symmetry)} %MDPI: please confirm if ir is assumption or list, if assumption, please use \begin{Assumption}...\end{Assumption}. It is list
 \textit{A single quantum system $\bsU$ is characterized at the fundamental level by a unitial $C^\ast$-algebra $\fA$ consisting of the energy operator $H$ and the $H$-adjoint generators of fundamental symmetry group $G_f$, which have a common with $H$ set of eigenfunctions.} %Is the italics necessary?

\item[\textbf{A.II}] \textrm{(States)} \textit{A physical state of $C^\ast$-algebra $\fA$ is determined by the cyclic vector $\left|\Phi\right\rangle$ of the $C^\ast$-algebra representation $\pi$ on separable Hilbert space $\sH_\infty$:
\[
\omega_\Phi(H)=\frac{\langle\Phi\mid\pi(H)\Phi\rangle}{\langle\Phi\mid\Phi\rangle}.
\]
The set $PS(\fA)$ of all pure states of $C^\ast$-algebra $\fA$ coincides with the set of all states $\omega_\Phi(H)$ associated with all irreducible cyclic representations $\pi$ of $\fA$, $\left|\Phi\right\rangle\in\sH_\infty$ (the Gelfand--Naimark--Segal construction)}.

\item[\textbf{A.III}] \textrm{(Space of rays)} \textit{The set of all positive pure states $\omega_\Phi(H)\geq 0$ forms a physical $\K$-Hilbert space $\bsH_{\rm phys}(\K)$ (a GNS-Hilbert space equipped with the $\K$-structure of $\ast$-ring). For each state vector $\left|\Psi\right\rangle\in\bsH_{\rm phys}$ there is a unit ray $\boldsymbol{\Psi}=e^{i\alpha}\left|\Psi\right\rangle$, where $\alpha$ runs through all real numbers and $\sqrt{\left\langle\Psi\right.\!\left|\Psi\right\rangle}=1$. The ray space is the quotient space $\hat{H}=\bsH_{\rm phys}/S^1$, that is, the projective space of one-dimensional subspaces of $\bsH_{\rm phys}$. All states of a single quantum system $\bsU$ are described by the unit rays}.

\item[\textbf{A.IV}] (Axiom of spectrality) \textit{Space $\hat{H}$ contains a complete set of states with non-negative energy}.

\item[\textbf{A.V}] (Superposition principle) \textit{The basic correspondence between physical states and elements of space $\hat{H}$ involves the superposition principle of quantum theory; that is, there is a set of basic states such that arbitrary states can be constructed from them using linear superpositions}.
\end{enumerate}

A-I. The first axiom defines basic ingredients of the formalism. According to von Neumann \cite{Neu64}, the primary (undefinable) concepts of quantum mechanics are system, observable, and state. As an observable, we take a $C^\ast$-algebra consisting of the energy operator $H$ and the adjoint generators of the fundamental symmetry group $G_f$. This group $G_f$ provides the structure of energy levels since the spaces of irreducible representations of $G_f$ coincide with the eigenspaces of energy operator $H$. This is precisely the symmetry which Heisenberg assigned to be in the beginning. In other words, group $G_f$ specifies a template for the spectrum-generating mechanism provided by GNS-construction. For $C^\ast$-algebra of observables $\fA$ and a set of states $S(\fA)$, we identify group $G_f$ as the fundamental symmetry group of $\fA$ if the map $g\rightarrow(\alpha_g,\alpha^\prime_g)$ of group $G_f$ into the group of all symmetries of $(\fA,S(\fA))$ is a homomorphism, where bijections $\alpha:\;\fA\rightarrow\fA$ and $\alpha^\prime:\;S(\fA)\rightarrow S(\fA)$ satisfy the matching condition $(\alpha^\prime\omega)(\alpha A)=\omega(A)$ for all $A\in\fA$, $\omega\in S(\fA)$. From here on we assume that $G_f$ is a non-compact Lie group (for example, a Lorentz group or a conformal group). Then, for any physical state $\omega\in S(\fA)$ and any fixed $A\in\fA$, the map $g\rightarrow\omega(\alpha_g(A))$  is continuous in $g$. Moreover, the group $G_f$  is unitarily or antiunitarily realized if there exists a continuous representation $g\rightarrow U_g$ of $G_f$ by unitary or antiunitary operators (according to whether $\alpha_g$ are algebraic automorphisms or antiautomorphisms) in the Hilbert space $\sH_\infty$ such that $\alpha_g(A)=U_gA^{(\ast)}U^{-1}_g$ for any $A\in\fA$, $g\in G$, and $A^{(\ast)}$ standing for $A$ or $A^\ast$ in case $U_g$ is unitary or antiunitary, respectively. In terms of the universal covering of fundamental group $\widetilde{G}_f=\pin(p,q)$, the $CPT$-group is given by automorphisms and antiautomorphisms of the Clifford algebra $\cl_{p,q}$ \cite{Var01,Var04,Var05,Var12,Var15}. The structure of automorphism $\cA\rightarrow\overline{A}$ specifies the charge conjugation $C$ and determines the charge distribution of physical states in the $\K$-Hilbert space.

A-II. The second axiom establishes the GNS mechanism of spectrum generation. The arising Hilbert space is an emergent construction whose explicit form depends on the choice of group $G_f$ (the so-called {\it dressing} of the operator algebra). For any state $\omega$ on a $C^\ast$-algebra $\fA$, GNS construction defines a cyclic representation $\pi_\omega$ of $\fA$ on the Hilbert space $\sH_\infty$ with a cyclic vector $\left|\Phi\right\rangle$ so that $\omega(\fa)=\langle\Phi\mid\pi_\omega(\fa)\mid\Phi\rangle$, $\forall\fa\in\fA$. These conditions define a representation $\pi_\omega$ that is unique up to unitary equivalence (which relates cyclic vectors of different representations). Further, each state $\omega$ defines a certain representation of algebra $\fA$, and the resulting representation $\pi_\omega$ is irreducible exactly when $\omega$ is {\it pure}.

A-III. The third axiom defines the physical $\K$-Hilbert space ($\K=\R,\C,\BH$), where the canonical correspondence $\omega\leftrightarrow\pi_\omega$ allows one to realize the Dyson “threefold way” \cite{Dys62,Baez}, which is the symmetry between quaternion, complex, and real representations of group $G_f$. With the respective representations we identify the charged ($\C$), neutral ($\BH$), and purely neutral ($\R$) states of the spectrum of matter, while the Dyson symmetry provides the dynamic relations between spin, charge, and mass in terms of tensor product \cite{Var20}.

Thus, the axioms {\bf A.I-A.III} define a single quantum system $\bsU$ consisting of a $C^\ast$-algebra (built on the energy operator $H$ and the group $G_f$ generators) and a $\K$-Hilbert space $\bsH_{\rm phys}(\K)$ generated by GNS construction. The cyclic vectors of space $\bsH_{\rm phys}(\K)$ represent all possible states of $\bsU$, thus giving the spectrum of matter. For a specific choice of group $G_f$, eigenvectors related to the discrete eigenvalues of $H$ define the stationary states of $\bsU$. In particular, for $G_f=\SO_0(1,3)$ (Lorentz group), the spectrum of matter is the spectrum of {\it elementary particles} \cite{Var17,Var17a}, whose complete list is given by the Particle Data Group \cite{PDG}. For $G_f=\SO_0(2,4)$, where $\SO_0(2,4)$ is a conformal group, the spectrum of matter is structured as the periodic system of {\it chemical elements} \cite{Fet,Var18b,Var19,Var19a,VPB22}. In this case, the stationary states of $\bsU_A$ correspond to the atoms of chemical elements. Restricting the group $\SO_0(2,4)$ to the subgroup $SO_0(1,3)$ we get the reduction $\bsU_A\rightarrow\bsU_E$, where the resulting subsystem $\bsU_E$ again corresponds to elementary particles, this time represented by the cyclic vectors of space $\bsH_{\rm phys}(\K)$. This fact supports the Weisskopf’s conjecture that the nuclear and particle physics are not separate disciplines but a single science. The quantum system $\bsU_A$ corresponds to a higher symmetry group $G_f$ than its subsystem $\bsU_E$ and, hence, describes a higher level of the organization of matter.

\subsection{Symmetries of the Spectrum of Matter\label{sec4_2}}

Note that the symmetries of the spectrum of matter can be classified into three types: (1) {\it fundamental symmetries} $G_f$ responsible for the pure states (rays) of the quantum system $\bsU$ and the corresponding coherent subspaces of $\bsH_{\rm phys}(\K)$; (2) {\it dynamic symmetries} $G_d$ describing transitions between the states of different coherent subspaces; and (3) {\it gauge symmetries} $G_g$ relating pure states within a coherent subspace. All the dynamical symmetries $G_d$ (the so-called {\it internal} symmetries of compact groups $\SU(n)$) can be lifted to the $\K$-Hilbert space by means of the central extension technique \cite{BR86,Var15b}. The result includes the $\SU(3)$-, $\SU(4)$-, and $\SU(5)$-systematics of hadronic spectra as special cases. The dynamic symmetries $G_d=\SU(n)$ relate different cyclic vectors of the $\K$-Hilbert space, i.e. set quantum transitions between states (levels of the spectrum of matter). It is natural to assume that the operators of group $G_d$  and its subgroups relate states with similar characteristics; for this reason, the approximate group $G_d$ symmetries should be called {\it external} rather than {\it internal} symmetries. In the case of atomic system $\bsU_A$, the similar relation is provided by operators $\Gamma_+$ and $\Gamma_-$ connecting the homological series of the periodic system of elements (see \cite{Var19,VPB22}). Obviously, such an {\it external} description of the $G_d$ groups turns the ({\it internal}) quark composition of hadrons into a fiction.

\subsection{The Hilbert Space structure\label{sec4_3}}
Below we show that a general vector of a physical $\K$-Hilbert space $\bsH_{\rm phys}(\K)$ can be represented as a product of elementary vectors of two types. An odd/even product yields a fermionic/bosonic state, correspondingly.
\begin{thm} ({\rm \cite{KspinorVar}})
A physical $\K$-Hilbert space $\bsH_{\rm phys}(\K)$ can be expanded into a direct sum of (non-zero) coherent subspaces
\[
\bsH_{\rm phys}(\K)=\underset{\nu\in N}{\bigoplus}\bsH^\nu_{\rm phys}(\K).
\]
In this case, the superposition principle takes place in a restricted form, i.e. for coherent subspaces $\bsH^\nu_{\rm phys}(\K)$. A non-zero linear combination of cyclic vectors of pure separable states is a cyclic vector of a pure separable state, provided that all the original vectors lie in the same coherent subspace $\bsH^\nu_{\rm phys}(\K)$. A superposition of cyclic vectors of pure separable states from different coherent subspaces is a cyclic vector of a mixed state.
\end{thm}
 In the state-of-the-art high-energy physics, the superselection rules can be completely described by electric $Q$ ($=Q_1$), baryon $B$ ($=Q_2$), and lepton $L$ ($=Q_3$) charges. The electric charge is taken into account by the $\K$-linear structure of the space $\bsH_{\rm phys}(\K)$, see \cite{KspinorVar}. In order to describe the entire spectrum of observed states (spectrum of matter), we introduce a 2-parameter gauge group $G=U(1)^2\equiv U(1)\times U(1)$ with respect to the baryon $B$ and lepton $L$ charges. Then the total space $\bsH_{\rm phys}(\K)$ is a sum of coherent subspaces
\begin{equation}\label{KS}
\bsH_{\rm phys}(\K)=\bigoplus_{b,\ell\in\dZ}\bsH^{(b,\ell)}_{\rm phys}(\K),\quad\K=\R,\C,\BH,
\end{equation}
with a certain baryon $b$ and lepton $\ell$ numbers. Consequently, the entire set of cyclic vectors of $\bsH_{\rm phys}(\K)$ can be divided into subspaces of vectors of the form
\begin{equation}\label{Vector}
\left|\K,b,\ell,s\right\rangle=\left|\K,b,\ell,|l-\dot{l}|\right\rangle
\end{equation}
with given values of charge, spin, mass, baryon and lepton numbers. Each vector (10) is associated with a $CPT$ group, see \cite{Var04}.

\begin{thm} ({\rm \cite{KspinorVar}}) All cyclic vectors $\left|\psi\right\rangle$ of the physical $\K$-Hilbert space $\bsH_{\rm phys}(\K)$, which define fermionic and bosonic states of $\R$-, $\C$-, and $\BH$-subspaces, are given by a composition of the fusion and doubling operations acting on the active $\left|\mathfrak{q}_a\right\rangle$ and inert $\left|\mathfrak{q}_s\right\rangle$ elementary states.
\end{thm}
Any even-dimensional algebra $\cl_{p,q}$ over the field $\F=\R$ is isomorphic to the tensor product of two-dimensional algebras $\cl_{0,2}$ ($\K\simeq\BH$) and $\cl_{1,1}$, $\cl_{2,0}$ ($\K\simeq\R$). Accordingly, we have \textit{\textbf{two types of elementary states}}
\begin{description}
\item[I.] \phantom{I}$\left|\mathfrak{q}_a\right\rangle=\left|\BH,0,1,\frac{1}{2}\right\rangle$,
$\left|\bar{\mathfrak{q}}_a\right\rangle=\left|\overline{\BH},0,-1,\frac{1}{2}\right\rangle$ active state, and
\item[II.] $\left|\mathfrak{q}_s\right\rangle=\left|\R,0,0,\frac{1}{2}\right\rangle$,
$\left|\bar{\mathfrak{q}}_s\right\rangle=\left|\mathfrak{q}_s\right\rangle$ inert state.
\end{description}

In terms of the binary structure, there are three basic operations on states with a minimal tensor dimension:\\
1) \textit{\textbf{Fusion}}.
\[
\left|\mathfrak{q}\right\rangle\otimes\left|\bar{\mathfrak{q}}\right\rangle=\left|\BH,0,1,\frac{1}{2}\right\rangle\otimes
\left|\overline{\BH},0,-1,\frac{1}{2}\right\rangle=\left|\BH\otimes\overline{\BH},0,0,1\right\rangle=
\left|\R,0,0,1\right\rangle=\left|\gamma\right\rangle.
\]
2) \textit{\textbf{Doubling}}.
\begin{eqnarray}
\left|e^-\right\rangle&=&\left|\mathfrak{q}\right\rangle\oplus\left|\bar{\mathfrak{q}}\right\rangle=
\left|\BH\oplus i\BH,0,1,\frac{1}{2}\right\rangle=\left|\C,0,1,\frac{1}{2}\right\rangle,\nonumber\\
\left|e^+\right\rangle&=&\left|\mathfrak{q}\right\rangle\ominus\left|\bar{\mathfrak{q}}\right\rangle=
\left|\BH\ominus i\BH,0,-1,\frac{1}{2}\right\rangle=\left|\overline{\C},0,-1,\frac{1}{2}\right\rangle.\nonumber
\end{eqnarray}
3) \textit{\textbf{Annihilation}}.
\begin{multline}
\left|e^-\right\rangle\otimes\left|e^+\right\rangle=\left|\C,0,1,\frac{1}{2}\right\rangle\otimes
\left|\overline{\C},0,-1,\frac{1}{2}\right\rangle=\\
=\left|\BH\oplus i\BH,0,1,\frac{1}{2}\right\rangle\otimes\left|\BH\ominus i\BH,0,-1,\frac{1}{2}\right\rangle=
\left|\BH\otimes\BH,0,0,1\right\rangle+i\left|\BH\otimes\BH,0,0,1\right\rangle-\\
-i\left|\BH\otimes\BH,0,0,1\right\rangle+\left|\BH\otimes\BH,0,0,1\right\rangle=2\left|\R,0,0,1\right\rangle=
2\left|\gamma\right\rangle.\nonumber
\end{multline}
The latter operation is an algebraic analog of the electron-positron annihilation: $e^-e^+\rightarrow 2\gamma$. For more details see \cite{KspinorVar}.

According to Theorem 2, the \textit{fermionic states} $F$ correspond to cyclic vectors $\boldsymbol{\tau}_{\frac{k}{2},0}\otimes\boldsymbol{\tau}_{0,\frac{r}{2}}\left|\omega\right\rangle$ with an odd number of cofactors  $\boldsymbol{\tau}_{\frac{1}{2},0}$ (resp. $\boldsymbol{\tau}_{0,\frac{1}{2}}$) in the tensor product. In turn, the \textit{bosonic states} $B$ correspond to cyclic vectors with an even number of cofactors. Therefore,
\begin{equation}\label{State}
\underbrace{\boldsymbol{\tau}_{\frac{1}{2},0}\otimes\boldsymbol{\tau}_{\frac{1}{2},0}\otimes\cdots\otimes
\boldsymbol{\tau}_{\frac{1}{2},0}\bigotimes
\boldsymbol{\tau}_{0,\frac{1}{2}}\otimes\boldsymbol{\tau}_{0,\frac{1}{2}}\otimes\cdots\otimes
\boldsymbol{\tau}_{0,\frac{1}{2}}}_{m\;\text{times}}\Rightarrow\left\{\begin{array}{lc}
F,& m\equiv 1\pmod{2};\\
B, & m\equiv 0\pmod{2}.
\end{array}\right.
\end{equation}
It follows that the cyclic vector of the $\K$-Hilbert space, representing bosons or fermions depending on $m\equiv 0,1\pmod{2}$, have a separabel structure.

\subsection{Lie algebra $\mathfrak{sl}(2,\C)$, Weyl diagrams, and the mass formula\label{sec4_4}}

The approach introduced above implies two levels of the organization of matter (a single quantum system $\bsU$): 1) $\bsU_E$ with $G_f=\SO_0(1,3)$ and 2) $\bsU_A$ with $G_f=\SO_0(2,4)$. Here we study the quantum system $\bsU_E$, in which case the spectrum of matter corresponds to the spectrum of elementary states (``particles'') and the fundamental symmetry is set by the Lorentz group.

As is known, the proper orthochronous Lorentz group $\SO_0(1,3)$
(rotation group of the Minkowski space-time $\R^{1,3}$) has a universal covering, that is the spinor group
\[\ar
\spin_+(1,3)\simeq\left\{\begin{pmatrix} \alpha & \beta \\ \gamma &
\delta
\end{pmatrix}\in\C_2:\;\;\det\begin{pmatrix}\alpha & \beta \\ \gamma & \delta
\end{pmatrix}=1\right\}=\SL(2,\C).
\]

Let $\fg\rightarrow T_{\fg}$ be an arbitrary linear
representation of the proper orthochronous Lorentz group
$\SO_0(1,3)$ and $\sA_i(t)=T_{a_i(t)}$ be an infinitesimal
operator corresponding to a rotation $a_i(t)\in\SO_0(1,3)$.
Analogously, let $\sB_i(t)=T_{b_i(t)}$, where $b_i(t)\in\SO_0(1,3)$ is
a hyperbolic rotation. The elements $\sA_i$ and $\sB_i$ form a basis of the group algebra
$\mathfrak{sl}(2,\C)$ and satisfy the commutation relations
\begin{equation}\label{Com1}
\left.\begin{array}{lll} \ld\sA_1,\sA_2\rd=\sA_3, &
\ld\sA_2,\sA_3\rd=\sA_1, &
\ld\sA_3,\sA_1\rd=\sA_2,\\[0.1cm]
\ld\sB_1,\sB_2\rd=-\sA_3, & \ld\sB_2,\sB_3\rd=-\sA_1, &
\ld\sB_3,\sB_1\rd=-\sA_2,\\[0.1cm]
\ld\sA_1,\sB_1\rd=0, & \ld\sA_2,\sB_2\rd=0, &
\ld\sA_3,\sB_3\rd=0,\\[0.1cm]
\ld\sA_1,\sB_2\rd=\sB_3, & \ld\sA_1,\sB_3\rd=-\sB_2, & \\[0.1cm]
\ld\sA_2,\sB_3\rd=\sB_1, & \ld\sA_2,\sB_1\rd=-\sB_3, & \\[0.1cm]
\ld\sA_3,\sB_1\rd=\sB_2, & \ld\sA_3,\sB_2\rd=-\sB_1. &
\end{array}\right\}
\end{equation}

Now construct a \textit{complex shell} of $\mathfrak{sl}(2,\C)$ through linear combinations
\begin{equation}\label{Shell}
\bsX=\frac{1}{2}i\left(\bsA+i\bsB\right),\quad\bsY=\frac{1}{2}i\left(\bsA-i\bsB\right),
\end{equation}
and commutation relations
\[
\ld\bsX,\bsY\rd=0,
\]
and
\[
\ld\sX_i,\sX_j\rd=i\varepsilon_{ijk}\sX_k,\quad\ld\sY_i,\sY_j\rd=i\varepsilon_{ijk}\sY_k.
\]
The generators $\bsX$ and $\bsY$ form a basis of two independent algebras $\mathfrak{so}(3)$. Thus, Lie algebra $\mathfrak{sl}(2,\C)$ is isomorphic to the direct sum of Lie algebras $\mathfrak{su}(2)$ (the so-called ``unitary trick'' of Weyl, see \cite[c.~28]{Knapp})
\begin{equation}\label{Sum2}
\mathfrak{sl}(2,\C)\simeq\mathfrak{su}(2)\oplus i\mathfrak{su}(2).
\end{equation}

We now define the \textit{Cartan subalgebra} $\fK$ and the corresponding  \textit{Weyl diagram} for the Lie algebra $\mathfrak{sl}(2,\C)$. To this end, it is necessary to move from the basis of the complex shell $\lf\sX_1,\sX_2,\sX_3,\sY_1,\sY_2,\sY_3\rf$ to the \textit{Cartan-Weyl basis}. The first step is to determine the maximal subset of mutually commuting generators of the algebra $\mathfrak{sl}(2,\C)$. Since each subalgebra $\mathfrak{su}(2)$ in the direct sum (\ref{Sum2}) is a Lie algebra of rank 1, it is natural to expect that the algebra $\mathfrak{sl}(2,\C)$ contains at most two commuting generators. Out of nine possible pairs of commuting generators $\lf\sX_i,\sY_j\rf$ ($i,j=1,2,3$), we choose a pair $\lf\sX_3,\sY_3\rf$ satisfying condition (\ref{Per0}) (see Appendix A), i.e.
\begin{equation}\label{Ksl2}
\ld\sX_3,\sY_3\rd=0.
\end{equation}
The set of Cartan generators $\lf\sX_3,\sY_3\rf$ forms a Cartan subalgebra $\fK\subset\mathfrak{sl}(2,\C)$ of dimension 2, which determines also the \textit{rank} of the Lie algebra $\mathfrak{sl}(2,\C)$.

Next, the remaining generators $\sX_1$, $\sX_2$, $\sY_1$, $\sY_2$ are used to form the \textit{rising} and \textit{lowering} operators (the \textit{Weyl generators}):
\begin{equation}\label{GenW2}
\left.\begin{array}{cc}
\sX_+=\sX_1+i\sX_2, & \sX_-=\sX_1-i\sX_2,\\[0.1cm]
\sY_+=\sY_1+i\sY_2, & \sY_-=\sY_1-i\sY_2,
\end{array}\right\}
\end{equation}
Four Weyl generators $\bsE_\alpha=\lf\sX_\pm,\sY_\pm\rf$ and two Cartan generators $\bsH_i=\lf\sX_3,\sY_3\rf$ of the subalgebra $\fK\subset\mathfrak{sl}(2,\C)$ form the \textit{Cartan-Weyl basis} of the algebra $\mathfrak{sl}(2,\C)$:
\[
\lf\sX_3,\sY_3,\sX_+,\sX_-,\sY_+,\sY_-\rf.
\]
These six operators satisfy the commutation relations
\begin{equation}\label{Per2s}
\ld\sX_3,\sX_+\rd=\sX_+,\quad\ld\sX_3,\sX_-\rd=-\sX_-,\quad\ld\sX_+,\sX_-\rd=2\sX_3,
\end{equation}
\begin{equation}\label{Per3s}
\ld\sY_3,\sY_+\rd=\sY_+,\quad\ld\sY_3,\sY_-\rd=-\sY_-,\quad\ld\sY_+,\sY_-\rd=2\sY_3,
\end{equation} where $\forall i,j=1,2;\;\alpha=1\rightarrow 4$:
In terms of root vectors (\ref{Per2}), the commutator $\ld\bsE_\alpha,\bsE_{-\alpha}\rd=\ld\sX_+,\sX_-\rd$ defines two roots $\pm 2$. Thus, in accordance with (\ref{Roots}) we have four different roots: $\alpha=\pm 1,\,\pm 2$.

Since $\mathfrak{sl}(2,\C)$ is a Lie algebra of rank 2, it follows from Racah Theorem (see Appendix A) that there are two independent \textit{Casimir invariants} $\bsC_\mu$ that commute with all generators of $\mathfrak{sl}(2,\C)$, including two Cartan elements $\bsH_i$:
\begin{equation}\label{Csl4}
\ld\bsC_\mu,\bsH_i\rd=0,\quad\forall\mu=1\rightarrow 2;\;i=1\rightarrow 2.
\end{equation}
The Casimir invariants $\bsC_1$ and $\bsC_2$ have the explicit form
\begin{eqnarray}
\bsC_1&\equiv&\bsX^2=\sX^2_1+\sX^2_2+\sX^2_3=\frac{1}{4}\left(\bsA^2-\bsB^2+2i\bsA\bsB\right),\nonumber\\
\bsC_2&\equiv&\bsY^2=\sY^2_1+\sY^2_2+\sY^2_3=\frac{1}{4}\left(\bsA^2-\bsB^2-2i\bsA\bsB\right).\nonumber
\end{eqnarray}
These invariants, better known as \textit{Laplace-Beltrami operators}, lead to the Fuchs class differential equations for hyperspherical functions \cite{Var}. The Laplace-Beltrami operators contain the Casimir operators of the Lorentz group $\bsA^2-\bsB^2$ and $\bsA\bsB$ as the real and imaginary parts.

By virtue of (\ref{Ksl2}) and (\ref{Csl4}), a complete set of states is defined in the proper subspace $\sH_E$ of the energy operator $H$, which are simultaneously the eigenstates of the operators $\bsX^2$, $\bsY^2$, $\sX_3$ and $\sY_3$. Let's make a ket-vector $\left|l,\dot{l};m,\dot{m}\right\rangle$. It should be noted that $l$ and $\dot{l}$ are not quantum numbers, but only \textit{set} them, and the real quantum numbers, i.e. the eigenvalues of the Casimir operators $\bsX^2$ and $\bsY^2$ are $l(l+1)$ and $\dot{l}(\dot{l}+1)$ according to the following relations:
\begin{eqnarray}
\bsX^2\left|l,\dot{l};m,\dot{m}\right\rangle&=&l(l+1)\left|l,\dot{l};m,\dot{m}\right\rangle,
\nonumber\\
\bsY^2\left|l,\dot{l};m,\dot{m}\right\rangle&=&\dot{l}(\dot{l}+1)\left|l,\dot{l};m,\dot{m}\right\rangle,
\nonumber
\end{eqnarray}
where $l,\dot{l}\in\lf 0,\frac{1}{2},1,\frac{3}{2},\ldots\rf$. Each subspace $\sH_E$ has dimension $(2l+1)(2\do{l}+1)$, which implies that
\begin{eqnarray}
\sX_3\left|l,\dot{l};m,\dot{m}\right\rangle&=&m\left|l,\dot{l};m,\dot{m}\right\rangle,\label{X3_2}\\
\sY_3\left|l,\dot{l};m,\dot{m}\right\rangle&=&\dot{m}\left|l,\dot{l};m,\dot{m}\right\rangle\label{Y3_2}
\end{eqnarray}
with $m\in\lf-l,-l+1,\ldots,l-1,l\rf$ and $\dot{m}\in\lf-\dot{l},-\dot{l}+1,\ldots,\dot{l}-1,\dot{l}\rf$. The eigenvalues $m$ and $\dot{m}$ are the \textit{weights} of Cartan generators $\sX_3$ and $\sY_3$.

The \textit{Weyl diagrams} of the group $\SL(2,\C)$ are plotted in a two-dimensional orthogonal coordinate system formed by the generators $\sX_3$ and $\sY_3$ of Cartan subalgebra $\fK=\lf\sX_3,\sY_3\rf$. Weights $m$ and $\dot{m}$ are used as coordinates to plot each state of the $\SL(2,\C)$-multiplet in the plane $(\sX_3,\sY_3)$, i.e. a two-dimensional \textit{weight vector} $\boldsymbol{h}=(m,\dot{m})$ goes from the origin of the coordinate system to the state $\left|l,\dot{l};m,\dot{m}\right\rangle$. Weyl generators $\bsE_\alpha=\lf\sX_\pm,\sY_\pm\rf$ act on a ket-vector $\left|l,\dot{l};m,\dot{m}\right\rangle$ by shifting the eigenvalues $m$ and $\dot{m}$ by an amount given by the roots $\alpha_1$ and $\alpha_2$ of the respective commutation relation between this Weyl generator and Cartan generators $\sX_3$ and $\sY_3$:
\[
\bsE_\alpha\left|l,\dot{l};m,\dot{m}\right\rangle\;\longrightarrow\;
\left|l,\dot{l};m+\alpha_1,\dot{m}+\alpha_2\right\rangle.
\]
As an example, consider the action of generator $\sX_+$ on a state $\left|l,\dot{l};m,\dot{m}\right\rangle$. By virtue of (\ref{Per1}), (\ref{Per2s}) and (\ref{X3_2}),
\begin{eqnarray}
\sX_3\sX_+\left|l,\dot{l};m,\dot{m}\right\rangle&=&\left(\ld\sX_3,\sX_+\rd+\sX_+\sX_3\right)
\left|l,\dot{l};m,\dot{m}\right\rangle\nonumber\\
&=&\left(\sX_++ m\sX_+\right)\left|l,\dot{l};m,\dot{m}\right\rangle\nonumber\\
&=&\left(m+1\right)\sX_+\left|l,\dot{l};m,\dot{m}\right\rangle.\nonumber
\end{eqnarray}
Thus, $\sX_+$ increases the eigenvalue $m$ by $+1$, which is equal to the root of the generator $\sX_+$ relative to the Cartan generator $\sX_3$ according to (\ref{Per2s}). Similarly, using (\ref{Y3_2}), we obtain
\begin{eqnarray}
\sY_3\sX_+\left|l,\dot{l};m,\dot{m}\right\rangle&=&\left(\ld\sY_3,\sX_+\rd+\sX_+\sY_3\right)
\left|l,\dot{l};m,\dot{m}\right\rangle\nonumber\\
&=&\left(0+\dot{m}\sX_+\right)\left|l,\dot{l};m,\dot{m}\right\rangle\nonumber\\
&=&\dot{m}\sX_+\left|l,\dot{l};m,\dot{m}\right\rangle,\nonumber
\end{eqnarray}
whence the generator $\sX_+$ leaves the eigenvalue $\dot{m}$ unchanged. Therefore,
\[
\sX_+\left|l,\dot{l};m,\dot{m}\right\rangle\;\longrightarrow\;\left|l,\dot{l};m+1,\dot{m}\right\rangle.
\]
The actions of the other three Weyl generators are defined similarly:
\[
\sX_-\left|l,\dot{l};m,\dot{m}\right\rangle\;\longrightarrow\;\left|l,\dot{l};m-1,\dot{m}\right\rangle,
\]
\[
\sY_+\left|l,\dot{l};m,\dot{m}\right\rangle\;\longrightarrow\;\left|l,\dot{l};m,\dot{m}+1\right\rangle,
\]
\[
\sY_-\left|l,\dot{l};m,\dot{m}\right\rangle\;\longrightarrow\;\left|l,\dot{l};m,\dot{m}-1\right\rangle.
\]
Let us graphically show the action of the generators $\bsE_\alpha$ on the \textit{root diagram}. To do this, we take the roots $\alpha_1$ and $\alpha_2$ of each Weyl element $\bsE_\alpha$ as components of a 2-dimensional \textit{root vector} $\boldsymbol{\alpha}=(\alpha_1,\alpha_2)$ and depict it in a 2-dimensional \textit{weight space} $(\sX_3,\sY_3)$. This gives the root diagram of the Lie algebra $\mathfrak{sl}(2,\C)$ shown on Fig. 1, where for simplicity we denote various root vectors $\boldsymbol{\alpha}$ by the corresponding symbol of the Weyl generator $\bsE_\alpha$.
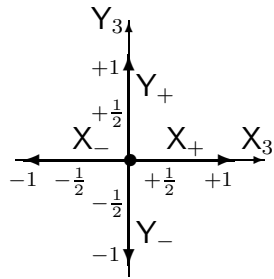
\begin{figure}[ht]
\unitlength=1mm
\begin{center}
\begin{picture}(20,30)
\put(10,-8){\vector(0,1){34}}
\put(9.15,6.5){$\bullet$}
%\put(2,7.15){\circle{4}}
%\put(1,7){$q$}
%\put(18,7.5){\circle{4}}
%\put(17,7){$\bar{q}$}
%\put(4,7.5){\line(1,0){12}}
\put(-5,7.5){\vector(1,0){33}}
\put(25,9){$\sX_3$}
\put(15,9){$\sX_+$}
\put(12,4){$\scriptstyle+\frac{1}{2}$}
\put(20,4){$\scriptstyle+1$}
\put(0,4){$\scriptstyle-\frac{1}{2}$}
\put(-6,4){$\scriptstyle-1$}
\put(2.5,9){$\sX_-$}
\put(5,25){$\sY_3$}
\put(5,13){$\scriptstyle+\frac{1}{2}$}
\put(5,1){$\scriptstyle-\frac{1}{2}$}
\put(5,19){$\scriptstyle+1$}
\put(5,-6){$\scriptstyle-1$}
\put(11,16){$\sY_+$}
\put(11,-3.5){$\sY_-$}
\thicklines
\put(10,7.5){\vector(1,0){14}}
\put(10,7.5){\vector(0,1){14}}
\put(10,7.5){\vector(-1,0){14}}
\put(10,7.5){\vector(0,-1){14}}
%\put(1,12){$\uparrow$}
%\put(17,12){$\downarrow$}
%\put(1,1){$\uparrow$}
%\put(17,1){$\uparrow$}
%\put(-7,5){$n$}
%\put(8,15){$l$}
%\put(20,11){$s=0$}
%\put(20,2){$s=1$}
%\put(10,10){\oval(3,2)}
\end{picture}
\end{center}
\vspace{0.3cm}
\caption{The root diagram of the Lie algebra $\mathfrak{sl}(2,\C)$. The action of each Weyl generator is shown in the ($\sX_3,\sY_3$)-plane.\label{fig1}}
\end{figure}

Obviously, generators $\sX_+$ and $\sX_-$ (roots $\alpha=-1$, $\alpha=+1$) shift a state one step to the \textit{left} and \textit{right}, while the generators $\sY_+$ and $\sY_-$ shift it \textit{up} and \textit{down}. All states of the $\SL(2,\C)$-multiplet can be translated into each other by a repeated action of these ladder operators. The generators of algebra $\mathfrak{sl}(2,\C)$ shown in Fig. 1 can be grouped into two manifolds: $(1,0)$ is formed by generators $\sX_3$, $\sX_+$ and $\sX_-$ (of the first subalgebra $\mathfrak{su}(2)$ in (\ref{Sum2})) and $(0,1)$ formed by generators $\sY_3$, $\sY_+$ and $\sY_-$ of the second subalgebra $i\mathfrak{su}(2)$. The Weyl diagrams for the first three $\SL(2,\C)$-multiplets are shown in Fig. 2.
%\newpage
\begin{figure}[h]
\unitlength=1.5mm
\begin{center}
\begin{picture}(30,30)(37,-10)
\put(50,0){$\overset{(0,0)}{\bullet}$}%\put(47,5.5){\line(1,0){10}}%\put(52.25,2.75){\line(0,1){7.25}}
%\put(47,-5.5){\line(1,0){10}}%\put(52.25,-7.15){\line(0,1){7.25}}
%\put(10,0.5){\line(1,0){12}}
\put(40,0.5){\vector(1,0){25}}\put(63,1.5){$\sX_3$}\put(30,0.5){$a)$}
\put(55,-1){$\scriptscriptstyle+\frac{1}{2}$}
\put(57,0.5){\line(0,1){0.5}}
\put(45,-1){$\scriptscriptstyle-\frac{1}{2}$}
\put(47,0.5){\line(0,1){0.5}}
\put(42,0.5){\line(0,1){0.5}}
\put(60.5,-1){$\scriptscriptstyle+1$}
\put(62,0.5){\line(0,1){0.5}}
\put(40.5,-1){$\scriptscriptstyle-1$}
\put(52,-12){\vector(0,1){26}}\put(48.5,13){$\sY_3$}
\put(49.25,5.25){$\scriptscriptstyle+\frac{1}{2}$}
\put(52,5.5){\line(1,0){0.5}}
\put(49.25,-5.75){$\scriptscriptstyle-\frac{1}{2}$}
\put(52,-5.5){\line(1,0){0.5}}
\put(49.25,10.25){$\scriptscriptstyle+1$}
\put(52,10.5){\line(1,0){0.5}}
\put(49.25,-10.75){$\scriptscriptstyle-1$}
\put(52,-10.5){\line(1,0){0.5}}
\put(55,5){$\overset{(\frac{1}{2},0)}{\bullet}$}
\put(55,-6){$\overset{(\frac{1}{2},0)}{\bullet}$}
%\put(53,8){$\scr(\tfrac{1}{2},0)$}
\put(45,5){$\overset{(0,\frac{1}{2})}{\bullet}$}
\put(45,-6){$\overset{(0,\frac{1}{2})}{\bullet}$}
\end{picture}
\end{center}
\end{figure}

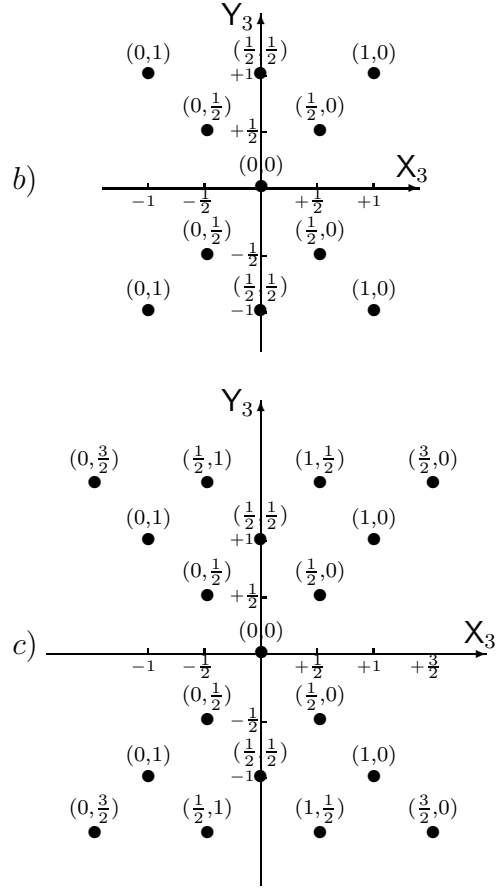
\begin{figure}[h]
\unitlength=1.5mm
\begin{center}
\begin{picture}(30,30)(37,-10)
\put(50,0){$\overset{(0,0)}{\bullet}$}%\put(47,5.5){\line(1,0){10}}%\put(52.25,2.75){\line(0,1){7.25}}
%\put(47,-5.5){\line(1,0){10}}%\put(52.25,-7.15){\line(0,1){7.25}}
%\put(10,0.5){\line(1,0){12}}
\put(38,0.5){\vector(1,0){28}}\put(64,1.5){$\sX_3$}\put(30,0.5){$b)$}
\put(55,-1){$\scriptscriptstyle+\frac{1}{2}$}
\put(57,0.5){\line(0,1){0.5}}
\put(45,-1){$\scriptscriptstyle-\frac{1}{2}$}
\put(47,0.5){\line(0,1){0.5}}
\put(42,0.5){\line(0,1){0.5}}
\put(60.5,-1){$\scriptscriptstyle+1$}
\put(62,0.5){\line(0,1){0.5}}
\put(40.5,-1){$\scriptscriptstyle-1$}
\put(52,-14){\vector(0,1){30}}\put(48.5,15){$\sY_3$}
\put(49.25,5.25){$\scriptscriptstyle+\frac{1}{2}$}
\put(52,5.5){\line(1,0){0.5}}
\put(49.25,-5.75){$\scriptscriptstyle-\frac{1}{2}$}
\put(52,-5.5){\line(1,0){0.5}}
\put(49.25,10.25){$\scriptscriptstyle+1$}
\put(52,10.5){\line(1,0){0.5}}
\put(49.25,-10.75){$\scriptscriptstyle-1$}
\put(52,-10.5){\line(1,0){0.5}}
\put(55,5){$\overset{(\frac{1}{2},0)}{\bullet}$}
\put(55,-6){$\overset{(\frac{1}{2},0)}{\bullet}$}
%\put(53,8){$\scr(\tfrac{1}{2},0)$}
\put(45,5){$\overset{(0,\frac{1}{2})}{\bullet}$}
\put(45,-6){$\overset{(0,\frac{1}{2})}{\bullet}$}
\put(49.5,10){$\overset{(\frac{1}{2},\frac{1}{2})}{\bullet}$}
\put(49.5,-11){$\overset{(\frac{1}{2},\frac{1}{2})}{\bullet}$}
%\put(52,10.5){\line(1,0){10}}\put(57.25,7.75){\line(0,1){7.25}}
%\put(52,-10.5){\line(1,0){10}}\put(57.25,-12.75){\line(0,1){7.25}}
\put(60,10){$\overset{(1,0)}{\bullet}$}
\put(60,-11){$\overset{(1,0)}{\bullet}$}
\put(40,10){$\overset{(0,1)}{\bullet}$}%\put(42,10.5){\line(1,0){10}}\put(47.25,7.75){\line(0,1){7.25}}
\put(40,-11){$\overset{(0,1)}{\bullet}$}%\put(42,-10.5){\line(1,0){10}}\put(47.25,-12.75){\line(0,1){7.25}}
\end{picture}
\end{center}
%\end{figure}

%\vspace{1.3cm}
%\begin{figure}[h]
\unitlength=1.5mm
\begin{center}
\begin{picture}(30,30)(37,-2)
\put(50,0){$\overset{(0,0)}{\bullet}$}%\put(47,5.5){\line(1,0){10}}%\put(52.25,2.75){\line(0,1){7.25}}
%\put(47,-5.5){\line(1,0){10}}%\put(52.25,-7.15){\line(0,1){7.25}}
%\put(10,0.5){\line(1,0){12}}
\put(33,0.5){\vector(1,0){39}}\put(70,1.5){$\sX_3$}\put(30,0.5){$c)$}
\put(55,-1){$\scriptscriptstyle+\frac{1}{2}$}
\put(57,0.5){\line(0,1){0.5}}
\put(45,-1){$\scriptscriptstyle-\frac{1}{2}$}
\put(47,0.5){\line(0,1){0.5}}
\put(42,0.5){\line(0,1){0.5}}
\put(60.5,-1){$\scriptscriptstyle+1$}
\put(62,0.5){\line(0,1){0.5}}
\put(65.25,-1){$\scriptscriptstyle+\frac{3}{2}$}
\put(40.5,-1){$\scriptscriptstyle-1$}
\put(52,-20){\vector(0,1){43}}\put(48.5,22){$\sY_3$}
\put(49.25,5.25){$\scriptscriptstyle+\frac{1}{2}$}
\put(52,5.5){\line(1,0){0.5}}
\put(49.25,-5.75){$\scriptscriptstyle-\frac{1}{2}$}
\put(52,-5.5){\line(1,0){0.5}}
\put(49.25,10.25){$\scriptscriptstyle+1$}
\put(52,10.5){\line(1,0){0.5}}
\put(49.25,-10.75){$\scriptscriptstyle-1$}
\put(52,-10.5){\line(1,0){0.5}}
\put(55,5){$\overset{(\frac{1}{2},0)}{\bullet}$}
\put(55,-6){$\overset{(\frac{1}{2},0)}{\bullet}$}
%\put(53,8){$\scr(\tfrac{1}{2},0)$}
\put(45,5){$\overset{(0,\frac{1}{2})}{\bullet}$}
\put(45,-6){$\overset{(0,\frac{1}{2})}{\bullet}$}
\put(49.5,10){$\overset{(\frac{1}{2},\frac{1}{2})}{\bullet}$}
\put(49.5,-11){$\overset{(\frac{1}{2},\frac{1}{2})}{\bullet}$}
%\put(52,10.5){\line(1,0){10}}\put(57.25,7.75){\line(0,1){7.25}}
%\put(52,-10.5){\line(1,0){10}}\put(57.25,-12.75){\line(0,1){7.25}}
\put(60,10){$\overset{(1,0)}{\bullet}$}
\put(60,-11){$\overset{(1,0)}{\bullet}$}
\put(40,10){$\overset{(0,1)}{\bullet}$}%\put(42,10.5){\line(1,0){10}}\put(47.25,7.75){\line(0,1){7.25}}
\put(40,-11){$\overset{(0,1)}{\bullet}$}%\put(42,-10.5){\line(1,0){10}}\put(47.25,-12.75){\line(0,1){7.25}}
\put(35,15){$\overset{(0,\frac{3}{2})}{\bullet}$}%\put(37,15.5){\line(1,0){10}}\put(42.25,12.75){\line(0,1){7.25}}
\put(35,-16){$\overset{(0,\frac{3}{2})}{\bullet}$}%\put(37,-15.5){\line(1,0){10}}\put(42.25,-17.75){\line(0,1){7.25}}
\put(45,15){$\overset{(\frac{1}{2},1)}{\bullet}$}%\put(47,15.5){\line(1,0){10}}\put(52.25,12.75){\line(0,1){7.25}}
\put(45,-16){$\overset{(\frac{1}{2},1)}{\bullet}$}%\put(47,-15.5){\line(1,0){10}}\put(52.25,-17.75){\line(0,1){7.25}}
\put(55,15){$\overset{(1,\frac{1}{2})}{\bullet}$}%\put(57,15.5){\line(1,0){10}}\put(62.25,12.75){\line(0,1){7.25}}
\put(55,-16){$\overset{(1,\frac{1}{2})}{\bullet}$}%\put(57,-15.5){\line(1,0){10}}\put(62.25,-17.75){\line(0,1){7.25}}
\put(65,15){$\overset{(\frac{3}{2},0)}{\bullet}$}
\put(65,-16){$\overset{(\frac{3}{2},0)}{\bullet}$}
\end{picture}
\end{center}
\vspace{2.4cm}
%\vspace{0.3cm}
\caption{The first three weight (Weyl) diagrams of the Lie algebra $\mathfrak{sl}(2,\C)$: a) $(\tfrac{1}{2},\tfrac{1}{2})$-manifold; b) $(1,1)$-manifold; c) $(\tfrac{3}{2},\tfrac{3}{2})$-manifold.\label{fig2}}
\end{figure}
%\newpage
%{\begin{minipage}{25pc}{\small {\bf Fig.\,2.}{The first three weight (Weyl) diagrams of the Lie algebra $\mathfrak{sl}(2,\C)$: a) $(\tfrac{1}{2},\tfrac{1}{2})$-manifold; b) $(1,1)$-manifold; c) $(\tfrac{3}{2},\tfrac{3}{2})$-manifold.\label{fig2}\end{minipage}}}
%\end{figure}

An extended version of the weight diagram for any $(l,\dot{l})$-manifold is shown in Figure 3.
\begin{figure}[h]
\unitlength=1.5mm
\begin{center}
\begin{picture}(100,85)(6,-35)
\put(50,0){$\overset{(0,0)}{\bullet}$}\put(47,5.5){\line(1,0){10}}\put(52.25,2.75){\line(0,1){7.25}}
\put(47,-5.5){\line(1,0){10}}\put(52.25,-7.15){\line(0,1){7.25}}
%\put(48,3){$\scr(0,0)$}
\put(55,5){$\overset{(\frac{1}{2},0)}{\bullet}$}
\put(55,-6){$\overset{(\frac{1}{2},0)}{\bullet}$}
%\put(53,8){$\scr(\tfrac{1}{2},0)$}
\put(45,5){$\overset{(0,\frac{1}{2})}{\bullet}$}
\put(45,-6){$\overset{(0,\frac{1}{2})}{\bullet}$}
\put(40,10){$\overset{(0,1)}{\bullet}$}\put(42,10.5){\line(1,0){10}}\put(47.25,7.75){\line(0,1){7.25}}
\put(40,-11){$\overset{(0,1)}{\bullet}$}\put(42,-10.5){\line(1,0){10}}\put(47.25,-12.75){\line(0,1){7.25}}
\put(50,10){$\overset{(\frac{1}{2},\frac{1}{2})}{\bullet}$}
\put(50,-11){$\overset{(\frac{1}{2},\frac{1}{2})}{\bullet}$}
\put(52,10.5){\line(1,0){10}}\put(57.25,7.75){\line(0,1){7.25}}
\put(52,-10.5){\line(1,0){10}}\put(57.25,-12.75){\line(0,1){7.25}}
\put(60,10){$\overset{(1,0)}{\bullet}$}
\put(60,-11){$\overset{(1,0)}{\bullet}$}
\put(35,15){$\overset{(0,\frac{3}{2})}{\bullet}$}\put(37,15.5){\line(1,0){10}}\put(42.25,12.75){\line(0,1){7.25}}
\put(35,-16){$\overset{(0,\frac{3}{2})}{\bullet}$}\put(37,-15.5){\line(1,0){10}}\put(42.25,-17.75){\line(0,1){7.25}}
\put(45,15){$\overset{(\frac{1}{2},1)}{\bullet}$}\put(47,15.5){\line(1,0){10}}\put(52.25,12.75){\line(0,1){7.25}}
\put(45,-16){$\overset{(\frac{1}{2},1)}{\bullet}$}\put(47,-15.5){\line(1,0){10}}\put(52.25,-17.75){\line(0,1){7.25}}
\put(55,15){$\overset{(1,\frac{1}{2})}{\bullet}$}\put(57,15.5){\line(1,0){10}}\put(62.25,12.75){\line(0,1){7.25}}
\put(55,-16){$\overset{(1,\frac{1}{2})}{\bullet}$}\put(57,-15.5){\line(1,0){10}}\put(62.25,-17.75){\line(0,1){7.25}}
\put(65,15){$\overset{(\frac{3}{2},0)}{\bullet}$}
\put(65,-16){$\overset{(\frac{3}{2},0)}{\bullet}$}
\put(30,20){$\overset{(0,2)}{\bullet}$}\put(32,20.5){\line(1,0){10}}\put(37.25,17.75){\line(0,1){7.25}}
\put(30,-21){$\overset{(0,2)}{\bullet}$}\put(32,-20.5){\line(1,0){10}}\put(37.25,-22.75){\line(0,1){7.25}}
\put(40,20){$\overset{(\frac{1}{2},\frac{3}{2})}{\bullet}$}
\put(40,-21){$\overset{(\frac{1}{2},\frac{3}{2})}{\bullet}$}
\put(42,20.5){\line(1,0){10}}\put(47.25,17.75){\line(0,1){7.25}}
\put(42,-20.5){\line(1,0){10}}\put(47.25,-22.75){\line(0,1){7.25}}
\put(50,20){$\overset{(1,1)}{\bullet}$}\put(52,20.5){\line(1,0){10}}\put(57.25,17.75){\line(0,1){7.25}}
\put(50,-21){$\overset{(1,1)}{\bullet}$}\put(52,-20.5){\line(1,0){10}}\put(57.25,-22.75){\line(0,1){7.25}}
\put(60,20){$\overset{(\frac{3}{2},\frac{1}{2})}{\bullet}$}
\put(60,-21){$\overset{(\frac{3}{2},\frac{1}{2})}{\bullet}$}
\put(62,20.5){\line(1,0){10}}\put(67.25,17.75){\line(0,1){7.25}}
\put(62,-20.5){\line(1,0){10}}\put(67.25,-22.75){\line(0,1){7.25}}
\put(70,20){$\overset{(2,0)}{\bullet}$}
\put(70,-21){$\overset{(2,0)}{\bullet}$}
\put(25,25){$\overset{(0,\frac{5}{2})}{\bullet}$}\put(27,25.5){\line(1,0){10}}\put(32.25,22.75){\line(0,1){7.25}}
\put(25,-26){$\overset{(0,\frac{5}{2})}{\bullet}$}\put(27,-25.5){\line(1,0){10}}\put(32.25,-27.75){\line(0,1){7.25}}
\put(35,25){$\overset{(\frac{1}{2},2)}{\bullet}$}\put(37,25.5){\line(1,0){10}}\put(42.25,22.75){\line(0,1){7.25}}
\put(35,-26){$\overset{(\frac{1}{2},2)}{\bullet}$}\put(37,-25.5){\line(1,0){10}}\put(42.25,-27.75){\line(0,1){7.25}}
\put(45,25){$\overset{(1,\frac{3}{2})}{\bullet}$}\put(47,25.5){\line(1,0){10}}\put(52.25,22.75){\line(0,1){7.25}}
\put(45,-26){$\overset{(1,\frac{3}{2})}{\bullet}$}\put(47,-25.5){\line(1,0){10}}\put(52.25,-27.75){\line(0,1){7.25}}
\put(55,25){$\overset{(\frac{3}{2},1)}{\bullet}$}\put(57,25.5){\line(1,0){10}}\put(62.25,22.75){\line(0,1){7.25}}
\put(55,-26){$\overset{(\frac{3}{2},1)}{\bullet}$}\put(57,-25.5){\line(1,0){10}}\put(62.25,-27.75){\line(0,1){7.25}}
\put(65,25){$\overset{(2,\frac{1}{2})}{\bullet}$}\put(67,25.5){\line(1,0){10}}\put(72.25,22.75){\line(0,1){7.25}}
\put(65,-26){$\overset{(2,\frac{1}{2})}{\bullet}$}\put(67,-25.5){\line(1,0){10}}\put(72.25,-27.75){\line(0,1){7.25}}
\put(75,25){$\overset{(\frac{5}{2},0)}{\bullet}$}
\put(75,-26){$\overset{(\frac{5}{2},0)}{\bullet}$}
\put(20,30){$\overset{(0,3)}{\bullet}$}\put(22,30.5){\line(1,0){10}}\put(27.25,27.75){\line(0,1){7.25}}
\put(20,-31){$\overset{(0,3)}{\bullet}$}\put(22,-30.5){\line(1,0){10}}\put(27.25,-32.75){\line(0,1){7.25}}
\put(30,30){$\overset{(\frac{1}{2},\frac{5}{2})}{\bullet}$}
\put(30,-31){$\overset{(\frac{1}{2},\frac{5}{2})}{\bullet}$}
\put(32,30.5){\line(1,0){10}}\put(37.25,27.75){\line(0,1){7.25}}
\put(32,-30.5){\line(1,0){10}}\put(37.25,-32.75){\line(0,1){7.25}}
\put(40,30){$\overset{(1,2)}{\bullet}$}\put(42,30.5){\line(1,0){10}}\put(47.25,27.75){\line(0,1){7.25}}
\put(40,-31){$\overset{(1,2)}{\bullet}$}\put(42,-30.5){\line(1,0){10}}\put(47.25,-32.75){\line(0,1){7.25}}
\put(50,30){$\overset{(\frac{3}{2},\frac{3}{2})}{\bullet}$}
\put(50,-31){$\overset{(\frac{3}{2},\frac{3}{2})}{\bullet}$}
\put(52,30.5){\line(1,0){10}}\put(57.25,27.75){\line(0,1){7.25}}
\put(52,-30.5){\line(1,0){10}}\put(57.25,-32.75){\line(0,1){7.25}}
\put(60,30){$\overset{(2,1)}{\bullet}$}\put(62,30.5){\line(1,0){10}}\put(67.25,27.75){\line(0,1){7.25}}
\put(60,-31){$\overset{(2,1)}{\bullet}$}\put(62,-30.5){\line(1,0){10}}\put(67.25,-32.75){\line(0,1){7.25}}
\put(70,30){$\overset{(\frac{5}{2},\frac{5}{2})}{\bullet}$}
\put(70,-31){$\overset{(\frac{5}{2},\frac{5}{2})}{\bullet}$}
\put(72,30.5){\line(1,0){10}}\put(77.25,27.75){\line(0,1){7.25}}
\put(72,-30.5){\line(1,0){10}}\put(77.25,-32.75){\line(0,1){7.25}}
\put(80,30){$\overset{(3,0)}{\bullet}$}
\put(80,-31){$\overset{(3,0)}{\bullet}$}
\put(15,35){$\overset{(0,\frac{7}{2})}{\bullet}$}\put(17,35.5){\line(1,0){10}}\put(22.25,32.75){\line(0,1){7.25}}
\put(15,-36){$\overset{(0,\frac{7}{2})}{\bullet}$}\put(17,-35.5){\line(1,0){10}}\put(22.25,-37.75){\line(0,1){7.25}}
\put(25,35){$\overset{(\frac{1}{2},3)}{\bullet}$}\put(27,35.5){\line(1,0){10}}\put(32.25,32.75){\line(0,1){7.25}}
\put(25,-36){$\overset{(\frac{1}{2},3)}{\bullet}$}\put(27,-35.5){\line(1,0){10}}\put(32.25,-37.75){\line(0,1){7.25}}
\put(35,35){$\overset{(1,\frac{5}{2})}{\bullet}$}\put(37,35.5){\line(1,0){10}}\put(42.25,32.75){\line(0,1){7.25}}
\put(35,-36){$\overset{(1,\frac{5}{2})}{\bullet}$}\put(37,-35.5){\line(1,0){10}}\put(42.25,-37.75){\line(0,1){7.25}}
\put(45,35){$\overset{(\frac{3}{2},2)}{\bullet}$}\put(47,35.5){\line(1,0){10}}\put(52.25,32.75){\line(0,1){7.25}}
\put(45,-36){$\overset{(\frac{3}{2},2)}{\bullet}$}\put(47,-35.5){\line(1,0){10}}\put(52.25,-37.75){\line(0,1){7.25}}
\put(55,35){$\overset{(2,\frac{3}{2})}{\bullet}$}\put(57,35.5){\line(1,0){10}}\put(62.25,32.75){\line(0,1){7.25}}
\put(55,-36){$\overset{(2,\frac{3}{2})}{\bullet}$}\put(57,-35.5){\line(1,0){10}}\put(62.25,-37.75){\line(0,1){7.25}}
\put(65,35){$\overset{(\frac{5}{2},1)}{\bullet}$}\put(67,35.5){\line(1,0){10}}\put(72.25,32.75){\line(0,1){7.25}}
\put(65,-36){$\overset{(\frac{5}{2},1)}{\bullet}$}\put(67,-35.5){\line(1,0){10}}\put(72.25,-37.75){\line(0,1){7.25}}
\put(75,35){$\overset{(3,\frac{1}{2})}{\bullet}$}\put(77,35.5){\line(1,0){10}}\put(82.25,32.75){\line(0,1){7.25}}
\put(75,-36){$\overset{(3,\frac{1}{2})}{\bullet}$}\put(77,-35.5){\line(1,0){10}}\put(82.25,-37.75){\line(0,1){7.25}}
\put(85,35){$\overset{(\frac{7}{2},0)}{\bullet}$}
\put(85,-36){$\overset{(\frac{7}{2},0)}{\bullet}$}
\put(10,40){$\overset{(0,4)}{\bullet}$}\put(12,40.5){\line(1,0){10}}
\put(10,-41){$\overset{(0,4)}{\bullet}$}\put(12,-40.5){\line(1,0){10}}
\put(20,40){$\overset{(\frac{1}{2},\frac{7}{2})}{\bullet}$}\put(22,40.5){\line(1,0){10}}
\put(20,-41){$\overset{(\frac{1}{2},\frac{7}{2})}{\bullet}$}\put(22,-40.5){\line(1,0){10}}
\put(30,40){$\overset{(1,3)}{\bullet}$}\put(32,40.5){\line(1,0){10}}
\put(30,-41){$\overset{(1,3)}{\bullet}$}\put(32,-40.5){\line(1,0){10}}
\put(40,40){$\overset{(\frac{3}{2},\frac{5}{2})}{\bullet}$}\put(42,40.5){\line(1,0){10}}
\put(40,-41){$\overset{(\frac{3}{2},\frac{5}{2})}{\bullet}$}\put(42,-40.5){\line(1,0){10}}
\put(50,40){$\overset{(2,2)}{\bullet}$}\put(52,40.5){\line(1,0){10}}
\put(50,-41){$\overset{(2,2)}{\bullet}$}\put(52,-40.5){\line(1,0){10}}
\put(60,40){$\overset{(\frac{5}{2},\frac{3}{2})}{\bullet}$}\put(62,40.5){\line(1,0){10}}
\put(60,-41){$\overset{(\frac{5}{2},\frac{3}{2})}{\bullet}$}\put(62,-40.5){\line(1,0){10}}
\put(70,40){$\overset{(3,1)}{\bullet}$}\put(72,40.5){\line(1,0){10}}
\put(70,-41){$\overset{(3,1)}{\bullet}$}\put(72,-40.5){\line(1,0){10}}
\put(80,40){$\overset{(\frac{7}{2},\frac{1}{2})}{\bullet}$}\put(82,40.5){\line(1,0){10}}
\put(80,-41){$\overset{(\frac{7}{2},\frac{1}{2})}{\bullet}$}\put(82,-40.5){\line(1,0){10}}
\put(90,40){$\overset{(4,0)}{\bullet}$}
\put(90,-41){$\overset{(4,0)}{\bullet}$}
\put(11.5,45){$\vdots$}
\put(21.5,45){$\vdots$}
\put(31.5,45){$\vdots$}
\put(41.5,45){$\vdots$}
%\put(51.5,45){$\vdots$}
\put(52,42){\vector(0,1){10}}\put(48,50){$\sY_3$}
\put(61.5,45){$\vdots$}
\put(71.5,45){$\vdots$}
\put(81.5,45){$\vdots$}
\put(91.5,45){$\vdots$}
\put(11.5,-45){$\vdots$}
\put(21.5,-45){$\vdots$}
\put(31.5,-45){$\vdots$}
\put(41.5,-45){$\vdots$}
\put(51.5,-45){$\vdots$}
\put(61.5,-45){$\vdots$}
\put(71.5,-45){$\vdots$}
\put(81.5,-45){$\vdots$}
\put(91.5,-45){$\vdots$}
\put(10,0.5){\line(1,0){42}}\put(50,0.5){\vector(1,0){42}}\put(90,2){$\sX_3$}
\put(16.5,32){$\vdots$}
\put(16.5,29){$\vdots$}
\put(16.5,26){$\vdots$}
\put(16.5,23){$\vdots$}
\put(16.5,20){$\vdots$}
\put(16.5,17){$\vdots$}
\put(16.5,14){$\vdots$}
\put(16.5,11){$\vdots$}
\put(16.5,9){$\vdots$}
\put(16.5,6){$\vdots$}
\put(16.5,3){$\vdots$}
\put(16.5,1.5){$\cdot$}
\put(16.5,0){$\cdot$}
\put(16.5,-32){$\vdots$}
\put(16.5,-29){$\vdots$}
\put(16.5,-26){$\vdots$}
\put(16.5,-23){$\vdots$}
\put(16.5,-20){$\vdots$}
\put(16.5,-17){$\vdots$}
\put(16.5,-14){$\vdots$}
\put(16.5,-11){$\vdots$}
\put(16.5,-9){$\vdots$}
\put(16.5,-6){$\cdot$}
%\put(16.5,-6){$\vdots$}
%\put(16.5,-3){$\vdots$}
\put(14.5,-3){$-\frac{7}{2}$}
\put(21.5,27){$\vdots$}
\put(21.5,24){$\vdots$}
\put(21.5,21){$\vdots$}
\put(21.5,18){$\vdots$}
\put(21.5,15){$\vdots$}
\put(21.5,13){$\vdots$}
\put(21.5,9){$\vdots$}
\put(21.5,6){$\vdots$}
\put(21.5,3){$\vdots$}
\put(21.5,1.5){$\cdot$}
\put(21.5,0){$\cdot$}
\put(21.5,-27){$\vdots$}
\put(21.5,-24){$\vdots$}
\put(21.5,-21){$\vdots$}
\put(21.5,-18){$\vdots$}
\put(21.5,-15){$\vdots$}
\put(21.5,-13){$\vdots$}
\put(21.5,-9){$\vdots$}
\put(21.5,-6){$\vdots$}
%\put(21.5,3){$\vdots$}
\put(19.5,-3){$-3$}
\put(26.5,22){$\vdots$}
\put(26.5,19){$\vdots$}
\put(26.5,16){$\vdots$}
\put(26.5,13){$\vdots$}
\put(26.5,10){$\vdots$}
\put(26.5,7){$\vdots$}
\put(26.5,4){$\vdots$}
\put(26.5,1){$\vdots$}
\put(26.5,-22){$\vdots$}
\put(26.5,-19){$\vdots$}
\put(26.5,-16){$\vdots$}
\put(26.5,-13){$\vdots$}
\put(26.5,-10){$\vdots$}
\put(26.5,-7){$\vdots$}
\put(26.5,-4){$\vdots$}
\put(26.5,-5){$\cdot$}
\put(24.5,-3){$-\frac{5}{2}$}
\put(31.5,17){$\vdots$}
\put(31.5,14){$\vdots$}
\put(31.5,11){$\vdots$}
\put(31.5,8){$\vdots$}
\put(31.5,5){$\vdots$}
\put(31.5,2){$\vdots$}
\put(31.5,0.5){$\cdot$}
\put(31.5,-17){$\vdots$}
\put(31.5,-14){$\vdots$}
\put(31.5,-11){$\vdots$}
\put(31.5,-8){$\vdots$}
\put(31.5,-5){$\vdots$}
\put(29.5,-3){$-2$}
\put(36.5,12){$\vdots$}
\put(36.5,9){$\vdots$}
\put(36.5,6){$\vdots$}
\put(36.5,3){$\vdots$}
\put(36.5,1.5){$\cdot$}
\put(36.5,0){$\cdot$}
\put(36.5,-12){$\vdots$}
\put(36.5,-9){$\vdots$}
\put(36.5,-6){$\vdots$}
\put(34.5,-3){$-\frac{3}{2}$}
\put(41.5,7){$\vdots$}
\put(41.5,4){$\vdots$}
\put(41.5,1){$\vdots$}
\put(39.5,-3){$-1$}
\put(46.5,2){$\vdots$}
\put(46.5,0.5){$\cdot$}
\put(41.5,-7){$\vdots$}
\put(41.5,-4.5){$\cdot$}
\put(44.5,-2){${\scriptstyle-\frac{1}{2}}$}
\put(51,-3){$0$}
\put(56.5,2){$\vdots$}
\put(56.5,0.5){$\cdot$}
\put(56.5,-2){${\scriptstyle\frac{1}{2}}$}
\put(61.5,7){$\vdots$}
\put(61.5,4){$\vdots$}
\put(61.5,1){$\vdots$}
\put(61.5,-7){$\vdots$}
\put(61.5,-4.5){$\cdot$}
\put(61.5,-3){$1$}
\put(66.5,12){$\vdots$}
\put(66.5,9){$\vdots$}
\put(66.5,6){$\vdots$}
\put(66.5,3){$\vdots$}
\put(66.5,1.5){$\cdot$}
\put(66.5,0){$\cdot$}
\put(66.5,-12){$\vdots$}
\put(66.5,-9){$\vdots$}
\put(66.5,-6){$\vdots$}
\put(66.5,-3){$\frac{3}{2}$}
\put(71.5,17){$\vdots$}
\put(71.5,14){$\vdots$}
\put(71.5,11){$\vdots$}
\put(71.5,8){$\vdots$}
\put(71.5,5){$\vdots$}
\put(71.5,2){$\vdots$}
\put(71.5,0.5){$\cdot$}
\put(71.5,-17){$\vdots$}
\put(71.5,-14){$\vdots$}
\put(71.5,-11){$\vdots$}
\put(71.5,-8){$\vdots$}
\put(71.5,-5){$\vdots$}
\put(71.5,-3){$2$}
\put(76.5,22){$\vdots$}
\put(76.5,19){$\vdots$}
\put(76.5,16){$\vdots$}
\put(76.5,13){$\vdots$}
\put(76.5,10){$\vdots$}
\put(76.5,7){$\vdots$}
\put(76.5,4){$\vdots$}
\put(76.5,1){$\vdots$}
\put(76.5,-22){$\vdots$}
\put(76.5,-19){$\vdots$}
\put(76.5,-16){$\vdots$}
\put(76.5,-13){$\vdots$}
\put(76.5,-10){$\vdots$}
\put(76.5,-7){$\vdots$}
\put(76.5,-5){$\cdot$}
\put(76.5,-3){$\frac{5}{2}$}
\put(81.5,27){$\vdots$}
\put(81.5,24){$\vdots$}
\put(81.5,21){$\vdots$}
\put(81.5,18){$\vdots$}
\put(81.5,15){$\vdots$}
\put(81.5,13){$\vdots$}
\put(81.5,9){$\vdots$}
\put(81.5,6){$\vdots$}
\put(81.5,3){$\vdots$}
\put(81.5,1.5){$\cdot$}
\put(81.5,0){$\cdot$}
\put(81.5,-27){$\vdots$}
\put(81.5,-24){$\vdots$}
\put(81.5,-21){$\vdots$}
\put(81.5,-18){$\vdots$}
\put(81.5,-15){$\vdots$}
\put(81.5,-13){$\vdots$}
\put(81.5,-9){$\vdots$}
\put(81.5,-6){$\vdots$}
\put(81.5,-3){$3$}
\put(86.5,32){$\vdots$}
\put(86.5,29){$\vdots$}
\put(86.5,26){$\vdots$}
\put(86.5,23){$\vdots$}
\put(86.5,20){$\vdots$}
\put(86.5,17){$\vdots$}
\put(86.5,14){$\vdots$}
\put(86.5,11){$\vdots$}
\put(86.5,9){$\vdots$}
\put(86.5,6){$\vdots$}
\put(86.5,3){$\vdots$}
\put(86.5,1.5){$\cdot$}
\put(86.5,0){$\cdot$}
\put(86.5,-32){$\vdots$}
\put(86.5,-29){$\vdots$}
\put(86.5,-26){$\vdots$}
\put(86.5,-23){$\vdots$}
\put(86.5,-20){$\vdots$}
\put(86.5,-17){$\vdots$}
\put(86.5,-14){$\vdots$}
\put(86.5,-11){$\vdots$}
\put(86.5,-9){$\vdots$}
\put(86.5,-6){$\cdot$}
\put(86.5,-3){$\frac{7}{2}$}
%\put(53.8,1.65){\line(5,6){3}}
\put(53.8,1.7){$\cdot$}\put(54.3,2.2){$\cdot$}\put(54.8,2.7){$\cdot$}\put(55.3,3.3){$\cdot$}\put(55.8,3.8){$\cdot$}
\put(56.3,4.3){$\cdot$}
\put(52.8,-0.7){$\cdot$}\put(53.3,-1.2){$\cdot$}\put(53.8,-1.7){$\cdot$}\put(54.3,-2.3){$\cdot$}\put(54.8,-2.8){$\cdot$}
\put(55.3,-3.3){$\cdot$}
%\put(58.75,6.75){\line(5,6){3}}
\put(58.8,6.8){$\cdot$}\put(59.3,7.3){$\cdot$}\put(59.8,7.8){$\cdot$}\put(60.3,8.3){$\cdot$}\put(60.8,8.8){$\cdot$}
\put(61.3,9.3){$\cdot$}
\put(57.8,-5.8){$\cdot$}\put(58.3,-6.3){$\cdot$}\put(58.8,-6.8){$\cdot$}\put(59.3,-7.3){$\cdot$}\put(59.8,-7.8){$\cdot$}
\put(60.3,-8.3){$\cdot$}
%\put(63.75,11.75){\line(5,6){3}}
\put(63.8,11.8){$\cdot$}\put(64.3,12.3){$\cdot$}\put(64.8,12.8){$\cdot$}\put(65.3,13.3){$\cdot$}\put(65.8,13.8){$\cdot$}
\put(66.3,14.3){$\cdot$}
\put(62.8,-10.8){$\cdot$}\put(63.3,-11.3){$\cdot$}\put(63.8,-11.8){$\cdot$}\put(64.3,-12.3){$\cdot$}\put(64.8,-12.8){$\cdot$}
\put(65.3,-13.3){$\cdot$}
%\put(68.75,16.75){\line(5,6){3}}
\put(68.8,16.8){$\cdot$}\put(69.3,17.3){$\cdot$}\put(69.8,17.8){$\cdot$}\put(70.3,18.3){$\cdot$}\put(70.8,18.8){$\cdot$}
\put(71.3,19.3){$\cdot$}
\put(67.8,-15.8){$\cdot$}\put(68.3,-16.3){$\cdot$}\put(68.8,-16.8){$\cdot$}\put(69.3,-17.3){$\cdot$}\put(69.8,-17.8){$\cdot$}
\put(70.3,-18.3){$\cdot$}
%\put(33.75,21.75){\line(5,6){3}}
\put(33.8,21.8){$\cdot$}\put(34.3,22.3){$\cdot$}\put(34.8,22.8){$\cdot$}\put(35.3,23.3){$\cdot$}\put(35.8,23.8){$\cdot$}
\put(36.3,24.3){$\cdot$}
\put(32.3,-21.8){$\cdot$}\put(32.8,-22.3){$\cdot$}\put(33.3,-22.8){$\cdot$}\put(33.8,-23.3){$\cdot$}\put(34.3,-23.8){$\cdot$}
\put(34.8,-24.3){$\cdot$}
%\put(38.75,26.75){\line(5,6){3}}
\put(38.8,26.8){$\cdot$}\put(39.3,27.3){$\cdot$}\put(39.8,27.8){$\cdot$}\put(40.3,28.3){$\cdot$}\put(40.8,28.8){$\cdot$}
\put(41.3,29.3){$\cdot$}
\put(37.3,-26.8){$\cdot$}\put(37.8,-27.3){$\cdot$}\put(38.3,-27.8){$\cdot$}\put(38.8,-28.3){$\cdot$}\put(39.3,-28.8){$\cdot$}
\put(39.8,-29.3){$\cdot$}
%\put(43.75,31.75){\line(5,6){3}}
\put(43.8,31.8){$\cdot$}\put(44.3,32.3){$\cdot$}\put(44.8,32.8){$\cdot$}\put(45.3,33.3){$\cdot$}\put(45.8,33.8){$\cdot$}
\put(46.3,34.3){$\cdot$}
\put(42.3,-31.8){$\cdot$}\put(42.8,-32.3){$\cdot$}\put(43.3,-32.8){$\cdot$}\put(43.8,-33.3){$\cdot$}\put(44.3,-33.8){$\cdot$}
\put(44.8,-34.3){$\cdot$}
%\put(48.75,36.75){\line(5,6){3}}
\put(48.8,36.8){$\cdot$}\put(49.3,37.3){$\cdot$}\put(49.8,37.8){$\cdot$}\put(50.3,38.3){$\cdot$}\put(50.8,38.8){$\cdot$}
\put(51.3,39.3){$\cdot$}
\put(47.3,-36.8){$\cdot$}\put(47.8,-37.3){$\cdot$}\put(48.3,-37.8){$\cdot$}\put(48.8,-38.3){$\cdot$}\put(49.3,-38.8){$\cdot$}
\put(49.8,-39.3){$\cdot$}
%\put(47.25,5.25){\line(5,-6){3}}
\put(47.3,4.4){$\cdot$}\put(47.8,3.9){$\cdot$}\put(48.3,3.4){$\cdot$}\put(48.8,2.9){$\cdot$}\put(49.3,2.4){$\cdot$}
\put(49.8,1.9){$\cdot$}
\put(50.8,-0.7){$\cdot$}\put(50.3,-1.2){$\cdot$}\put(49.8,-1.7){$\cdot$}\put(49.3,-2.2){$\cdot$}\put(48.8,-2.7){$\cdot$}
\put(48.3,-3.2){$\cdot$}
%\put(42.25,10.25){\line(5,-6){3}}
\put(42.3,9.4){$\cdot$}\put(42.8,8.9){$\cdot$}\put(43.3,8.4){$\cdot$}\put(43.8,7.9){$\cdot$}\put(44.3,7.4){$\cdot$}
\put(44.8,6.9){$\cdot$}
\put(45.8,-5.8){$\cdot$}\put(45.3,-6.3){$\cdot$}\put(44.8,-6.8){$\cdot$}\put(44.3,-7.3){$\cdot$}\put(43.8,-7.8){$\cdot$}
\put(43.3,-8.3){$\cdot$}
%\put(37.25,15.25){\line(5,-6){3}}
\put(37.3,14.4){$\cdot$}\put(37.8,13.9){$\cdot$}\put(38.3,13.4){$\cdot$}\put(38.8,12.9){$\cdot$}\put(39.3,12.4){$\cdot$}
\put(39.8,11.9){$\cdot$}
\put(40.8,-10.8){$\cdot$}\put(40.3,-11.3){$\cdot$}\put(39.8,-11.8){$\cdot$}\put(39.3,-12.3){$\cdot$}\put(38.8,-12.8){$\cdot$}
\put(38.3,-13.3){$\cdot$}
%\put(32.25,20.25){\line(5,-6){3}}
\put(32.3,19.4){$\cdot$}\put(32.8,18.9){$\cdot$}\put(33.3,18.4){$\cdot$}\put(33.8,17.9){$\cdot$}\put(34.3,17.4){$\cdot$}
\put(34.8,16.9){$\cdot$}
\put(35.8,-15.8){$\cdot$}\put(35.3,-16.3){$\cdot$}\put(34.8,-16.8){$\cdot$}\put(34.3,-17.3){$\cdot$}\put(33.8,-17.8){$\cdot$}
\put(33.3,-18.3){$\cdot$}
%\put(67.25,25.25){\line(5,-6){3}}
\put(67.3,24.4){$\cdot$}\put(67.8,23.9){$\cdot$}\put(68.3,23.4){$\cdot$}\put(68.8,22.9){$\cdot$}\put(69.3,22.4){$\cdot$}
\put(69.8,21.9){$\cdot$}
\put(71.3,-21.8){$\cdot$}\put(70.8,-22.3){$\cdot$}\put(70.3,-22.8){$\cdot$}\put(69.8,-23.3){$\cdot$}\put(69.3,-23.8){$\cdot$}
\put(68.8,-24.3){$\cdot$}
%\put(62.25,30.25){\line(5,-6){3}}
\put(62.3,29.4){$\cdot$}\put(62.8,28.9){$\cdot$}\put(63.3,28.4){$\cdot$}\put(63.8,27.9){$\cdot$}\put(64.3,27.4){$\cdot$}
\put(64.8,26.9){$\cdot$}
\put(66.3,-26.8){$\cdot$}\put(65.8,-27.3){$\cdot$}\put(65.3,-27.8){$\cdot$}\put(64.8,-28.3){$\cdot$}\put(64.3,-28.8){$\cdot$}
\put(63.8,-29.3){$\cdot$}
%\put(57.25,35.25){\line(5,-6){3}}
\put(57.3,34.4){$\cdot$}\put(57.8,33.9){$\cdot$}\put(58.3,33.4){$\cdot$}\put(58.8,32.9){$\cdot$}\put(59.3,32.4){$\cdot$}
\put(59.8,31.9){$\cdot$}
\put(61.3,-31.8){$\cdot$}\put(60.8,-32.3){$\cdot$}\put(60.3,-32.8){$\cdot$}\put(59.8,-33.3){$\cdot$}\put(59.3,-33.8){$\cdot$}
\put(58.8,-34.3){$\cdot$}
%\put(52.25,40.25){\line(5,-6){3}}
\put(52.3,39.4){$\cdot$}\put(52.8,38.9){$\cdot$}\put(53.3,38.4){$\cdot$}\put(53.8,37.9){$\cdot$}\put(54.3,37.4){$\cdot$}
\put(54.8,36.9){$\cdot$}
\put(56.3,-36.8){$\cdot$}\put(55.8,-37.3){$\cdot$}\put(55.3,-37.8){$\cdot$}\put(54.8,-38.3){$\cdot$}\put(54.3,-38.8){$\cdot$}
\put(53.8,-39.3){$\cdot$}
\end{picture}
\end{center}
\vspace{1.3cm}
\begin{center}\begin{minipage}{30pc}{\small {\bf Figure\,3:} Extended weight (Weyl) diagram of the group $\mathfrak{sl}(2,\C)$. Each node $(l,\dot{l})$ of the diagram is associated with a state whose mass is determined by the formula (\ref{Mass2}).}\end{minipage}\end{center}
\end{figure}

{\it Mass formula}.
We are now  in one step to obtain the formula for the mass spectrum which also explains the meaning of number $N$ in the Balmer-like formulas (\ref{Numbu})-(\ref{Alpha}).

Recall that the basic observable of the spectrum of matter is {\it energy} given by the Hermitian operator $H$. Let $G_f=\SO_0(1,3)\simeq\SL(2,\C)/\dZ_2$ be the fundamental symmetry group, where $\SO_0(1,3)$ is the Lorentz group. Due to the isomorphism $\SL(2,\C)\simeq\spin_+(1,3)$, we can use double covering of the fundamental symmetry group $\widetilde{G}_f\simeq\spin_+(1,3)$. Let energy operator $H$ be defined on a separable Hilbert space $\sH_\infty$ with its eigenvalues giving the possible values of energy (states). If $E_1\neq E_2$ are two eigenvalues of $H$ and $\left|\Phi_1\right\rangle$, $\left|\Phi_2\right\rangle$ are their eigenvectors in space $\sH_\infty$, then $\langle\Phi_1\mid\Phi_2\rangle=0$. The set of all eigenvectors related to a given eigenvalue $E$ with the zero vector added comprises an eigensubspace $\sH_E$ of $\sH_\infty$. All such eigensubspaces $\sH_E\in\sH_\infty$ are finite-dimensional.

Consider the group algebra $\mathfrak{sl}(2,\C)$ of the double covering group $\widetilde{G}_f$ and its complex shell with infinitesimal operators $\sX_l$, $\sY_l$, where $l=1,2,3$ (formulas (\ref{Shell})). It is known that the energy operator $H$ commutes with all the operators on $\sH_\infty$ which represent the Lie algebra of group $\widetilde{G}_f$. Let us consider an arbitrary eigensubspace $\sH_E$ of $H$. For three commuting operators $\sX_l$, $\sY_l$, and $H$, a common system of eigenfunctions can be constructed. This means that the subspace $\sH_E$ is invariant under the action of $\sX_l$ and $\sY_l$ (moreover, the operators $\sX_l$, $\sY_l$ can only be defined on $\sH_E$).

Let a certain {\it local representation} of group $\widetilde{G}_f$ be given by operators acting on $\sH_\infty$. We require that all the representing operators commute with $H$. Then each eigen subspace $\sH_E$ of the energy operator is invariant under the action of the complex shell operators $\sX_l$, $\sY_l$. Therefore, we can identify the subspaces $\sH_E$ with the symmetric spaces $\Sym_{(k,r)}$ of interlocking representations $\boldsymbol{\tau}_{k/2,r/2}$ of the Lorentz group and thus obtain a concrete realization (``dressing'') of the operator algebra $\pi(\fA)\rightarrow\pi(H)$, where $\pi\equiv\boldsymbol{\tau}_{k/2,r/2}$. It follows that each possible energy value (energy level) is a vector state of the form (see Axiom \textbf{A.II} in section \ref{sec4_1})
\begin{equation}\label{VectState2}
\omega_\Phi(H)=\frac{\langle\Phi\mid\pi(H)\Phi\rangle}{\langle\Phi\mid\Phi\rangle}=
\frac{\langle\Phi\mid\boldsymbol{\tau}_{k/2,r/2}(H)\Phi\rangle}{\langle\Phi\mid\Phi\rangle},
\end{equation}
associated with representation $\pi\equiv\boldsymbol{\tau}_{k/2,r/2}$ and corresponding to a nonzero cyclic vector $\left|\Phi\right\rangle\in\sH_\infty$.

Since $\sH_E\simeq\Sym_{(k,r)}$ is an eigen subspace of the energy operator $H$, the energy level (term) $E$ is proportional to the dimension of space $\sH_E$: $E\sim\dim\sH_E$. Introducing a factor to account for physical units, in compliance with the quantum nature of energy, we obtain $E\sim E_0\dim\sH_E$, where $E_0$ is a minimum portion of energy (energy quantum). The value of $\dim\sH_E$ is equal to
\[
\dim\Sym_{(k,r)}=(k+1)(r+1).
\]
Consequently,
\begin{equation}\label{Energy}
E\;\simeq\;E_0\left(\frac{k}{2}+\frac{1}{2}\right)\left(\frac{r}{2}+\frac{1}{2}\right),
\end{equation}
where $k/2=l$, $r/2=\dot{l}$; and $k$ and $r$ count the factors $\C_2$ and $\overset{\ast}{\C}_2$ (biquaternionic algebras) in the tensor product
\[
\underbrace{\C_2\otimes\C_2\otimes\cdots\otimes\C_2}_{k\;\text{times}}\bigotimes
\underbrace{\overset{\ast}{\C}_2\otimes\overset{\ast}{\C}_2\otimes\cdots\otimes
\overset{\ast}{\C}_2}_{r\;\text{times}},
\]
or the factors $\dS_2$ and $\dot{\dS}_2$ (2-dim spin-spaces) in the tensor product
\[
\underbrace{\dS_2\otimes\dS_2\otimes\cdots\otimes\dS_2}_{k\;\text{times}}\bigotimes
\underbrace{\dot{\dS}_2\otimes\dot{\dS}_2\otimes\cdots\otimes\dot{\dS}_2}_{r\;\text{times}},
\]
as well as the factors $\boldsymbol{\tau}_{\frac{1}{2},0}$ and $\boldsymbol{\tau}_{0,\frac{1}{2}}$ (fundamental representations) in the tensor product
\[
\underbrace{\boldsymbol{\tau}_{\frac{1}{2},0}\otimes\boldsymbol{\tau}_{\frac{1}{2},0}\otimes\cdots\otimes
\boldsymbol{\tau}_{\frac{1}{2},0}}_{k\;\text{times}}\bigotimes
\underbrace{\boldsymbol{\tau}_{0,\frac{1}{2}}\otimes\boldsymbol{\tau}_{0,\frac{1}{2}}\otimes\cdots\otimes
\boldsymbol{\tau}_{0,\frac{1}{2}}}_{r\;\text{times}}.
\]
Substitution of the explicit expressions $E=mc^2$ and $E_0=m_ec^2$, where $m_e$ is the rest mass of electron, into (\ref{Energy}) yields
\begin{equation}\label{Mass}
m\sim m_e\left(l+\frac{1}{2}\right)\left(\dot{l}+\frac{1}{2}\right).
\end{equation}
Using the fact that Eq. (\ref{Mass}) becomes an identity $m_e\equiv m_e$ when $l=1/2$, $\dot{l}=0$ or $l=0$, $\dot{l}=1/2$, we finally arrive at the mass formula
\begin{equation}\label{Mass2}
m=2m_e\left(l+\frac{1}{2}\right)\left(\dot{l}+\frac{1}{2}\right).
\end{equation}
For the earliest mention of formula (\ref{Mass2}), see \cite{Var15b}.

Now we can get an expression for $N$, which follows from comparing (\ref{Mass2}) and (\ref{Alpha}):
\begin{equation}\label{N}
N=4\alpha\left(l+\frac{1}{2}\right)\left(\dot{l}+\frac{1}{2}\right).
\end{equation}
Thus, we can assign a group-theoretic meaning to the empirical number $N$ in the Nambu formula (\ref{Numbu}), {\it that is a function of the quantum numbers $l$ and $\dot{l}$, which defines eigenvalues $l(l+1)$ and $\dot{l}(\dot{l}+1)$ of the Casimir operators $\bsX^2$ and $\bsY^2$ of the Lorentz group}.

Ground state $(0,0)$ (center of the weight diagram, see Figure 3) corresponds to the cyclic vector $\mid\psi_0\rangle=\boldsymbol{\tau}_{0,0}\mid\omega\rangle$, i.e. the vacuum vector, where $\boldsymbol{\tau}_{0,0}$ is the unit representation of the group $\SL(2,\C)\simeq\spin_+(1,3)$. The vacuum vector corresponds to the stationary state of the spectrum of matter with the lowest energy. It follows from the formula (\ref{Mass2}) that $m_{ground}=m_e/2$. All other states in the weight diagram are reached from the ground state $(0,0)$ with energy $m_{ground}$ through the action of ladder operators (Weyl generators $\sX_\pm$, $\sY_\pm$).

\section{Mass Formula: Verification by PDG\label{sec5}}
The computed mass values were compared to the observation data according to the latest version of PDG available \cite{PDG}. The results are presented in the following order: we start with leptons, then give detailed data for baryons (charged and neutral) and mesons (charged, neutral, and truly neutral).
\subsection{Leptons\label{sec5_1}}
We determine the masses of charged leptons according to the formula (\ref{Mass2}). The lowest rest mass corresponds to the electron
\[
m_e=0,511\;\;\text{MeV}.
\]
This experimental value of the electron mass is included as a constant in formula (\ref{Mass2}), i.e. the base value of the mass, relative to which the theoretical masses of states (``elementary particles'') are further calculated. From the mass formula (\ref{Mass2}) at $l=1/2$ and $\dot{l}=0$ it follows that
\[
m=2m_e\left(\frac{1}{2}+\frac{1}{2}\right)\left(0+\frac{1}{2}\right)=m_e,
\]
as well as at $l=0$ and $\dot{l}=1/2$:
\[
m=2m_e\left(0+\frac{1}{2}\right)\left(\frac{1}{2}+\frac{1}{2}\right)=m_e.
\]
In the weight diagram (Figure 3), the electron corresponds to a \textit{fundamental doublet}
\begin{equation}\label{Fdoublet}
\unitlength=1mm
%\begin{center}
\begin{picture}(20,13)
\put(0,5){$\overset{(0,\frac{1}{2})}{\bullet}$}
\put(20,5){$\overset{(\frac{1}{2},0)}{\bullet}$}
\put(3,6){\line(1,0){20}}
\put(-0.5,0){$-\frac{1}{2}$}
\put(22.5,0){$\frac{1}{2}$}
\put(6,0){$\cdots$}
\put(11.5,0){$\cdots$}
\put(17.5,0){$\cdots$}
%\put(19,0){$\cdots$}
%\put(21,0){$\cdot$}
\end{picture}
\end{equation}
which is the simplest multiplet of the group $\SL(2,\C)$ or \textit{primary splitting (doubling) of states}. The doublet (\ref{Fdoublet}) corresponds to the direct sum of representations of the group $\SL(2,\C)$:
\[
\boldsymbol{\tau}_{\frac{1}{2},0}\oplus\boldsymbol{\tau}_{0,\frac{1}{2}},
\]
within which the so-called \textit{electronic field} \cite{RF70}
\[
(1/2,0)\oplus(0,1/2)
\]
is defined.

The next largest lepton is the muon $\mu^-$ with an experimental mass value \cite{PDG}
\begin{equation}\label{Mexp}
m_\mu=105,66\;\;\text{MeV},
\end{equation}
which is approximately 207 (206,77) times the mass of an electron. This value corresponds to the cyclic representation $\boldsymbol{\tau}_{l,\dot{l}}$ on the spin 1/2 line at $l=10$ and $\dot{l}=19/2$:
\begin{equation}\label{Mtheor}
m_\mu=2m_e\left(10+\frac{1}{2}\right)\left(\frac{19}{2}+\frac{1}{2}\right)=107,31.
\end{equation}
The absolute error between the experimental (\ref{Mexp}) and calculated (\ref{Mtheor}) values is 1,65; the relative error is $1,56\%$. Accordingly, the direct sum $\boldsymbol{\tau}_{10,\frac{19}{2}}\oplus\boldsymbol{\tau}_{\frac{19}{2},10}$ corresponds to two nodes of the spin 1/2 and -1/2 lines in the weight diagram forming the (muon) doublet
\[
\unitlength=1mm
%\begin{center}
\begin{picture}(20,13)
\put(0,5){$\overset{(\frac{19}{2},10)}{\bullet}$}
\put(20,5){$\overset{(10,\frac{19}{2})}{\bullet}$}
\put(3,6){\line(1,0){20}}
\put(-0.5,0){$-\frac{1}{2}$}
\put(22.5,0){$\frac{1}{2}$}
\put(6,0){$\cdots$}
\put(11.5,0){$\cdots$}
\put(17.5,0){$\cdots$}
%\put(19,0){$\cdots$}
%\put(21,0){$\cdot$}
\end{picture}
\]
as well as the ``muon field'' $(10,19/2)\oplus(19/2,10)$.

Similarly, for a $\tau$-lepton with an experimental mass value \cite{PDG}
\begin{equation}\label{Texp}
m_\tau=1776,86\pm 0,12\;\;\text{MeV}
\end{equation}
we have a cyclic representation $\boldsymbol{\tau}_{l,\dot{l}}$ on the spin 1/2 line at $l=83/2$ and $\dot{l}=41$ with a theoretical mass
\begin{equation}\label{Ttheor}
m_\tau=2m_e\left(\frac{83}{2}+\frac{1}{2}\right)\left(41+\frac{1}{2}\right)=1781,35.
\end{equation}
The absolute error between (\ref{Texp}) and (\ref{Ttheor}) is 4,49; the relative error is 0,25\%. The corresponding $\tau$-lepton doublet has the form
\[
\unitlength=1mm
%\begin{center}
\begin{picture}(20,13)
\put(0,5){$\overset{(41,\frac{83}{2})}{\bullet}$}
\put(20,5){$\overset{(\frac{83}{2},41)}{\bullet}$}
\put(3,6){\line(1,0){20}}
\put(-0.5,0){$-\frac{1}{2}$}
\put(22.5,0){$\frac{1}{2}$}
\put(6,0){$\cdots$}
\put(11.5,0){$\cdots$}
\put(17.5,0){$\cdots$}
%\put(19,0){$\cdots$}
%\put(21,0){$\cdot$}
\end{picture}
\]
with a cyclic representation $\boldsymbol{\tau}_{\frac{83}{2},10}\oplus\boldsymbol{\tau}_{10,\frac{19}{2}}$ of the group $\SL(2,\C)$.

All charged leptons form a coherent subspace $\bsH^{(0,1,1/2)}_{\rm phys}(\C)$, where the baryon number $b=0$, the lepton number $\ell=1$ (see Table 1).
\begin{center}
{\textbf{Table 1. Charged leptons, spin 1/2, $\K=\C$}\\
(Subspace $\bsH^{(0,1,1/2)}_{\rm phys}(\C)$, vectors $\left|\K,b,\ell,l-\dot{l}\right\rangle=\left|\C,0,1,1/2\right\rangle$).}
\vspace{0.1cm}
{\renewcommand{\arraystretch}{1.0}
\begin{tabular}{|c||l|l|c|l|}\hline
   & State and mass & $m_\omega$ (theor.) & Error \% & $(l,\dot{l})$ \\ \hline\hline
1. & $e^\pm$ -- $0,511$ & $0,511$ & $-$ & $(1/2,0)$ \\
2. & $\mu^\pm$ -- $105,66$ & $107,31$ & $+1,56$ & $(10,19/2)$ \\
3. & $\tau^\pm$ -- $1776,86\pm 0,12$ & $1781,35$ & $+0,25$ & $(83/2,41)$ \\
\hline
\end{tabular}
}
\end{center}

\subsection{Baryon Sector\label{sec5_2}}
All states of the spectrum of matter forming a physical $\K$-Hilbert space are divided into two main classes: 1) fermi-states, i.e. states of half-integer spin (leptons and baryons); 2) bose-states, i.e. states of integer spin (photon and mesons). In this section, baryons (hadrons of half-integer spin) are considered, forming, according to modern data \cite{PDG}, a fairly numerous sector of states (Tables 2--17). Further, according to the presence or absence of charge, all baryons are grouped into coherent subspaces of $\K$-Hilbert space: charged states form $\C$-subspaces, neutral states belong to $\BH$-subspaces, the absence of $\R$-subspaces in the baryon sector is explained by the corresponding absence of truly neutral (Majorana) baryons in nature.

The first column of Tables 2--17 shows the standard designation of the state and its mass in MeV. The second column contains the mass value of the corresponding state $\omega$, the mass of which is calculated according to the formula (\ref{Mass2}). The third column contains the relative error between the experimental and calculated value. The fourth column shows the parameters $l$ and $\dot{l}$ of the corresponding cyclic representation of group $\spin_+(1,3)$. The fifth column indicates the ``quark composition'' of the state. The quantum numbers of the state according to the quark model are also given here (see Appendix B): $I(J^P)$, where $I$ is the isotopic spin, $J$ is the ``spin'' of the state, $P$ is parity.\\
\textit{Remark}. Paying tribute to the important role played by the quark model in the formation of the theory of hadron spectra, we add the quark composition of the state to the fifth column of the tables. The fifth column serves only one purpose: to establish a connection with the classification adopted in Particle Data Group \cite{PDG}. A brief overview of the quark model is provided in Appendix B.

\subsubsection{$\C$-subspaces/charged baryons\label{sec5_2_1}}

All charged states of the baryon sector belong to coherent subspaces $\bsH^{(b,\ell,s)}_{\rm phys}(\C)$ of the physical $\K$-Hilbert space $\bsH_{\rm phys}(\K)$, where $b$ is the baryon number, $\ell$ is the lepton number ($\ell=0$ for baryons), $s=l-\dot{l}$ is the spin of the state. Tables 2--9 show all charged states having at least a three-star status according the PDG classification \cite{PDG} for the baryon sector. To this should be added anti-states (antimatter), which form conjugate coherent subspaces $\bsH^{(b,\ell,s)}_{\rm phys}(\overline{\C})$.
\begin{center}
{\textbf{Table 2. Charged baryons, spin 1/2, $\K=\C$}\\
(Subspace $\bsH^{(1,0,1/2)}_{\rm phys}(\C)$, vectors $\left|\K,b,\ell,l-\dot{l}\right\rangle=\left|\C,1,0,1/2\right\rangle$).}
\vspace{0.1cm}
{\renewcommand{\arraystretch}{1.0}
\begin{tabular}{|c||l|l|c|l|l|}\hline
   & State and mass & $m_\omega$ (theor.) & Error \% & $(l,\dot{l})$ & $qqq$, $I(J^P)$\\ \hline\hline
   &                 &          &         &             & $N^+=uud$\\
1. & $p$ -- $938,27$ & $935,13$ & $-0,33$ & $(30,59/2)$ & $\frac{1}{2}({\frac{1}{2}}^+)$\\
2. & $N(1440)^+\approx 1370$     & $1380,21$ & $-0,74$ & $(73/2,36)$ & $\frac{1}{2}({\frac{1}{2}}^+)$\\
3. & $N(1535)^+\approx 1510$ & $1495,17$ & $-0,98$ & $(38,75/2)$ & $\frac{1}{2}({\frac{1}{2}}^-)$\\
4. & $N(1650)^+\approx 1655$ & $1655,54$ & $+0,03$ & $(40,79/2)$ & $\frac{1}{2}({\frac{1}{2}}^-)$\\
5. & $N(1710)^+\approx 1700$ & $1697,03$ & $-0,17$ & $(81/2,40)$ & $\frac{1}{2}({\frac{1}{2}}^+)$\\
6. & $N(1880)^+\approx 1860$ & $1867,70$ & $+0,41$ & $(85/2,42)$ & $\frac{1}{2}({\frac{1}{2}}^+)$\\
7. & $N(1895)^+\approx 1910$ & $1911,65$ & $+0,08$ & $(43,85/2)$ & $\frac{1}{2}({\frac{1}{2}}^-)$\\
8. & $N(2100)^+\approx 2100$ & $2092,54$ & $-0,35$ & $(45,89/2)$ & $\frac{1}{2}({\frac{1}{2}}^+)$\\
9. & $N(2300)^+$ -- $2300^{+40}_{-30}$ & $2281,61$ & $-0,79$ & $(47,93/2)$ & $\frac{1}{2}({\frac{1}{2}}^+)$\\
\hline
   &             &           &         &              & $\Delta^+=uud$\\
10. & $\Delta(1620)^+\approx 1600$ & $1614,76$ & $+0,92$ & $(79/2,39)$ & $\frac{3}{2}({\frac{1}{2}}^-)$\\
11. & $\Delta(1750)^+\approx 1748$ & $1738,93$ & $-0,52$ & $(41,81/2)$ & $\frac{3}{2}({\frac{1}{2}}^+)$\\
12. & $\Delta(1900)^+\approx 1865$ & $1867,70$ & $+0,14$ & $(85/2,42)$ & $\frac{3}{2}({\frac{1}{2}}^-)$\\
13. & $\Delta(1910)^+\approx 1860$ & $1867,70$ & $+0,41$ & $(85/2,42)$ & $\frac{3}{2}({\frac{1}{2}}^+)$\\
14. & $\Delta(2150)^+$ -- $2140\pm 80$ & $2139,05$ & $-0,04$ & $(91/2,45)$ & $\frac{3}{2}({\frac{1}{2}}^-)$\\
\hline
\end{tabular}
}
\end{center}
\begin{center}
{\renewcommand{\arraystretch}{1.0}
\begin{tabular}{|c||l|l|c|l|l|}\hline
 & State and mass & $m_\omega$ (theor.) & Error \% & $(l,\dot{l})$ & $qqq$, $I(J^P)$\\ \hline\hline
   &             &           &         &              & $\Sigma^+=uus$\\
   &             &           &         &              & $\Sigma^-=dds$\\
15. & $\Sigma^+$ -- $1189,37$ & $1198,80$ & $+0,79$ & $(34,67/2)$ & $1({\frac{1}{2}}^+)$\\
16. & $\Sigma^-$ -- $1197,45$ & $1198,80,39$ & $+0,11$ & $(34,67/2)$ & $1({\frac{1}{2}}^+)$\\
\hline
17. & $\Sigma(1620)^\pm$ -- $1680\pm 8$ & $1697,03$ & +1,01 & (81/2,40) &  $1({\frac{1}{2}}^-)$\\
18. & $\Sigma(1660)^\pm$ -- $1585\pm 20$ & $1574,80$ & $-0,67$ & $(39,77/2)$ & $1({\frac{1}{2}}^+)$\\
19. & $\Sigma(1750)^\pm$ -- $1689\pm 11$ & $1697,03$ & $+0,47$ & $(81/2,40)$ & $1({\frac{1}{2}}^-)$\\
20. & $\Sigma(1880)^\pm\approx 1880$ & $1867,70$ & $-0,65$ & $(85/2,42)$ & $1({\frac{1}{2}}^+)$\\
21. & $\Sigma(1900)^\pm$ -- $1936\pm 10$ & $1956,11$ & $+1,04$ & $(87/2,43)$ & $1({\frac{1}{2}}^-)$\\
22. & $\Sigma(2110)^\pm$ -- $2158\pm 25$ & $2139,05$ & $-0,88$ & $(91/2,45)$ & $1({\frac{1}{2}}^-)$\\
\hline
   &             &           &         &              & $\Xi^-=dss$\\
23. & $\Xi^-$ -- $1321,71$ & $1306,12$ & $-1,18$ & $(71/2,35)$ & $\frac{1}{2}({\frac{1}{2}}^+)$\\
\hline
   &             &           &         &              & $\Lambda^+_c=udc$\\
24. & $\Lambda^+_c$ -- $2286,46\pm 0,14$ & $2281,61$ & $-0,21$ & $(47,93/2)$ & $0({\frac{1}{2}}^+)$\\
25. & $\Lambda_c(2595)^+$ -- $2592,25\pm 0,28$ & $2580,55$ & $-0,45$ & $(50,99/2)$ & $0({\frac{1}{2}}^-)$\\
\hline
   &             &           &         &              & $\Sigma^+_c=udc$\\
26. & $\Sigma_c(2455)^+$ -- $2452,65^{+0,22}_{-0,16}$ & $2428,78$ & $-0,97$ & $(97/2,48)$ & $1({\frac{1}{2}}^+)$\\
27. & $\Sigma_c(2455)^{++}$ -- $2453,97\pm 0,14$ & $2428,78$ & $-1,03$ & $(97/2,48)$ & $1({\frac{1}{2}}^+)$\\
\hline
   &             &           &         &              & $\Xi^+_c=usc$\\
28. & $\Xi^+_c$ -- $2467,71\pm 0,23$ & $2478,86$ & $+0,45$ & $(49,97/2)$ & $\frac{1}{2}({\frac{1}{2}}^+)$\\
29. & ${\Xi^\prime}^+_c$ -- $2578,2\pm 0,5$ & $2580,55$ & $+0,45$ & $(50,99/2)$ & $\frac{1}{2}({\frac{1}{2}}^+)$\\
30. & $\Xi_c(2790)^+$ -- $2791,9\pm 0,5$  & $2790,06$ & $+0,09$ & $(52,103/2)$ & $\frac{1}{2}({\frac{1}{2}}^-)$\\
31. & $\Xi_c(2970)^+$ -- $2964,3\pm 1,5$  & $2952,56$ & $-0,37$ & $(107/2,57)$ & $\frac{1}{2}({\frac{1}{2}}^-)$\\
\hline
   &             &           &         &              & $\Sigma^+_b=uub$\\
   &             &           &         &              & $\Sigma^-_b=ddb$\\
32. & $\Sigma^+_b$ -- $5810,56\pm 0,25$ & $5787,07$ & $-0,40$ & $(75,149/2)$ & $1({\frac{1}{2}}^+)$\\
33. & $\Sigma^-_b$ -- $5815,64\pm 0,25$ & $5787,07$ & $-0,49$ & $(75,149/2)$ & $1({\frac{1}{2}}^+)$\\
\hline
   &             &           &         &              & $\Xi^-_b=dsb$\\
34. & $\Xi^-_b$ -- $5797,0\pm 0,6$ & $5787,07$ & $-0,17$ & $(76,149/2)$ & $\frac{1}{2}({\frac{1}{2}}^+)$\\
35. & $\Xi^\prime_b(5935)^-$ -- $5935,02\pm 0,05$ & $5941,91$ & $+0,12$ & $(76,151/2)$ & $?({\frac{1}{2}}^+)$\\
\hline
   &             &           &         &              & $\Omega^-_b=ssb$\\
36. & $\Omega^-_b$ -- $6045,2\pm 1,2$ & $6020,09$ & $-0,41$ & $(153/2,76)$ & $0({\frac{1}{2}}^+)$\\
\hline
\end{tabular}
}
\end{center}
\newpage
\begin{center}
{\textbf{Table 3. Charged baryons, spin 3/2, $\K=\C$}\\
(Subspace $\bsH^{(1,0,3/2)}_{\rm phys}(\C)$, vectors $\left|\K,b,\ell,l-\dot{l}\right\rangle=\left|\C,1,0,3/2\right\rangle$).}
\vspace{0.1cm}
{\renewcommand{\arraystretch}{1.0}
\begin{tabular}{|c||l|l|c|l|l|}\hline
 & State and mass & $m_\omega$ (theor.) & Error \% & $(l,\dot{l})$ & $qqq$, $I(J^P)$\\ \hline\hline
   &                 &          &         &             & $N^+=uud$\\
1. & $N(1520)^+\approx 1510$     & $1494,67$ & $-1,01$ & $(79/2,37)$ & $\frac{1}{2}({\frac{3}{2}}^-)$\\
2. & $N(1700)^+\approx 1700$     & $1696,52$ & $-0,20$ & $(41,79/2)$ & $\frac{1}{2}({\frac{3}{2}}^-)$\\
3. & $N(1720)^+\approx 1675$     & $1655,13$ & $-1,00$ & $(81/2,39)$ & $\frac{1}{2}({\frac{3}{2}}^+)$\\
4. & $N(1875)^+\approx 1900$     & $1911,14$ & $+0,59$ & $(87/2,42)$ & $\frac{1}{2}({\frac{3}{2}}^-)$\\
5. & $N(1900)^+\approx 1920$     & $1911,14$ & $-0,46$ & $(87/2,42)$ & $\frac{1}{2}({\frac{3}{2}}^+)$\\
6. & $N(2040)^+$ -- $2040^{+4}_{-3}$ & $2046,04$ & $+0,29$ & $(45,87/2)$ & $\frac{1}{2}({\frac{3}{2}}^+)$\\
7. & $N(2120)^+\approx 2100$     & $2092,03$ & $-0,38$ & $(91/2,44)$ & $\frac{1}{2}({\frac{3}{2}}^-)$\\
\hline
   &             &           &         &              & $\Delta^+=uud$\\
8. & $\Delta(1232)^+\approx 1210$ & $1198,29$ & $-0,97$ & $(69/2,33)$ & $\frac{3}{2}({\frac{3}{2}}^+)$\\
9. & $\Delta(1600)^+\approx 1510$ & $1494,67$ & $-1,01$ & $(77/2,37)$ & $\frac{3}{2}({\frac{3}{2}}^+)$\\
10. & $\Delta(1700)^+\approx 1665$ & $1655,13$ & $-0,59$ & $(81/2,39)$ & $\frac{3}{2}({\frac{3}{2}}^-)$\\
11. & $\Delta(1920)^+\approx 1900$ & $1911,14$ & $-0,59$ & $(87/2,42)$ & $\frac{3}{2}({\frac{3}{2}}^+)$\\
12. & $\Delta(1940)^+\approx 1950$ & $1955,59$ & $+0,29$ & $(44,85/2)$ & $\frac{3}{2}({\frac{3}{2}}^-)$\\
\hline
   &             &           &         &              & $\Sigma^+=uus$\\
   &             &           &         &              & $\Sigma^-=dds$\\
13. & $\Sigma(1385)^+$ -- $1382,83$ & $1379,70$ & $-0,23$ & $(37,71/2)$ & $1({\frac{3}{2}}^+)$\\
14. & $\Sigma(1385)^-$ -- $1387,2$ & $1379,70$ & $-0,54$ & $(37,71/2)$ & $1({\frac{3}{2}}^+)$\\
15. & $\Sigma(1580)^\pm$ -- $1607^{+13}_{-11}$ & $1614,24$ & $+0,45$ & $(40,77/2)$ & $1({\frac{3}{2}}^-)$\\
16. & $\Sigma(1670)^\pm\approx 1662$ & $1655,13$ & $-0,41$ & $(81/2,39)$ & $1({\frac{3}{2}}^-)$\\
17. & $\Sigma(1780)^\pm\approx 1780$ & $1780,83$ & $+0,05$ & $(42,81/2)$ & $1({\frac{3}{2}}^+)$\\
18. & $\Sigma(1910)^\pm\approx 1910$ & $1911,14$ & $+0,06$ & $(87/2,42)$ & $1({\frac{3}{2}}^-)$\\
19. & $\Sigma(1940)^\pm\approx 1940$ & $1955,59$ & $+0,80$ & $(44,85/2)$ & $1({\frac{3}{2}}^+)$\\
20. & $\Sigma(2010)^\pm$ -- $1995\pm 12$ & $2000,56$ & $+0,28$ & $(89/2,43)$ & $1({\frac{3}{2}}^-)$\\
21. & $\Sigma(2080)^\pm\approx 2090$ & $2092,03$ & $+0,09$ & $(91/2,44)$ & $1({\frac{3}{2}}^+)$\\
22. & $\Sigma(2230)^\pm$ -- $2234\pm 25$ & $2233,07$ & $-0,04$ & $(47,91/2)$ & $1({\frac{3}{2}}^+)$\\
\hline
   &             &           &         &              & $\Xi^-=dss$\\
23. & $\Xi(1530)^-$ -- $1535,0\pm 0,6$ & $1534,02$ & $-0,06$ & $(39,75/2)$ & $\frac{1}{2}({\frac{3}{2}}^+)$\\
24. & $\Xi(1820)^-$ -- $1823\pm 5$ & $1823,76$ & $+0,04$ & $(85/2,41)$ & $\frac{1}{2}({\frac{3}{2}}^-)$\\
\hline
   &             &           &         &              & $\Omega^-=sss$\\
25. & $\Omega^-$ -- $1672,45$ & $1655,13$ & $-1,03$ & $(81/2,39)$ & $0({\frac{3}{2}}^+)$\\
\hline
   &             &           &         &              & $\Lambda^+_c=udc$\\
26. & $\Lambda_c(2625)^+$ -- $2628,11\pm 0,19$ & $2631,65$ & $+0,13$ & $(51,99/2)$ & $0({\frac{3}{2}}^-)$\\
27. & $\Lambda_c(2860)^+$ -- $2856,1^{+2,3}_{-6,0}$ & $2843,20$ & $-0,43$ & $(53,103/2)$ & $0({\frac{3}{2}}^+)$\\
28. & $\Lambda_c(2940)^+$ -- $2939,6^{+1,3}_{-1,5}$ & $2952,04$ & $+0,42$ & $(54,105/2)$ & $0({\frac{3}{2}}^-)$\\
\hline
\end{tabular}
}
\end{center}
\begin{center}
{\renewcommand{\arraystretch}{1.0}
\begin{tabular}{|c||l|l|c|l|l|}\hline
 & State and mass & $m_\omega$ (theor.) & Error \% & $(l,\dot{l})$ & $qqq$, $I(J^P)$\\ \hline\hline
   &             &           &         &              & $\Sigma^{++}_c=uuc$\\
29. & $\Sigma_c(2520)^{++}$ -- $2518,41^{+0,22}_{-0,18}$ & $2528,94$ & $+0,42$ & $(50,97/2)$ & $1({\frac{3}{2}}^+)$\\
   &             &           &         &              & $\Sigma^{+}_c=udc$\\
30. & $\Sigma_c(2520)^{+}$ -- $2517,4^{+0,7}_{-0,5}$ & $2528,94$ & $+0,46$ & (50,97/2) &  $1({\frac{3}{2}}^+)$\\
\hline
   &             &           &         &              & $\Xi^+_c=usc$\\
31. & $\Xi_c(2645)^+$ -- $2645,10\pm 0,3$ & $2631,65$ & $-0,51$ & $(51,99/2)$ & $\frac{1}{2}({\frac{3}{2}}^+)$\\
32. & $\Xi_c(2815)^+$ -- $2816,51\pm 0,25$ & $2789,55$ & $-0,96$ & $(105/2,51)$ & $\frac{1}{2}({\frac{3}{2}}^-)$\\
\hline
33. & ${\Sigma^\ast_b}^+$ -- $5830,32\pm 0,27$ & $5863,72$ & $+0,57$ & $(76,149/2)$ & $1({\frac{3}{2}}^+)$\\
34. & ${\Sigma^\ast_b}^-$ -- $5834,71\pm 0,30$ & $5863,72$ & $+0,49$ & $(76,149/2)$ & $1({\frac{1}{2}}^+)$\\
\hline
   &             &           &         &              & $\Xi^-_b=dsb$\\
35. & $\Xi_b(5955)^-$ -- $5955,33\pm 0,13$ & $5941,40$ & $-0,23$ & $(153/2,75)$ & $?({\frac{1}{2}}^+)$\\
36. & $\Xi_b(6100)^-$ -- $6100,3\pm 0,6$ & $6098,27$ & $-0,03$ & $(155/2,76)$ & $?({\frac{1}{2}}^+)$\\
\hline
\end{tabular}
}
\end{center}
\begin{center}
{\textbf{Table 4. Charged baryons, spin 5/2, $\K=\C$}\\
(Subspace $\bsH^{(1,0,5/2)}_{\rm phys}(\C)$, vectors $\left|\K,b,\ell,l-\dot{l}\right\rangle=\left|\C,1,0,5/2\right\rangle$).}
\vspace{0.1cm}
{\renewcommand{\arraystretch}{1.0}
\begin{tabular}{|c||l|l|c|l|l|}\hline
 & State and mass & $m_\omega$ (theor.) & Error \% & $(l,\dot{l})$ & $qqq$, $I(J^P)$\\ \hline\hline
   &                 &          &         &             & $N^+=uud$\\
1. & $N(1675)^+\approx 1660$    & $1654,11$ & $-0,35$ & $(41,77/2)$ & $\frac{1}{2}({\frac{5}{2}}^-)$\\
2. & $N(1680)^+\approx 1675$    & $1695,49$ & $+1,22$ & $(83/2,39)$ & $\frac{1}{2}({\frac{5}{2}}^+)$\\
3. & $N(1860)^+$ -- $1834\pm 19$ & $1822,74$ & $-0,61$ & $(48,81/2)$ & $\frac{1}{2}({\frac{5}{2}}^+)$\\
4. & $N(2000)^+$ -- $2030\pm 40$ & $2045,02$ & $+0,74$ & $(91/2,43)$ & $\frac{1}{2}({\frac{5}{2}}^+)$\\
5. & $N(2060)^+\approx 2070$    & $2091,01$ & $+1,01$ & $(46,87/2)$ & $\frac{1}{2}({\frac{5}{2}}^-)$\\
6. & $N(2570)^+$ -- $2570^{+19}_{-10}$ & $2579,02$ & $+0,35$ & $(51,97/2)$ & $\frac{1}{2}({\frac{5}{2}}^-)$\\
\hline
   &             &           &         &              & $\Delta^+=uud$\\
7. & $\Delta(1905)^+\approx 1800$ & $1779,81$ & $-1,12$ & $(85/2,40)$ & $\frac{3}{2}({\frac{5}{2}}^+)$\\
8. & $\Delta(1930)^+\approx 1880$ & $1866,07$ & $-0,73$ & $(87/2,41)$ & $\frac{3}{2}({\frac{5}{2}}^-)$\\
9. & $\Delta(2000)^+$ -- $1998\pm 4$ & $1999,54$ & $+0,07$ & $(45,85/2)$ & $\frac{3}{2}({\frac{5}{2}}^+)$\\
10. & $\Delta(2350)^+$ -- $2400\pm 125$ & $2427,25$ & $+1,13$ & $(99/2,47)$ & $\frac{3}{2}({\frac{5}{2}}^-)$\\
\hline
   &             &           &         &              & $\Sigma^+=uus$\\
   &             &           &         &              & $\Sigma^-=dds$\\
11. & $\Sigma(1775)^\pm\approx 1770$ & $1779,84$ & $+0,55$ & $(85/2,40)$ & $1({\frac{5}{2}}^-)$\\
12. & $\Sigma(1915)^\pm\approx 1900$ & $1910,12$ & $+0,53$ & $(44,83/2)$ & $1({\frac{5}{2}}^+)$\\
13. & $\Sigma(2070)^\pm\approx 2060$ & $2045,02$ & $-0,73$ & $(91/2,43)$ & $1({\frac{5}{2}}^+)$\\
\hline
   &             &           &         &              & $\Xi^-=dss$\\
14. & $\Xi(2030)^-$ -- $2025\pm 5$  & $2045,02$ & $+0,98$ & $(91/2,43)$ & $\frac{1}{2}({\frac{5}{2}}^?)$\\
\hline
   &             &           &         &              & $\Lambda^+_c=udc$\\
15. & $\Lambda_c(2880)^+$ -- $2881,63\pm 0,24$ & $2896,35$ & $+0,51$ & $(54,103/2)$ & $0({\frac{5}{2}}^+)$\\
\hline
\end{tabular}
}
\end{center}
\begin{center}
\newpage
{\textbf{Table 5. Charged baryons, spin 7/2, $\K=\C$}\\
(Subspace $\bsH^{(1,0,7/2)}_{\rm phys}(\C)$, vectors $\left|\K,b,\ell,l-\dot{l}\right\rangle=\left|\C,1,0,7/2\right\rangle$).}
\vspace{0.1cm}
{\renewcommand{\arraystretch}{1.0}
\begin{tabular}{|c||l|l|c|l|l|}\hline
 & State and mass & $m_\omega$ (theor.) & Error \% & $(l,\dot{l})$ & $qqq$, $I(J^P)$\\ \hline\hline
   &                 &          &         &             & $N^+=uud$\\
1. & $N(1990)^+$ -- $2030\pm 65$ & $2043,49$ & $+0,66$ & $(46,85/2)$ & $\frac{1}{2}({\frac{7}{2}}^+)$\\
2. & $N(2190)^+\approx 2100$    & $2089,48$ & $-0,50$ & $(93/2,43)$ & $\frac{1}{2}({\frac{7}{2}}^-)$\\
\hline
   &             &           &         &              & $\Delta^+=uud$\\
3. & $\Delta(1950)^+\approx 1880$ & $1864,64$ & $-0,82$ & $(44,81/2)$ & $\frac{3}{2}({\frac{7}{2}}^+)$\\
4. & $\Delta(2200)^+\approx 2100$ & $2089,48$ & $-0,50$ & $(93/2,43)$ & $\frac{3}{2}({\frac{7}{2}}^-)$\\
5. & $\Delta(2390)^+$ -- $2223\pm 19$ & $2230,51$ & $+0,34$ & $(48,89/2)$ & $\frac{3}{2}({\frac{7}{2}}^+)$\\
\hline
   &             &           &         &              & $\Sigma^+=uus$\\
   &             &           &         &              & $\Sigma^-=dds$\\
6. & $\Sigma(2030)^\pm\approx 2020$ & $1998,01$ & $-1,08$ & $(91/2,42)$ & $1({\frac{7}{2}}^+)$\\
7. & $\Sigma(2100)^\pm$ -- $2093\pm 16$ & $2089,48$ & $-0,17$ & $(93/2,43)$ & $1({\frac{7}{2}}^-)$\\
\hline
\end{tabular}
}
\end{center}
\begin{center}
{\textbf{Table 6. Charged baryons, spin 9/2, $\K=\C$}\\
(Subspace $\bsH^{(1,0,9/2)}_{\rm phys}(\C)$, vectors $\left|\K,b,\ell,l-\dot{l}\right\rangle=\left|\C,1,0,9/2\right\rangle$).}
\vspace{0.1cm}
{\renewcommand{\arraystretch}{1.0}
\begin{tabular}{|c||l|l|c|l|l|}\hline
 & State and mass & $m_\omega$ (theor.) & Error \% & $(l,\dot{l})$ & $qqq$, $I(J^P)$\\ \hline\hline
   &                 &          &         &             & $N^+=uud$\\
1. & $N(2220)^+\approx 2170$    & $2180,95$ & $+0,5$ & $(48,87/2)$ & $\frac{1}{2}({\frac{9}{2}}^+)$\\
2. & $N(2250)^+\approx 2200$    & $2228,47$ & $+1,29$ & $(97/2,44)$ & $\frac{1}{2}({\frac{9}{2}}^-)$\\
\hline
   &             &           &         &              & $\Delta^+=uud$\\
3. & $\Delta(2300)^+$ -- $2370\pm 80$ & $2374,11$ & $+0,17$ & $(50,91/2)$ & $\frac{3}{2}({\frac{9}{2}}^+)$\\
4. & $\Delta(2400)^+$ -- $2260\pm 60$ & $2276,50$ & $+0,73$ & $(49,89/2)$ & $\frac{3}{2}({\frac{9}{2}}^-)$\\
\hline
\end{tabular}
}
\end{center}
\begin{center}
{\textbf{Table 7. Charged baryons, spin 11/2, $\K=\C$}\\
(Subspace $\bsH^{(1,0,11/2)}_{\rm phys}(\C)$, vectors $\left|\K,b,\ell,l-\dot{l}\right\rangle=\left|\C,1,0,11/2\right\rangle$).}
\vspace{0.1cm}
{\renewcommand{\arraystretch}{1.0}
\begin{tabular}{|c||l|l|c|l|l|}\hline
 & State and mass & $m_\omega$ (theor.) & Error \% & $(l,\dot{l})$ & $qqq$, $I(J^P)$\\ \hline\hline
   &                 &          &         &             & $N^+=uud$\\
1. & $N(2600)^+\approx 2600$    & $2590,26$ & $+0,94$ & $(53,95/2)$ & $\frac{1}{2}({\frac{11}{2}}^-)$\\
\hline
   &             &           &         &              & $\Delta^+=uud$\\
2. & $\Delta(2420)^+\approx 2400$ & $2421,12$ & $+0,88$ & $(51,91/2)$ & $\frac{3}{2}({\frac{11}{2}}^+)$\\
\hline
\end{tabular}
}
\end{center}
\begin{center}
{\textbf{Table 8. Charged baryons, spin 13/2, $\K=\C$}\\
(Subspace $\bsH^{(1,0,13/2)}_{\rm phys}(\C)$, vectora $\left|\K,b,\ell,l-\dot{l}\right\rangle=\left|\C,1,0,13/2\right\rangle$).}
\vspace{0.1cm}
{\renewcommand{\arraystretch}{1.0}
\begin{tabular}{|c||l|l|c|l|l|}\hline
 & State and mass & $m_\omega$ (theor.) & Error \% & $(l,\dot{l})$ & $qqq$, $I(J^P)$\\ \hline\hline
   &                 &          &         &             & $N^+=uud$\\
1. & $N(2700)^+$ -- $2612\pm 45$ & $2621,43$ & $+0,36$ & $(107/2,47)$ & $\frac{1}{2}({\frac{13}{2}}^+)$\\
\hline
   &             &           &         &              & $\Delta^+=uud$\\
2. & $\Delta(2750)^+$ -- $2794\pm 50$ & $2779,33$ & $-0,52$ & $(55,97/2)$ & $\frac{3}{2}({\frac{13}{2}}^-)$\\
\hline
\end{tabular}
}
\end{center}
\newpage
\begin{center}
{\textbf{Table 9. Charged baryons, spin 15/2, $\K=\C$}\\
(Subspace $\bsH^{(1,0,15/2)}_{\rm phys}(\C)$, vectors $\left|\K,b,\ell,l-\dot{l}\right\rangle=\left|\C,1,0,15/2\right\rangle$).}
\vspace{0.1cm}
{\renewcommand{\arraystretch}{1.0}
\begin{tabular}{|c||l|l|c|l|l|}\hline
 & State and mass & $m_\omega$ (theor.) & Error \% & $(l,\dot{l})$ & $qqq$, $I(J^P)$\\ \hline\hline
   &                 &          &         &             & $N^+=uud$\\
1. & $\Delta(2950)^+$ -- $2990\pm 100$ & $2993,44$ & $+0,11$ & $(115/2,50)$ & $\frac{1}{2}({\frac{15}{2}}^+)$\\
\hline
\end{tabular}
}
\end{center}

\subsubsection{$\BH$-subspaces/neutral baryons\label{sec5_2_2}}
The neutral states of the baryon sector form coherent subspaces $\bsH^{(b,0,s)}_{\rm phys}(\BH)$, i.e. quaternionic subspaces of the $\K$-Hilbert space (Tables 10--17). As in the case of $\C$-subspaces, $\BH$-subspaces have counterparts (antimatter): $\bsH^{(b,0,s)}_{\rm phys}(\overline{\BH})$. Unlike the meson sector, the baryon sector does not contain $\R$-subspaces due to the absence of truly neutral (Majorana) fermi-states in nature.
\begin{center}
{\textbf{Table 10. Neutral baryons, spin 1/2, $\K=\BH$}\\
(Subspace $\bsH^{(1,0,1/2)}_{\rm phys}(\BH)$, vectors $\left|\K,b,\ell,l-\dot{l}\right\rangle=\left|\BH,1,0,1/2\right\rangle$).}
\vspace{0.1cm}
{\renewcommand{\arraystretch}{1.0}
\begin{tabular}{|c||l|l|c|l|l|}\hline
 & State and mass & $m_\omega$ (theor.) & Error \% & $(l,\dot{l})$ & $qqq$, $I(J^P)$\\ \hline\hline
   &                 &          &         &             & $N^0=udd$\\
1. & $n$ -- $939,56$ & $935,13$ & $-0,47$ & $(30,59/2)$ & $\frac{1}{2}({\frac{1}{2}}^+)$\\
2. & $N(1440)^0\approx 1370$    & $1380,21$ & $-0,74$ & $(73/2,36)$ & $\frac{1}{2}({\frac{1}{2}}^+)$\\
3. & $N(1535)^0\approx 1510$    & $1495,17$ & $-0,98$ & $(38,75/2)$ & $\frac{1}{2}({\frac{1}{2}}^-)$\\
4. & $N(1650)^0\approx 1655$    & $1655,54$ & $+0,03$ & $(40,79/2)$ & $\frac{1}{2}({\frac{1}{2}}^-)$\\
5. & $N(1710)^0\approx 1700$    & $1697,03$ & $-0,17$ & $(81/2,40)$ & $\frac{1}{2}({\frac{1}{2}}^+)$\\
6. & $N(1880)^0\approx 1860$    & $1867,70$ & $+0,41$ & $(85/2,42)$ & $\frac{1}{2}({\frac{1}{2}}^+)$\\
7. & $N(1895)^0\approx 1910$    & $1911,65$ & $+0,08$ & $(43,85/2)$ & $\frac{1}{2}({\frac{1}{2}}^-)$\\
8. & $N(2100)^0\approx 2100$    & $2092,54$ & $-0,35$ & $(45,89/2)$ & $\frac{1}{2}({\frac{1}{2}}^+)$\\
9. & $N(2300)^0$ -- $2300^{+40}_{-30}$ & $2281,61$ & $-0,79$ & $(47,93/2)$ & $\frac{1}{2}({\frac{1}{2}}^+)$\\
\hline
   &             &           &         &              & $\Delta^0=udd$\\
10. & $\Delta(1620)^0\approx 1600$ & $1514,76$ & $+0,92$ & $(79/2,39)$ & $\frac{3}{2}({\frac{1}{2}}^-)$\\
11. & $\Delta(1750)^0\approx 1748$ & $1738,93$ & $-0,52$ & $(41,81/2)$ & $\frac{3}{2}({\frac{1}{2}}^+)$\\
12. & $\Delta(1900)^0\approx 1865$ & $1867,70$ & $+0,14$ & $(85/2,42)$ & $\frac{3}{2}({\frac{1}{2}}^-)$\\
13. & $\Delta(1910)^0\approx 1860$ & $1867,70$ & $+0,41$ & $(85/2,42)$ & $\frac{3}{2}({\frac{1}{2}}^+)$\\
14. & $\Delta(2150)^0$ -- $2140\pm 80$ & $2134,05$ & $-0,04$ & $(91/2,45)$ & $\frac{3}{2}({\frac{1}{2}}^-)$\\
\hline
   &             &           &         &              & $\Lambda^0=uds$\\
15. & $\Lambda^0$ -- $1115,68$ & $1129,82$ & $+1,26$ & $(33,65/2)$ & $0({\frac{1}{2}}^+)$\\
16. & $\Lambda(1380)^0$ -- $1325\pm 15$ & $1342,90$ & $+1,35$ & $(36,71/2)$ & $0({\frac{1}{2}}^-)$\\
17. & $\Lambda(1405)^0$ -- $1429^{+8}_{-7}$ & $1418,02$ & $-0,77$ & $(37,73/2)$ & $0({\frac{1}{2}}^-)$\\
18. & $\Lambda(1600)^0\approx 1550$ & $1534,53$ & $-0,99$ & $(77/2,38)$ & $0({\frac{1}{2}}^+)$\\
19. & $\Lambda(1670)^0\approx 1674$ & $1654,54$ & $-1,10$ & $(40,79/2)$ & $0({\frac{1}{2}}^-)$\\
\hline
\end{tabular}
}
\end{center}
\begin{center}
{\renewcommand{\arraystretch}{1.0}
\begin{tabular}{|c||l|l|c|l|l|}\hline
 & State and mass & $m_\omega$ (theor.) & Error \% & $(l,\dot{l})$ & $qqq$, $I(J^P)$\\ \hline\hline
20. & $\Lambda(1710)^0$ -- $1713\pm 13$ & $1697,03$ & $-0,93$ & $(81/2,40)$ & $0({\frac{1}{2}}^+)$\\
21. & $\Lambda(1800)^0$ -- $1809\pm 9$ & $1824,27$ & $+0,84$ & $(42,83/2)$ & $0({\frac{1}{2}}^-)$\\
22. & $\Lambda(1810)^0$ -- $1773\pm 7$ & $1781,35$ & $+0,47$ & $(83/2,41)$ & $0({\frac{1}{2}}^+)$\\
23. & $\Lambda(2000)^0\approx 2000$ & $2001,07$ & $+0,05$ & $(44,87/2)$ & $0({\frac{1}{2}}^-)$\\
\hline
   &             &           &         &              & $\Sigma^0=uds$\\
24. & $\Sigma^0$ -- $1192,64$ & $1198,80$ & $+0,51$ & $(34,67/2)$ & $1({\frac{1}{2}}^+)$\\
25. & $\Sigma(1620)^0$ -- $1680\pm 8$ & $1697,03$ & $+1,01$ & $(81/2,40)$ & $1({\frac{1}{2}}^-)$\\
26. & $\Sigma(1660)^0$ -- $1585\pm 20$ & $1574,39$ & $-0,67$ & $(39,77/2)$ & $1({\frac{1}{2}}^+)$\\
27. & $\Sigma(1750)^0$ -- $1689\pm 11$ & $1697,03$ & $+0,47$ & $(81/2,40)$ & $1({\frac{1}{2}}^-)$\\
28. & $\Sigma(1880)^0\approx 1880$ & $1867,70$ & $-0,65$ & $(85/2,42)$ & $1({\frac{1}{2}}^+)$\\
29. & $\Sigma(1900)^0$ -- $1936\pm 10$ & $1956,11$ & $+1,04$ & $(87/2,43)$ & $1({\frac{1}{2}}^-)$\\
30. & $\Sigma(2110)^0$ -- $2158\pm 25$ & $2139,05$ & $-0,88$ & $(91/2,45)$ & $1({\frac{1}{2}}^-)$\\
\hline
   &             &           &         &              & $\Xi^0=uss$\\
31. & $\Xi^0$ -- $1314,86$ & $1306,12$ & $-0,66$ & $(71/2,35)$ & $\frac{1}{2}({\frac{1}{2}}^+)$\\
\hline
   &             &           &         &              & $\Sigma^0_c=ddc$\\
32. & $\Sigma_c(2455)^0$ -- $2453,75\pm 0,14$ & $2428,78$ & $-1,02$ & $(97/2,48)$ & $1({\frac{1}{2}}^+)$\\
\hline
   &             &           &         &              & $\Xi^0_c=dsc$\\
33. & $\Xi^0_c$ -- $2470,44\pm 0,28$ & $2478,86$ & $+0,34$ & $(49,97/2)$ & $\frac{1}{2}({\frac{1}{2}}^+)$\\
34. & ${\Xi^\prime}^0_c$ -- $2578,7\pm 0,5$ & $2580,55$ & $+0,07$ & $(50,99/2)$ & $\frac{1}{2}({\frac{1}{2}}^+)$\\
35. & $\Xi_c(2790)^0$ -- $2793,9\pm 0,5$ & $2790,06$ & $-0,14$ & $(52,103/2)$ & $\frac{1}{2}({\frac{1}{2}}^-)$\\
36. & $\Xi_c(2970)^0$ -- $2967,1\pm 1,7$ & $2952,56$ & $-0,49$ & $(107/2,53)$ & $\frac{1}{2}({\frac{1}{2}}^+)$\\
\hline
   &             &           &         &              & $\Omega^0_c=ssc$\\
37. & $\Omega^0_c$ -- $2695,2\pm 1,7$ & $2684,28$ & $-0,40$ & $(51,101/2)$ & $0({\frac{1}{2}}^+)$\\
\hline
   &             &           &         &              & $\Lambda^0_b=udb$\\
38. & $\Lambda^0_b$ -- $5619,6\pm 0,17$ & $5634,29$ & $+0,26$ & $(74,147/2)$ & $0({\frac{1}{2}}^+)$\\
39. & $\Lambda_b(5912)^0$ -- $5912,19\pm 0,17$ & $5941,91$ & $+0,50$ & $(76,151/2)$ & $0({\frac{1}{2}}^-)$\\
40. & $\Lambda_b(6070)^0$ -- $6072,3\pm 2,9$ & $6098,78$ & $+0,44$ & $(77,153/2)$ & $0({\frac{1}{2}}^+)$\\
\hline
   &             &           &         &              & $\Xi^0_b=usb$\\
41. & $\Xi^0_b$ -- $5791,9\pm 0,5$ & $5787,07$ & $-0,08$ & $(75,149/2)$ & $\frac{1}{2}({\frac{1}{2}}^+)$\\
\hline
\end{tabular}
}
\end{center}
\begin{center}
{\textbf{Table 11. Neutral baryons, spin 3/2, $\K=\BH$}\\
(Subspace $\bsH^{(1,0,3/2)}_{\rm phys}(\BH)$, vectors $\left|\K,b,\ell,l-\dot{l}\right\rangle=\left|\BH,1,0,3/2\right\rangle$).}
\vspace{0.1cm}
{\renewcommand{\arraystretch}{1.0}
\begin{tabular}{|c||l|l|c|l|l|}\hline
 & State and mass & $m_\omega$ (theor.) & Error \% & $(l,\dot{l})$ & $qqq$, $I(J^P)$\\ \hline\hline
   &                 &          &         &             & $N^0=udd$\\
1. & $N(1520)^0\approx 1510$    & $1494,67$ & $-1,01$ & $(77/2,37)$ & $\frac{1}{2}({\frac{3}{2}}^-)$\\
2. & $N(1700)^0\approx 1700$    & $1696,52$ & $-0,20$ & $(41,79/2)$ & $\frac{1}{2}({\frac{3}{2}}^-)$\\
3. & $N(1720)^0\approx 1675$    & $1655,13$ & $-1,00$ & $(81/2,39)$ & $\frac{1}{2}({\frac{3}{2}}^+)$\\
4. & $N(1875)^0\approx 1900$    & $1911,14$ & $+0,59$ & $(87/2,42)$ & $\frac{1}{2}({\frac{3}{2}}^-)$\\
5. & $N(1900)^0\approx 1920$    & $1911,14$ & $-0,46$ & $(87/2,42)$ & $\frac{1}{2}({\frac{3}{2}}^+)$\\
6. & $N(2040)^0$ -- $2040^{+4}_{-3}$    & $2046,04$ & $+0,29$ & $(45,87/2)$ & $\frac{1}{2}({\frac{3}{2}}^+)$\\
7. & $N(2120)^0\approx 2100$    & $2092,03$ & $-0,38$ & $(91/2,44)$ & $\frac{1}{2}({\frac{3}{2}}^-)$\\
\hline
\end{tabular}
}
\end{center}
\begin{center}
{\renewcommand{\arraystretch}{1.0}
\begin{tabular}{|c||l|l|c|l|l|}\hline
 & State and mass & $m_\omega$ (theor.) & Error \% & $(l,\dot{l})$ & $qqq$, $I(J^P)$\\ \hline\hline
\hline
   &             &           &         &              & $\Delta^0=udd$\\
8. & $\Delta(1232)^0\approx 1210$ & $1198,29$ & $-0,97$ & $(69/2,33)$ & $\frac{3}{2}({\frac{3}{2}}^+)$\\
9. & $\Delta(1600)^0\approx 1510$ & $1494,67$ & $-1,01$ & $(77/2,37)$ & $\frac{3}{2}({\frac{3}{2}}^+)$\\
10. & $\Delta(1700)^0\approx 1665$ & $1655,13$ & $-0,59$ & $(81/2,39)$ & $\frac{3}{2}({\frac{3}{2}}^-)$\\
11. & $\Delta(1920)^0\approx 1900$ & $1911,14$ & $+0,59$ & $(87/2,42)$ & $\frac{3}{2}({\frac{3}{2}}^+)$\\
12. & $\Delta(1940)^0\approx 1950$ & $1955,59$ & $+0,29$ & $(44,85/2)$ & $\frac{3}{2}({\frac{3}{2}}^-)$\\
\hline
   &             &           &         &              & $\Lambda^0=uds$\\
13. & $\Lambda(1520)^0\approx 1517,5$ & $1534,02$ & $+1,08$ & $(39,75/2)$ & $0({\frac{3}{2}}^-)$\\
14. & $\Lambda(1690)^0\approx 1690$ & $1696,52$ & $+0,38$ & $(41,79/2)$ & $0({\frac{3}{2}}^-)$\\
15. & $\Lambda(1890)^0$ -- $1872\pm 5$ & $1867,19$ & $-0,25$ & $(43,83/2)$ & $0({\frac{3}{2}}^+)$\\
16. & $\Lambda(2050)^0$ -- $2056\pm 22$ & $2046,04$ & $-0,48$ & $(45,87/2)$ & $0({\frac{3}{2}}^-)$\\
17. & $\Lambda(2070)^0$ -- $2044\pm 20$ & $2046,04$ & $+0,09$ & $(45,87/2)$ & $0({\frac{3}{2}}^+)$\\
18. & $\Lambda(2325)^0\approx 2325$ & $2329,65$ & $+0,20$ & $(48,93/2)$ & $0({\frac{3}{2}}^-)$\\
\hline
   &             &           &         &              & $\Sigma^0=uds$\\
19. & $\Sigma(1385)^0$ -- $1383,7\pm 1,0$ & $1379,70$ & $-0,29$ & $(37,71/2)$ & $1({\frac{3}{2}}^+)$\\
20. & $\Sigma(1580)^0$ -- $1607^{+13}_{11}$  & $1614,24$ & $+0,45$ & $(40,77/2)$ & $1({\frac{3}{2}}^-)$\\
21. & $\Sigma(1670)^0\approx 1662$  & $1655,13$ & $-0,41$ & $(81/2,39)$ & $1({\frac{3}{2}}^-)$\\
22. & $\Sigma(1780)^0\approx 1780$  & $1780,83$ & $+0,05$ & $(42,81/2)$ & $1({\frac{3}{2}}^+)$\\
23. & $\Sigma(1910)^0\approx 1910$  & $1911,14$ & $+0,06$ & $(87/2,42)$ & $1({\frac{3}{2}}^-)$\\
24. & $\Sigma(1940)^0\approx 1940$  & $1955,59$ & $+0,80$ & $(44,85/2)$ & $1({\frac{3}{2}}^+)$\\
25. & $\Sigma(2010)^0$ -- $1995\pm 12$  & $2000,56$ & $+0,28$ & $(89/2,43)$ & $1({\frac{3}{2}}^-)$\\
26. & $\Sigma(2080)^0\approx 2090$  & $2093,03$ & $+0,09$ & $(91/2,44)$ & $1({\frac{3}{2}}^+)$\\
27. & $\Sigma(2230)^0$ -- $2234\pm 25$  & $2233,07$ & $-0,04$ & $(47,91/2)$ & $1({\frac{3}{2}}^+)$\\
\hline
   &             &           &         &              & $\Xi^0=uss$\\
28. & $\Xi(1530)^0$ -- $1531,80\pm 0,32$ & $1534,02$ & $+0,14$ & $(39,75/2)$ & $\frac{1}{2}({\frac{3}{2}}^+)$\\
29. & $\Xi(1820)^0$ -- $1823\pm 5$  & $1823,76$ & $+0,04$ & $(85/2,41)$ & $\frac{1}{2}({\frac{3}{2}}^-)$\\
\hline
   &             &           &         &              & $\Sigma^0_c=ddc$\\
30. & $\Sigma_c(2520)^0$ -- $2518,48\pm 20$ & $2528,94$ & $+0,41$ & $(50,97/2)$ & $1({\frac{3}{2}}^+)$\\
\hline
   &             &           &         &              & $\Xi^0_c=dsc$\\
31. & $\Xi_c(2645)^0$ -- $2646,16\pm 0,25$ & $2631,65$ & $-0,55$ & $(51,999/2)$ & $\frac{1}{2}({\frac{3}{2}}^+)$\\
32. & $\Xi_c(2815)^0$ -- $2819,79\pm 0,3$ & $2843,20$ & $+0,83$ & $(53,103/2)$ & $\frac{1}{2}({\frac{3}{2}}^-)$\\
\hline
   &             &           &         &              & $\Omega^0_c=ssc$\\
33. & $\Omega_c(2770)^0$ -- $2765,9\pm 2$ & $2789,55$ & $+0,85$ & $(105/2,51)$ & $0({\frac{3}{2}}^+)$\\
\hline
   &             &           &         &              & $\Lambda^0_b=udb$\\
34. & $\Lambda_b(5920)^0$ -- $5920,09$ & $5941,40$ & $+0,36$ & $(153/2,75)$ & $0({\frac{3}{2}}^-)$\\
35. & $\Lambda_b(6146)^0$ -- $6146,2\pm 0,4$ & $6177,48$ & $+0,51$ & $(78,153/2)$ & $0({\frac{3}{2}}^+)$\\
\hline
   &             &           &         &              & $\Xi^0_b=usb$\\
36. & $\Xi_b(5945)^0$ -- $5952,3\pm 0,6$ & $5941,40$ & $-0,18$ & $(153/2,75)$ & $?({\frac{1}{2}}^+)$\\
\hline
\end{tabular}
}
\end{center}
\newpage
\begin{center}
{\textbf{Table 12. Neutral baryons, spin 5/2, $\K=\BH$}\\
(Subspace $\bsH^{(1,0,5/2)}_{\rm phys}(\BH)$, vectors $\left|\K,b,\ell,l-\dot{l}\right\rangle=\left|\BH,1,0,5/2\right\rangle$).}
\vspace{0.1cm}
{\renewcommand{\arraystretch}{1.0}
\begin{tabular}{|c||l|l|c|l|l|}\hline
 & State and mass & $m_\omega$ (theor.) & Error \% & $(l,\dot{l})$ & $qqq$, $I(J^P)$\\ \hline\hline
   &                 &          &         &             & $N^0=udd$\\
1. & $N(1675)^0\approx 1660$    & $1654,11$ & $-0,35$ & $(41,77/2)$ & $\frac{1}{2}({\frac{5}{2}}^-)$\\
2. & $N(1680)^0\approx 1685$    & $1695,49$ & $+1,22$ & $(83/2,39)$ & $\frac{1}{2}({\frac{5}{2}}^+)$\\
3. & $N(1860)^0$ -- $1834\pm 19$ & $1822,74$ & $-0,61$ & $(48,81/2)$ & $\frac{1}{2}({\frac{5}{2}}^+)$\\
4. & $N(2000)^0$ -- $2030\pm 40$ & $2045,02$ & $+0,74$ & $(91/2,43)$ & $\frac{1}{2}({\frac{5}{2}}^+)$\\
5. & $N(2060)^0\approx 2070$    & $2091,01$ & $+1,01$ & $(46,87/2)$ & $\frac{1}{2}({\frac{5}{2}}^-)$\\
6. & $N(2570)^0$ -- $2570^{+19}_{-10}$ & $2579,02$ & $+0,35$ & $(51,97/2)$ & $\frac{1}{2}({\frac{5}{2}}^-)$\\
\hline
   &             &           &         &              & $\Delta^0=udd$\\
7. & $\Delta(1905)^0\approx 1800$ & $1779,81$ & $-1,12$ & $(85/2,40)$ & $\frac{3}{2}({\frac{5}{2}}^+)$\\
8. & $\Delta(1930)^0\approx 1880$ & $1866,17$ & $-0,73$ & $(87/2,41)$ & $\frac{3}{2}({\frac{5}{2}}^-)$\\
9. & $\Delta(2000)^0$ -- $1998\pm 4$ & $1999,54$ & $+0,07$ & $(45,85/2)$ & $\frac{3}{2}({\frac{5}{2}}^+)$\\
10. & $\Delta(2350)^0$ -- $2400\pm 125$ & $2427,25$ & $+1,13$ & $(99/2,47)$ & $\frac{3}{2}({\frac{5}{2}}^-)$\\
\hline
   &             &           &         &              & $\Lambda^0=uds$\\
11. & $\Lambda(1820)^0\approx 1818$ & $1822,74$ & $+0,26$ & $(43,81/2)$ & $0({\frac{5}{2}}^+)$\\
12. & $\Lambda(1830)^0\approx 1830$ & $1822,74$ & $-0,39$ & $(43,81/2)$ & $0({\frac{5}{2}}^-)$\\
13. & $\Lambda(2080)^0$ -- $2070\pm 15$ & $2091,01$ & $+1,01$ & $(46,87/2)$ & $0({\frac{5}{2}}^-)$\\
14. & $\Lambda(2110)^0$ -- $2048\pm 10$ & $2045,02$ & $-0,14$ & $(91/2,43)$ & $0({\frac{5}{2}}^+)$\\
\hline
   &             &           &         &              & $\Sigma^0=uds$\\
15. & $\Sigma(1775)^0\approx 1770$  & $1779,84$ & $+0,55$ & $(85/2,40)$ & $1({\frac{5}{2}}^-)$\\
16. & $\Sigma(1915)^0\approx 1900$  & $1910,12$ & $+0,53$ & $(44,83/2)$ & $1({\frac{5}{2}}^+)$\\
17. & $\Sigma(2070)^0\approx 2060$  & $2045,02$ & $-0,73$ & $(91/2,43)$ & $1({\frac{5}{2}}^+)$\\
\hline
   &             &           &         &              & $\Xi^0=uss$\\
18. & $\Xi(2030)^0$ -- $2025\pm 5$  & $2045,02$ & $+0,98$ & $(91/2,43)$ & $\frac{1}{2}({\frac{5}{2}}^?)$\\
\hline
   &             &           &         &              & $\Lambda^0_b=udb$\\
19. & $\Lambda_b(6152)^0$ -- $6152,5\pm 0,4$ & $6176,46$ & $+0,39$ & $(157/2,76)$ & $?({\frac{5}{2}}^+)$\\
\hline
\end{tabular}
}
\end{center}
\begin{center}
{\textbf{Table 13. Neutral baryons, spin 7/2, $\K=\BH$}\\
(Subspace $\bsH^{(1,0,7/2)}_{\rm phys}(\BH)$, vectors $\left|\K,b,\ell,l-\dot{l}\right\rangle=\left|\BH,1,0,7/2\right\rangle$).}
\vspace{0.1cm}
{\renewcommand{\arraystretch}{1.0}
\begin{tabular}{|c||l|l|c|l|l|}\hline
 & State and mass & $m_\omega$ (theor.) & Error \% & $(l,\dot{l})$ & $qqq$, $I(J^P)$\\ \hline\hline
   &                 &          &         &             & $N^0=udd$\\
1. & $N(1990)^0$ -- $2030\pm 65$    & $2043,49$ & $+0,66$ & $(46,85/2)$ & $\frac{1}{2}({\frac{7}{2}}^+)$\\
2. & $N(2190)^0\approx 2100$    & $2089,48$ & $-0,50$ & $(93/2,43)$ & $\frac{1}{2}({\frac{7}{2}}^-)$\\
\hline
   &             &           &         &              & $\Delta^0=udd$\\
3. & $\Delta(1950)^0\approx 1880$ & $1864,64$ & $-0,82$ & $(44,81/2)$ & $\frac{3}{2}({\frac{7}{2}}^+)$\\
4. & $\Delta(2200)^0\approx 2100$ & $2089,48$ & $-0,50$ & $(93/2,43)$ & $\frac{3}{2}({\frac{7}{2}}^-)$\\
5. & $\Delta(2390)^0$ -- $2223\pm 19$ & $2230,51$ & $+0,34$ & $(48,89/2)$ & $\frac{3}{2}({\frac{7}{2}}^-)$\\
\hline
   &             &           &         &              & $\Lambda^0=uds$\\
6. & $\Lambda(2085)^0$ -- $2041^{+80}_{-82}$ & $2043,49$ & $+0,12$ & $(46,85/2)$ & $0({\frac{7}{2}}^+)$\\
7. & $\Lambda(2100)^0$ -- $2040\pm 14$ & $2043,49$ & $+0,17$ & $(46,85/2)$ & $0({\frac{7}{2}}^-)$\\
\hline
\end{tabular}
}
\end{center}
\begin{center}
{\renewcommand{\arraystretch}{1.0}
\begin{tabular}{|c||l|l|c|l|l|}\hline
 & State and mass & $m_\omega$ (theor.) & Error \% & $(l,\dot{l})$ & $qqq$, $I(J^P)$\\ \hline\hline
   &             &           &         &              & $\Sigma^0=uds$\\
8. & $\Sigma(2030)^0\approx 2020$  & $1998,01$ & $-1,08$ & $(91/2,42)$ & $1({\frac{7}{2}}^+)$\\
9. & $\Sigma(2100)^0$ -- $2093\pm 16$  & $2089,48$ & $-0,17$ & $(93/2,43)$ & $1({\frac{7}{2}}^-)$\\
\hline
\end{tabular}
}
\end{center}
\begin{center}
{\textbf{Table 14. Neutral baryons, spin 9/2, $\K=\BH$}\\
(Subspace $\bsH^{(1,0,9/2)}_{\rm phys}(\BH)$, vectors $\left|\K,b,\ell,l-\dot{l}\right\rangle=\left|\BH,1,0,9/2\right\rangle$).}
\vspace{0.1cm}
{\renewcommand{\arraystretch}{1.0}
\begin{tabular}{|c||l|l|c|l|l|}\hline
 & State and mass & $m_\omega$ (theor.) & Error \% & $(l,\dot{l})$ & $qqq$, $I(J^P)$\\ \hline\hline
   &                 &          &         &             & $N^0=udd$\\
1. & $N(2220)^0\approx 2170$ & $2180,95$ & $+0,50$ & $(48,87/2)$ & $\frac{1}{2}({\frac{9}{2}}^+)$\\
2. & $N(2250)^0\approx 2200$    & $2228,47$ & $+1,29$ & $(97/2,44)$ & $\frac{1}{2}({\frac{9}{2}}^-)$\\
\hline
   &             &           &         &              & $\Delta^0=udd$\\
3. & $\Delta(2300)^0$ -- $2370\pm 80$ & $2374,11$ & $+0,17$ & $(50,91/2)$ & $\frac{3}{2}({\frac{9}{2}}^+)$\\
4. & $\Delta(2400)^0$ -- $2260\pm 60$ & $2276,50$ & $+0,73$ & $(49,89/2)$ & $\frac{3}{2}({\frac{9}{2}}^-)$\\
\hline
   &             &           &         &              & $\Lambda^0=uds$\\
5. & $\Lambda(2350)^0\approx 2350$ & $2374,11$ & $+1,02$ & $(50,91/2)$ & $0({\frac{9}{2}}^+)$\\
\hline
\end{tabular}
}
\end{center}
%\newpage
\begin{center}
{\textbf{Table 15. Neutral baryons, spin 11/2, $\K=\BH$}\\
(Subspace $\bsH^{(1,0,11/2)}_{\rm phys}(\BH)$, vectors $\left|\K,b,\ell,l-\dot{l}\right\rangle=\left|\BH,1,0,11/2\right\rangle$).}
\vspace{0.1cm}
{\renewcommand{\arraystretch}{1.0}
\begin{tabular}{|c||l|l|c|l|l|}\hline
 & State and mass & $m_\omega$ (theor.) & Error \% & $(l,\dot{l})$ & $qqq$, $I(J^P)$\\ \hline\hline
   &                 &          &         &             & $N^0=udd$\\
1. & $N(2600)^0\approx 2600$    & $2624,49$ & $+0,94$ & $(53,95/2)$ & $\frac{1}{2}({\frac{11}{2}}^-)$\\
\hline
   &             &           &         &              & $\Delta^0=udd$\\
2. & $\Delta(2420)^0\approx 2400$ & $2421,12$ & $+0,88$ & $(51,91/2)$ & $\frac{3}{2}({\frac{11}{2}}^+)$\\
\hline
\end{tabular}
}
\end{center}
\begin{center}
{\textbf{Table 16. Neutral baryons, spin 13/2, $\K=\BH$}\\
(Subspace $\bsH^{(1,0,13/2)}_{\rm phys}(\BH)$, vectors $\left|\K,b,\ell,l-\dot{l}\right\rangle=\left|\BH,1,0,13/2\right\rangle$).}
\vspace{0.1cm}
{\renewcommand{\arraystretch}{1.0}
\begin{tabular}{|c||l|l|c|l|l|}\hline
 & State and mass & $m_\omega$ (theor.) & Error \% & $(l,\dot{l})$ & $qqq$, $I(J^P)$\\ \hline\hline
   &                 &          &         &             & $N^0=udd$\\
1. & $N(2700)^0$ -- $2612\pm 45$ & $2621,43$ & $+0,36$ & $(107/2,47)$ & $\frac{1}{2}({\frac{13}{2}}^+)$\\
\hline
   &             &           &         &              & $\Delta^0=udd$\\
2. & $\Delta(2750)^0$ -- $2794\pm 50$ & $2779,33$ & $-0,52$ & $(55,97/2)$ & $\frac{3}{2}({\frac{13}{2}}^-)$\\
\hline
\end{tabular}
}
\end{center}
\begin{center}
{\textbf{Table 17. Neutral baryons, spin 15/2, $\K=\BH$}\\
(Subspace $\bsH^{(1,0,15/2)}_{\rm phys}(\BH)$, vectors $\left|\K,b,\ell,l-\dot{l}\right\rangle=\left|\BH,1,0,15/2\right\rangle$).}
\vspace{0.1cm}
{\renewcommand{\arraystretch}{1.0}
\begin{tabular}{|c||l|l|c|l|l|}\hline
 & State and mass & $m_\omega$ (theor.) & Error \% & $(l,\dot{l})$ & $qqq$, $I(J^P)$\\ \hline\hline
   &             &           &         &              & $\Delta^0=udd$\\
1. & $\Delta(2950)^0$ -- $2990\pm 100$ & $2993,44$ & $+0,11$ & $(115/2,50)$ & $\frac{3}{2}({\frac{15}{2}}^+)$\\
\hline
\end{tabular}
}
\end{center}

The entire spectrum of states of the baryon sector is not limited to Tables 2--17. As noted above, all states of the baryon sector have their charge-conjugated counterparts (antimatter), which leads to a doubling of the number of Tables 2--17. In addition, each table corresponds to only one spin projection, for example, Table 2 (subspace $\bsH^{(1,0,1/2)}_{\rm phys}(\C)$) contains spin 1/2 states with projection 1/2, for the second projection -1/2 we get a duplicate of Table 2 (subspace $\bsH^{(1,0,-1/2)}_{\rm phys}(\C)$). Similarly, for spin 3/2 we have a quadruplet of states -3/2, -1/2, 1/2, 3/2, etc.

For example, for $N$-states we have the following quadruplet: $N(1520)^+$ (state 1 from Table 3), $N(1535)^+$ (state 3 from Table 2) with projections of spin 3/2 and 1/2, respectively, as well as two states $\overline{N}(1520)^+$ and $\overline{N}(1535)^+$ with projections of spin -3/2 and -1/2. Graphically, this can be represented by a diagram (one of the spin chains belonging to the representation cone, see Figure 3)
\begin{equation}\label{3Chain}
\unitlength=1mm
%\begin{center}
\begin{picture}(60,13)
\put(-3.5,12){$\overset{-}{N}(1520)^+$}
\put(-1.5,5){$\overset{(\frac{107}{2},55)}{\bullet}$}
\put(16.5,12){$\overset{-}{N}(1535)^+$}
\put(18.5,5){$\overset{(54,\frac{109}{2})}{\bullet}$}
\put(36.5,12){$\overset{+}{N}(1535)^+$}
\put(38.5,5){$\overset{(\frac{109}{2},54)}{\bullet}$}
\put(56.5,12){$\overset{+}{N}(1520)^+$}
\put(58.5,5){$\overset{(55,\frac{107}{2})}{\bullet}$}
\put(3,6){\line(1,0){20}}
\put(23,6){\line(1,0){20}}
\put(43,6){\line(1,0){20}}
\put(-0.5,0){$-\frac{3}{2}$}
\put(19.5,0){$-\frac{1}{2}$}
\put(42.5,0){$\frac{1}{2}$}
\put(62.5,0){$\frac{3}{2}$}
\put(6,0){$\cdots$}
\put(11.5,0){$\cdots$}
%\put(17.5,0){$\cdots$}
\put(17.5,0){$\cdot$}
\put(26,0){$\cdots$}
\put(31.5,0){$\cdots$}
\put(37.5,0){$\cdots$}
%\put(21,0){$\cdot$}
\put(46,0){$\cdots$}
\put(51.5,0){$\cdots$}
\put(57.5,0){$\cdots$}
\end{picture}
%\end{center}
\end{equation}

Further, three $N$-states $N(1675)^+$ (state 1 from Table 3), $N(1700)^+$ (state 2 from Table 2), $N(1650)^+$ (state 4 from Table 1) with spin projections 5/2, 3/2, 1/2 and the corresponding states $\overline{N}(1675)^+$, $\overline{N}(1700)^+$, $\overline{N}(1660)^+$ with projections -5/2, -3/2, -1/2 form the following sextet:
\begin{equation}\label{Six1}
\unitlength=1mm
%\begin{center}
\begin{picture}(100,13)
\put(-3.5,12){$\overset{-}{N}(1675)^+$}
\put(-1.5,5){$\overset{(55,\frac{115}{2})}{\bullet}$}
\put(16.5,12){$\overset{-}{N}(1700)^+$}
\put(18.5,5){$\overset{(\frac{111}{2},57)}{\bullet}$}
\put(36.5,12){$\overset{-}{N}(1650)^+$}
\put(38.5,5){$\overset{(56,\frac{113}{2})}{\bullet}$}
\put(56.5,12){$\overset{+}{N}(1650)^+$}
\put(58.5,5){$\overset{(\frac{113}{2},56)}{\bullet}$}
\put(76.5,12){$\overset{+}{N}(1700)^+$}
\put(78.5,5){$\overset{(57,\frac{111}{2})}{\bullet}$}
\put(96.5,12){$\overset{+}{N}(1675)^+$}
\put(98.5,5){$\overset{(\frac{115}{2},55)}{\bullet}$}
\put(3,6){\line(1,0){20}}
\put(23,6){\line(1,0){20}}
\put(43,6){\line(1,0){20}}
\put(63,6){\line(1,0){20}}
\put(83,6){\line(1,0){20}}
\put(-0.5,0){$-\frac{5}{2}$}
\put(19.5,0){$-\frac{3}{2}$}
\put(39.5,0){$-\frac{1}{2}$}
\put(62.5,0){$\frac{1}{2}$}
\put(82.5,0){$\frac{3}{2}$}
\put(102.5,0){$\frac{5}{2}$}
\put(6,0){$\cdots$}
\put(11.5,0){$\cdots$}
%\put(17.5,0){$\cdots$}
\put(17.5,0){$\cdot$}
\put(26,0){$\cdots$}
\put(31.5,0){$\cdots$}
%\put(37.5,0){$\cdots$}
\put(37.5,0){$\cdot$}
\put(46,0){$\cdots$}
\put(51.5,0){$\cdots$}
\put(57.5,0){$\cdots$}
\put(66,0){$\cdots$}
\put(71.5,0){$\cdots$}
\put(77.5,0){$\cdots$}
\put(86,0){$\cdots$}
\put(91.5,0){$\cdots$}
\put(97.5,0){$\cdots$}
\end{picture}
\end{equation}

For $N$-states we have two more sextets
\begin{equation}\label{Six2}
\unitlength=1mm
%\begin{center}
\begin{picture}(100,13)
\put(-3.5,12){$\overset{-}{N}(1680)^+$}
\put(-1.5,5){$\overset{(56,\frac{117}{2})}{\bullet}$}
\put(16.5,12){$\overset{-}{N}(1720)^+$}
\put(18.5,5){$\overset{(\frac{113}{2},58)}{\bullet}$}
\put(36.5,12){$\overset{-}{N}(1710)^+$}
\put(38.5,5){$\overset{(57,\frac{115}{2})}{\bullet}$}
\put(56.5,12){$\overset{+}{N}(1710)^+$}
\put(58.5,5){$\overset{(\frac{115}{2},57)}{\bullet}$}
\put(76.5,12){$\overset{+}{N}(1720)^+$}
\put(78.5,5){$\overset{(58,\frac{113}{2})}{\bullet}$}
\put(96.5,12){$\overset{+}{N}(1680)^+$}
\put(98.5,5){$\overset{(\frac{117}{2},56)}{\bullet}$}
\put(3,6){\line(1,0){20}}
\put(23,6){\line(1,0){20}}
\put(43,6){\line(1,0){20}}
\put(63,6){\line(1,0){20}}
\put(83,6){\line(1,0){20}}
\put(-0.5,0){$-\frac{5}{2}$}
\put(19.5,0){$-\frac{3}{2}$}
\put(39.5,0){$-\frac{1}{2}$}
\put(62.5,0){$\frac{1}{2}$}
\put(82.5,0){$\frac{3}{2}$}
\put(102.5,0){$\frac{5}{2}$}
\put(6,0){$\cdots$}
\put(11.5,0){$\cdots$}
%\put(17.5,0){$\cdots$}
\put(17.5,0){$\cdot$}
\put(26,0){$\cdots$}
\put(31.5,0){$\cdots$}
%\put(37.5,0){$\cdots$}
\put(37.5,0){$\cdot$}
\put(46,0){$\cdots$}
\put(51.5,0){$\cdots$}
\put(57.5,0){$\cdots$}
\put(66,0){$\cdots$}
\put(71.5,0){$\cdots$}
\put(77.5,0){$\cdots$}
\put(86,0){$\cdots$}
\put(91.5,0){$\cdots$}
\put(97.5,0){$\cdots$}
\end{picture}
%\end{center}
\end{equation}
\begin{equation}\label{Six3}
\unitlength=1mm
%\begin{center}
\begin{picture}(100,13)
\put(-3.5,12){$\overset{-}{N}(1860)^+$}
\put(-1.5,5){$\overset{(59,\frac{123}{2})}{\bullet}$}
\put(16.5,12){$\overset{-}{N}(1900)^+$}
\put(18.5,5){$\overset{(\frac{119}{2},61)}{\bullet}$}
\put(36.5,12){$\overset{-}{N}(1880)^+$}
\put(38.5,5){$\overset{(60,\frac{121}{2})}{\bullet}$}
\put(56.5,12){$\overset{+}{N}(1880)^+$}
\put(58.5,5){$\overset{(\frac{121}{2},60)}{\bullet}$}
\put(76.5,12){$\overset{+}{N}(1900)^+$}
\put(78.5,5){$\overset{(61,\frac{119}{2})}{\bullet}$}
\put(96.5,12){$\overset{+}{N}(1860)^+$}
\put(98.5,5){$\overset{(\frac{123}{2},59)}{\bullet}$}
\put(3,6){\line(1,0){20}}
\put(23,6){\line(1,0){20}}
\put(43,6){\line(1,0){20}}
\put(63,6){\line(1,0){20}}
\put(83,6){\line(1,0){20}}
\put(-0.5,0){$-\frac{5}{2}$}
\put(19.5,0){$-\frac{3}{2}$}
\put(39.5,0){$-\frac{1}{2}$}
\put(62.5,0){$\frac{1}{2}$}
\put(82.5,0){$\frac{3}{2}$}
\put(102.5,0){$\frac{5}{2}$}
\put(6,0){$\cdots$}
\put(11.5,0){$\cdots$}
%\put(17.5,0){$\cdots$}
\put(17.5,0){$\cdot$}
\put(26,0){$\cdots$}
\put(31.5,0){$\cdots$}
%\put(37.5,0){$\cdots$}
\put(37.5,0){$\cdot$}
\put(46,0){$\cdots$}
\put(51.5,0){$\cdots$}
\put(57.5,0){$\cdots$}
\put(66,0){$\cdots$}
\put(71.5,0){$\cdots$}
\put(77.5,0){$\cdots$}
\put(86,0){$\cdots$}
\put(91.5,0){$\cdots$}
\put(97.5,0){$\cdots$}
\end{picture}
%\end{center}
\end{equation}

Sextet (\ref{Six2}) is formed by states $N(1790)^+$ (state 5 from Table 1), $N(1720)^+$ (state 3 from Table 2), $N(1680)^+$ (state 2 from Table 3) with spin projections 1/2, 3/2, 5/2 and corresponding dual projections -1/2, -3/2, -5/2. In turn, the sextet (\ref{Six3}) is formed by states $N(1860)^+$ (state 6 from Table 1), $N(1990)^+$ (state from Table 2), $N(1860)^+$ (state, Table 3) with spin projections 1/2, 3/2, 5/2 and dual projections -1/2,-3/2, -5/2.

\subsection{Meson sector\label{sec5_2}}
The meson sector of the matter spectrum contains coherent subspaces  $\bsH^{(b,\ell,s)}_{\rm phys}(\K)$ with $b=0$, $\ell=0$ and integer spin $s=l-\dot{l}=0,\pm 1,\pm 2,\pm 3,\pm 4\pm 5, \pm 6$. At the same time, unlike the baryon sector, all three charge realizations take place: 1) $\K=\C$ -- charged states; 2) $\K=\BH$ -- neutral states; 3) $\K=\R$ -- truly neutral states.

\subsubsection{$\C$-subspaces/charged mesons\label{sec5_2_1}}

All charged states (vectors $\left|\C,0,0,s\right\rangle$) of the meson sector belong to the coherent subspaces $\bsH^{(0,0,s)}_{\rm phys}(\C)$ of the physical $\K$-Hilbert space $\bsH_{\rm phys}(\K)$ (the spectrum of matter).
\newpage
\begin{center}
{\textbf{Table 18. Charged mesons, spin 0, $\K=\C$}\\
(Subspace $\bsH^{(0,0,0)}_{\rm phys}(\C)$, vectors $\left|\K,b,\ell,l-\dot{l}\right\rangle=\left|\C,0,0,0\right\rangle$).}
\vspace{0.1cm}
{\renewcommand{\arraystretch}{1.0}
\begin{tabular}{|c||l|l|c|l|l|}\hline
 & State and mass & $m_\omega$ & Error \% & $(l,\dot{l})$ & $q\bar{q}$, $I^G(J^{PC})$\\ \hline\hline
   &                 &          &         &             & $\pi^+,a_0^+=u\bar{d}$\\
   &                 &          &         &             & $\pi^-,a_0^-=\bar{u}d$\\
1. & $\pi^\pm$ -- $139,57$ & $135,16$ & $-3,16$ & $(11,11)$ & $1^-(0^-)$\\
2. & $a_0(980)^\pm$ -- $980\pm 20$ & $982,14$ & $+0,29$ & $(61/2,61/2)$ & $1^-(0^{++})$\\
3. & $\pi(1300)^\pm$ -- $1300\pm 100$ & $1287,97$ & $-0,92$ & $(35,35)$ & $1^-(0^{-+})$\\
4. & $a_0(1450)^\pm$ -- $1474\pm 19$  & $1475,77$ & $+0,13$ & $(75/2,75/2)$ & $1^-(0^{++})$\\
5. & $a_0(1700)^\pm$ -- $1704\pm 5$  & $1717,98$ & $+0,82$ & $(81/2,81/2)$ & $1^-(0^{++})$\\
6. & $\pi(1800)^\pm$ -- $1800^{+9}_{-11}$ & $1802,81$ & $-0,39$ & $(83/2,83/2)$ & $1^-(0^{-+})$\\
7. & $a_0(1950)^\pm$ -- $1931\pm 26$ & $1933,88$ & $+0,15$ & $(43,43)$ & $1^-(0^{++})$\\
8. & $a_0(2020)^\pm$ -- $2025\pm 30$ & $2023,81$ & $-0,06$ & $(44,44)$ & $1^-(0^{++})$\\
9. & $\pi(2070)^\pm$ -- $2070\pm 35$ & $2069,55$ & $-0,02$ & $(89/2,89/2)$ & $1^-(0^{-+})$\\
10. & $\pi(2360)^\pm$ -- $2060\pm 60$ & $2354,69$ & $-0,22$ & $(95/2,95/2)$ & $1^-(0^{-+})$\\
\hline
   &                 &          &         &             & $K^\pm=u\bar{s},\bar{u}s$\\
   &                 &          &         &             & $K^\ast_0{}^\pm=d\bar{s},\bar{d}s$\\
11. & $K^\pm$ -- $493,677\pm 0,016$ & $494,65$ & $+0,19$ & $(43/2,43/2)$ & $\frac{1}{2}(0^-)$\\
12. & $K^\ast_0(700)^\pm$ -- $845\pm 17$ & $859,50$ & $-1,71$ & $(57/2,57/2)$ & $\frac{1}{2}(0^+)$\\
13. & $K^\ast_0(1430)^\pm$ -- $1425\pm 50$ & $1437,18$ & $+0,85$ & $(37,37)$ & $\frac{1}{2}(0^+)$\\
14. & $K(1460)^\pm$ -- $1482,40\pm 3,58$ & $1475,77$ & $-0,45$ & $(75/2,75/2)$ & $\frac{1}{2}(0^-)$\\
15. & $K(1830)^\pm$ -- $1874\pm 43$ & $1889,68$ & $+0,83$ & $(85/2,85/2)$ & $\frac{1}{2}(0^-)$\\
16. & $K^\ast_0(1950)^\pm$ -- $1944\pm 18$ & $1933,88$ & $-0,52$ & $(43,43)$ & $\frac{1}{2}(0^+)$\\
\hline   &                 &          &         &             & $D^\pm=c\bar{d},\bar{c}d$\\
17. & $D^\pm$ -- $1869,65\pm 0,05$ & $1889,68$ & $+1,07$ & $(85/2,85/2)$ & $\frac{1}{2}(0^-)$\\
18. & $D^\ast_0(2300)^\pm$ -- $2343\pm 10$ & $2354,69$ & $+0,49$ & $(95/2,95/2)$ & $\frac{1}{2}(0^+)$\\
\hline
   &                 &          &         &             & $D^\pm=c\bar{s},\bar{c}s$\\
19. & $D^\pm_s$ -- $1968,35\pm 0,07$ & $1978,59$ & $+0,52$ & $(87/2,87/2)$ & $0(0^-)$\\
20. & $D^\ast_{s0}(2317)^\pm$ -- $2317,8\pm 0,5$ & $2305,89$ & $-0,51$ & $(47,47)$ & $0(0^+)$\\
21. & $D_{s0}(2590)^+$ -- $2591\pm 9$ & $2606,35$ & $+0,61$ & $(50,50)$ & $0(0^-)$\\
\hline
   &                 &          &         &             & $B^\pm=u\bar{b},\bar{u}b$\\
22. & $B^\pm$ -- $5279,34\pm 0,12$ & $5298,05$ & $+0,35$ & $(143/2,143/2)$ & $\frac{1}{2}(0^-)$\\
\hline
   &                 &          &         &             & $B^\pm_c=c\bar{b},\bar{c}b$\\
23. & $B^+_c$ -- $6274,47\pm 0,32$ & $6217,85$ & $-0,90$ & $(78,78)$ & $0(0^-)$\\
24. & $B_c(2S)^\pm$ -- $6871,2\pm 1,0$ & $6871,93$ & $+0,01$ & $(163/2,163/2)$ & $0(0^-)$\\
\hline
\end{tabular}
}
\end{center}
\newpage
\begin{center}
{\textbf{Table 19. Charged mesons, spin 1, $\K=\C$}\\
(Subspace $\bsH^{(0,0,1)}_{\rm phys}(\C)$, vectors $\left|\K,b,\ell,l-\dot{l}\right\rangle=\left|\C,0,0,1\right\rangle$).}
\vspace{0.1cm}
{\renewcommand{\arraystretch}{1.0}
\begin{tabular}{|c||l|l|c|l|l|}\hline
 & State and mass & $m_\omega$ & Error \% & $(l,\dot{l})$ & $q\bar{q}$, $I^G(J^{PC})$\\ \hline\hline
   &                 &          &         &             & $\rho^\pm=u\bar{d},\bar{u}d$\\
1. & $\rho(770)^\pm$ -- $775,26\pm 0,23$ & $772,63$ & $-0,34$ & $(55/2,53/2)$ & $1^+(1^{--})$\\
2. & $b_1(1235)^\pm$ -- $1229,5\pm 3,2$ & $1216,18$ & $-1,08$ & $(69/2,67/2)$ & $1^+(1^{+-})$\\
3. & $a_1(1260)^\pm$ -- $1230\pm 40$ & $1251,69$ & $+1,76$ & $(35,34)$ & $1^-(1^{++})$\\
4. & $\pi_1(1400)^\pm$ -- $1354\pm 25$ & $1361,30$ & $+0,54$ & $(73/2,71/2)$ & $1^-(1^{-+})$\\
5. & $\rho(1450)^\pm$ -- $1465\pm 25$ & $1475,51$ & $+0,72$ & $(38,37)$ & $1^+(1^{--})$\\
6. & $\rho(1570)^\pm$ --$1570\pm 70$ & $1554,21$ & $-1,00$ & $(39,38)$ & $1^+(1^{--})$\\
7. & $\pi_1(1600)^\pm$ -- $1661^{+15}_{-11}$ & $1676,08$ & $+0,91$ & $(81/2,79/2)$ & $1^-(1^{-+})$\\
8. & $a_1(1640)^\pm$ -- $1655\pm 16$ & $1634,94$ & $-1,21$ & $(40,39)$ & $1^-(1^{++})$\\
9. & $\rho(1700)^\pm$ -- $1720\pm 20$ & $1717,73$ & $-0,13$ & $(41,40)$ & $1^+(1^{--})$\\
10. & $\rho(1900)^\pm$ -- $1909\pm 17$ & $1889,42$ & $-1,02$ & $(43,42)$ & $1^+(1^{--})$\\
11. & $a_1(1930)^\pm$ -- $1930^{+30}_{-70}$ & $1933,62$ & $+0,19$ & $(87/2,85/2)$ & $1^-(1^{++})$\\
12. & $b_1(1960)^\pm$ -- $1960\pm 35$ & $1978,33$ & $+0,93$ & $(44,43)$ & $1^+(1^{+-})$\\
13. & $\rho(2000)^\pm$ -- $2000\pm 30$ & $1978,33$ & $-1,08$ & $(44,43)$ & $1^+(1^{--})$\\
14. & $\pi_1(2015)^\pm$ -- $2014\pm 20$ & $2023,56$ & $+0,47$ & $(89/2,87/2)$ & $1^-(1^{-+})$\\
15. & $a_1(2095)^\pm$ -- $2096\pm 17$ & $2069,29$ & $-1,27$ & $(45,44)$ & $1^-(1^{++})$\\
16. & $\rho(2150)^\pm$ -- $2111\pm 43$ & $2115,54$ & $+0,21$ & $(91/2,89/2)$ & $1^+(1^{--})$\\
17. & $b_1(2240)^\pm$ -- $2240\pm 35$ & $2257,34$ & $+0,77$ & $(47,46)$ & $1^+(1^{+-})$\\
18. & $\rho(2270)^\pm$ -- $2265\pm 40$ & $2257,34$ & $-0,34$ & $(47,46)$ & $1^+(1^{--})$\\
19. & $a_1(2270)^\pm$ -- $2270^{+50}_{-40}$ & $2257,34$ & $-0,56$ & $(47,46)$ & $1^-(1^{++})$\\
20. & $Z_{cs}(4000)^+$ -- $4003\pm 6$ & $3991,03$ & $-0,29$ & $(125/2,123/2)$ & $\frac{1}{2}(1^+)$, \\
&&&&& exotic state\\
21. & $Z_{cs}(4220)^+$ -- $4216\pm 24$ & $4251,52$ & $+0,84$ & $(129/2,127/2)$ & $\frac{1}{2}(1^+)$, \\
&&&&& exotic state\\
\hline
   &                 &          &         &             & $K^\pm=u\bar{s},\bar{u}s$\\
22. & $K^\ast(892)^\pm$ -- $891,67\pm 0,26$ & $889,14$ & $-0,28$ & $(59/2,57/2)$ & $\frac{1}{2}(1^-)$\\
23. & $K_1(1270)^\pm$ -- $1253\pm 7$ & $1251,69$ & $-0,10$ & $(35,34)$ & $\frac{1}{2}(1^+)$\\
24. & $K_1(1400)^\pm$ -- $1403\pm 7$ & $1398,86$ & $-0,29$ & $(37,36)$ & $\frac{1}{2}(1^+)$\\
25. & $K^\ast(1410)^\pm$ -- $1414\pm 15$ & $1398,86$ & $-1,07$ & $(37,36)$ & $\frac{1}{2}(1^-)$\\
26. & $K_1(1650)^\pm$ -- $1650\pm 50$ & $1634,94$ & $-0,91$ & $(40,39)$ & $\frac{1}{2}(1^+)$\\
27. & $K^\ast(1680)^\pm$ $1718\pm 18$ & $1717,73$ & $-0,01$ & $(41,40)$ & $\frac{1}{2}(1^-)$\\
\hline
   &                 &          &         &             & $D^\pm=c\bar{d},\bar{c}d$\\
28. & $D^\ast(2010)^\pm$ -- $2010,26\pm 0,05$ & $2023,56$ & $+0,66$ & $(89/2,87/2)$ & $\frac{1}{2}(1^-)$\\
29. & $D_1(2420)^\pm$ -- $2422,1\pm 0,6$ & $2403,74$ & $-0,76$ & $(97/2,95/2)$ & $\frac{1}{2}(1^+)$\\
\hline
   &                 &          &         &             & $D^\pm=c\bar{s},\bar{c}s$\\
30. & $D_{s1}(2460)^\pm$ -- $2459,5\pm 0,6$ & $2453,56$ & $-0,24$ & $(139/2,137/2)$ & $0(1^+)$\\
31. & $D_{s1}(2536)^\pm$ -- $2535,11\pm 0,06$ & $2554,74$ & $+0,77$ & $(50,49)$ & $0(1^+)$\\
32. & $D^\ast_{s1}(2700)^\pm$ -- $2714\pm 5$ & $2710,34$ & $-0,13$ & $(103/2,101/2)$ & $0(1^-)$\\
33. & $D^\ast_{s1}(2860)^\pm$ -- $2859\pm 27$ & $2870,54$ & $+0,40$ & $(53,52)$ & $0(1^-)$\\
\hline
   &                 &          &         &             & $B^\pm=u\bar{b},\bar{u}b$\\
34. & $B^\ast{}^\pm$ -- $5324,71\pm 0,21$ & $5297,79$ & $-0,50$ & $(72,71)$ & $\frac{1}{2}(1^-)$\\
35. & $B_1(5721)^+$ -- $5725,9^{+2,5}_{-2,7}$ & $5748,49$ & $+0,39$ & $(75,74)$ & $\frac{1}{2}(1^+)$\\
\hline
36. & $W^\pm$ -- $80379,9\pm 1,2$ & $80410,96$ & $+0,04$ & $(561/2,559/2)$ & \\
\hline
\end{tabular}
}
\end{center}
\newpage
\begin{center}
{\textbf{Table 20. Charged mesons, spin 2, $\K=\C$}\\
(Subspace $\bsH^{(0,0,2)}_{\rm phys}(\C)$, vectors $\left|\K,b,\ell,l-\dot{l}\right\rangle=\left|\C,0,0,2\right\rangle$).}
\vspace{0.1cm}
{\renewcommand{\arraystretch}{1.0}
\begin{tabular}{|c||l|l|c|l|l|}\hline
 & State and mass & $m_\omega$ & Error \% & $(l,\dot{l})$ & $q\bar{q}$, $I^G(J^{PC})$\\ \hline\hline
1. & $a_2(1320)^\pm$ -- $1318,2\pm 0,6$ & $1323,49$ & $+0,40$ & $(73/2,69/2)$ & $1^-(2^{++})$\\
2. & $\pi_2(1670)^\pm$ -- $1670,6^{+2,9}_{-1,2}$ & $1675,31$ & $+0,28$ & $(41,39)$ & $1^-(2^{-+})$\\
3. & $a_2(1700)^\pm$ -- $1698\pm 40$ & $1716,96$ & $+1,11$ & $(83/2,79/2)$ & $1^-(2^{++})$\\
4. & $\pi_2(1880)^\pm$ -- $1874^{+26}_{-5}$ & $1888,65$ & $+0,78$ & $(87/2,53/2)$ & $1^-(2^{-+})$\\
5. & $\rho_2(1940)^\pm$ -- $1940\pm 40$ & $1932,86$ & $-0,37$ & $(44,42)$ & $1^+(2^{--})$\\
6. & $a_2(1950)^\pm$ -- $1950^{+30}_{-70}$ & $1932,86$ & $-0,88$ & $(44,42)$ & $1^-(2^{++})$\\
7. & $a_2(1990)^\pm$ -- $2050\pm 10$ & $2068,53$ & $+0,90$ & $(91/2,87/2)$ & $1^-(2^{++})$\\
8. & $\pi_2(2005)^\pm$ -- $1953^{+17}_{-27}$ & $1977,57$ & $+0,74$ & $(89/2,85/2)$ & $1^-(2^{-+})$\\
9. & $a_2(2030)^\pm$ -- $2030\pm 20$ & $2022,79$ & $-0,35$ & $(45,43)$ & $1^-(2^{++})$\\
10. & $\pi_2(2100)^\pm$ -- $2090\pm 29$ & $2114,77$ & $+1,18$ & $(46,44)$ & $1^-(2^{-+})$\\
11. & $a_2(2175)^\pm$ -- $2175\pm 40$ & $2161,53$ & $-0,62$ & $(93/2,89/2)$ & $1^-(2^{++})$\\
12. & $\rho_2(2225)^\pm$ -- $2225\pm 35$ & $2208,80$ & $-0,73$ & $(47,45)$ & $1^+(2^{--})$\\
13. & $a_2(2255)^\pm$ -- $2255\pm 20$ & $2256,57$ & $+0,07$ & $(95/2,91/2)$ & $1^-(2^{++})$\\
14. & $\pi_2(2285)^\pm$ -- $2285\pm 20$ & $2304,86$ & $+0,87$ & $(48,46)$ & $1^-(2^{-+})$\\
\hline
   &                 &          &         &             & $K^\pm=u\bar{s},\bar{u}s$\\
15. & $K^\ast_2(1430)^\pm$ -- $1427,3\pm 1,5$ & $1436,16$ & $+0,62$ & $(38,36)$ & $\frac{1}{2}(2^+)$\\
16. & $K_2(1580)^\pm\approx 1580$ & $1593,55$ & $+0,85$ & $(40,38)$ & $\frac{1}{2}(2^-)$\\
17. & $K_2(1770)^\pm$ -- $1773\pm 8$ & $1759,17$ & $-0,78$ & $(42,40)$ & $\frac{1}{2}(2^-)$\\
18. & $K_2(1820)^\pm$ -- $1819\pm 12$ & $1801,78$ & $-0,94$ & $(85/2,81/2)$ & $\frac{1}{2}(2^-)$\\
19. & $K^\ast_2(1980)^\pm$ -- $1994^{+60}_{-50}$ & $1977,57$ & $-0,82$ & $(89/2,85/2)$ & $\frac{1}{2}(2^+)$\\
20. & $K_2(2250)^\pm$ -- $2247\pm 17$ & $2256,57$ & $+0,42$ & $(95/2,91/2)$ & $\frac{1}{2}(2^-)$\\
\hline
   &                 &          &         &             & $D^\pm=c\bar{d},\bar{c}d$\\
21. & $D^\ast_2(2460)^\pm$ -- $2461,1^{+0,7}_{-0,8}$ & $2452,75$ & $-0,34$ & $(99/2,95/2)$ & $\frac{1}{2}(2^+)$\\
\hline
   &                 &          &         &             & $D^\pm=c\bar{s},\bar{c}s$\\
22. & $D_{s2}(2573)^\pm$ -- $2569,1\pm 0,8$ & $2553,98$ & $-0,59$ & $(101/2,97/2)$ & $0(2^+)$\\
\hline
   &                 &          &         &             & $B^\pm=u\bar{b},\bar{u}b$\\
23. & $B^\ast_2(5747)^+$ -- $5737,2\pm 0,7$ & $5747,73$ & $+0,18$ & $(151/2,147/2)$ & $\frac{1}{2}(2^+)$\\
\hline
\end{tabular}
}
\end{center}
\begin{center}
{\textbf{Table 21. Charged mesons, spin 3, $\K=\C$}\\
(Subspace $\bsH^{(0,0,3)}_{\rm phys}(\C)$, vectors $\left|\K,b,\ell,l-\dot{l}\right\rangle=\left|\C,0,0,3\right\rangle$).}
\vspace{0.1cm}
{\renewcommand{\arraystretch}{1.0}
\begin{tabular}{|c||l|l|c|l|l|}\hline
 & State and mass & $m_\omega$ & Error \% & $(l,\dot{l})$ & $q\bar{q}$, $I^G(J^{PC})$\\ \hline\hline
1. & $\rho_3(1690)^\pm$ -- $1688,8\pm 2,1$ & $1674,04$ & $-0,87$ & $(83/2,77/2)$ & $1^+(3^{--})$\\
2. & $a_3(1875)^\pm$ -- $1874\pm 43$ & $1887,38$ & $+0,71$ & $(44,41)$ & $1^-(3^{++})$\\
3. & $\rho_3(1990)^\pm$ -- $1982\pm 14$ & $1976,29$ & $-0,29$ & $(45,42)$ & $1^+(3^{--})$\\
4. & $b_3(2030)^\pm$ -- $2032\pm 12$ & $2021,52$ & $-0,51$ & $(91/2,85/2)$ & $1^+(3^{+-})$\\
5. & $a_3(2030)^\pm$ -- $2031 \pm 12$ & $2021,52$ & $-0,47$ & $(91/2,85/2)$ & $1^-(3^{++})$\\
6. & $b_3(2245)^\pm$ -- $2245\pm 50$ & $2255,29$ & $+0,46$ & $(48,45)$ & $1^+(3^{+-})$\\
7. & $\rho_3(2250)^\pm$ -- $2248\pm 17$ & $2255,29$ & $+0,32$ & $(48,45)$ & $1^+(3^{--})$\\
8. & $a_3(2275)^\pm$ -- $2275\pm 35$ & $2303,59$ & $+1,25$ & $(97/2,91/2)$ & $1^-(3^{++})$\\
\hline
\end{tabular}
}
\end{center}
\begin{center}
{\renewcommand{\arraystretch}{1.0}
\begin{tabular}{|c||l|l|c|l|l|}\hline
 & State and mass & $m_\omega$ & Error \% & $(l,\dot{l})$ & $q\bar{q}$, $I^G(J^{PC})$\\ \hline\hline
   &                 &          &         &             & $K^\pm=u\bar{s},\bar{u}s$\\
9. & $K^\ast_3(1780)^\pm$ -- $1778\pm 8$ & $1757,84$ & $-1,13$ & $(85/2,79/2)$ & $\frac{1}{2}(3^-)$\\
10. & $K_3(2320)^\pm$ -- $2324\pm 24$ & $2303,59$ & $-0,88$ & $(97/2,91/2)$ & $\frac{1}{2}(3^+)$\\
\hline
   &                 &          &         &             & $D^\pm=c\bar{d},\bar{c}d$\\
11. & $D^\ast_3(2750)^\pm$ -- $2763,1\pm 3,2$ & $2761,19$ & $-0,07$ & $(53,50)$ & $\frac{1}{2}(3^-)$\\
\hline
   &                 &          &         &             & $D^\pm=c\bar{s},\bar{c}s$\\
12. & $D^\ast_{s3}(2860)^\pm$ -- $2860\pm 7$  & $2868,49$ & $+0,29$ & $(54,51)$ & $0(3^-)$\\
\hline
\end{tabular}
}
\end{center}
\begin{center}
%\newpage
{\textbf{Table 22. Charged mesons, spin 4, $\K=\C$}\\
(Subspace $\bsH^{(0,0,4)}_{\rm phys}(\C)$, vectors $\left|\K,b,\ell,l-\dot{l}\right\rangle=\left|\C,0,0,4\right\rangle$).}
\vspace{0.1cm}
{\renewcommand{\arraystretch}{1.0}
\begin{tabular}{|c||l|l|c|l|l|}\hline
 & State and mass & $m_\omega$ & Error \% & $(l,\dot{l})$ & $q\bar{q}$, $I^G(J^{PC})$\\ \hline\hline
1. & $a_4(1970)^\pm$ -- $1967\pm 16$ & $1974,50$ & $+0,38$ & $(91/2,83/2)$ & $1^-(4^{++})$\\
2. & $\rho_4(2230)^\pm$ -- $2230\pm 25$ & $2253,51$ & $+1,05$ & $(97/2,89/2)$ & $1^+(4^{--})$\\
3. & $\pi_4(2250)^\pm$ -- $2250\pm 15$ & $2253,51$ & $+0,15$ & $(97/2,89/2)$ & $1^-(4^{-+})$\\
4. & $a_4(2255)^\pm$ -- $2237\pm 5$ & $2253,51$ & $+0,74$ & $(97/2,89/2)$ & $1^-(4^{++})$\\
\hline
   &                 &          &         &             & $K^\pm=u\bar{s},\bar{u}s$\\
5. & $K^\ast_4(2045)^\pm$ -- $2048^{+8}_{-9}$ & $2065,46$ & $+0,85$ & $(93/2,85/2)$ & $\frac{1}{2}(4^+)$\\
6. & $K_4(2500)^\pm$ -- $2247\pm 17$ & $2253,51$ & $+0,29$ & $(97/2,89/2)$ & $\frac{1}{2}(4^-)$\\
\hline
\end{tabular}
}
\end{center}
\begin{center}
{\textbf{Table 23. Charged mesons, spin 5, $\K=\C$}\\
(Subspace $\bsH^{(0,0,5)}_{\rm phys}(\C)$, vectors $\left|\K,b,\ell,l-\dot{l}\right\rangle=\left|\C,0,0,5\right\rangle$).}
\vspace{0.1cm}
{\renewcommand{\arraystretch}{1.0}
\begin{tabular}{|c||l|l|c|l|l|}\hline
 & State and mass & $m_\omega$ & Error \% & $(l,\dot{l})$ & $q\bar{q}$, $I^G(J^{PC})$\\ \hline\hline
1. & $\rho_5(2350)^\pm$ -- $2330\pm 35$ & $2348,30$ & $+0,78$ & $(50,45)$ & $1^+(5^{--})$\\
\hline
   &                 &          &         &             & $K^\pm=u\bar{s},\bar{u}s$\\
2. & $K^\ast_5(2380)^\pm$ -- $2382\pm 24$ & $2397,61$ & $+0,65$ & $(101/2,91/2)$ & $\frac{1}{2}(5^-)$\\
\hline
\end{tabular}
}
\end{center}
\begin{center}
{\textbf{Table 24. Charged mesons, spin 6, $\K=\C$}\\
(Subspace $\bsH^{(0,0,6)}_{\rm phys}(\C)$, vectors $\left|\K,b,\ell,l-\dot{l}\right\rangle=\left|\C,0,0,6\right\rangle$).}
\vspace{0.1cm}
{\renewcommand{\arraystretch}{1.0}
\begin{tabular}{|c||l|l|c|l|l|}\hline
 & State and mass & $m_\omega$ & Error \% & $(l,\dot{l})$ & $q\bar{q}$, $I^G(J^{PC})$\\ \hline\hline
1. & $a_6(2450)^\pm$ -- $2450\pm 130$ & $2444,62$ & $-0,22$ & $(103/2,91/2)$ & $1^-(6^{++})$\\
\hline
\end{tabular}
}
\end{center}

\subsubsection{$\BH$-subspaces/neutral mesons\label{sec5_2_2}}

The neutral states of the meson sector form coherent subspaces $\bsH^{(0,0,s)}_{\rm phys}(\BH)$, i.e. quaternionic subspaces of the $\K$-Hilbert space (Tables 25--30). As in the case of $\C$-subspaces, $\BH$-subspaces have counterparts (antimatter): $\bsH^{(0,0,s)}_{\rm phys}(\overline{\BH})$.
\newpage
\begin{center}
{\textbf{Table 25. Neutral mesons, spin 0, $\K=\BH$}\\
(Subspace $\bsH^{(0,0,0)}_{\rm phys}(\BH)$, vectors $\left|\K,b,\ell,l-\dot{l}\right\rangle=\left|\BH,0,0,0\right\rangle$).}
\vspace{0.1cm}
{\renewcommand{\arraystretch}{1.0}
\begin{tabular}{|c||l|l|c|l|l|}\hline
 & State and mass & $m_\omega$ & Error \% & $(l,\dot{l})$ & $q\bar{q}$, $I^G(J^{PC})$\\ \hline\hline
   &                 &          &         &             & $K^0=d\bar{s}$,\\
   &&&&&$\bar{K}^0=\bar{d}s$\\
1. & $K^0$ -- $497,611\pm 0,013$ & $494,65$ & $-0,59$ & $(43/2,43/2)$ & $\frac{1}{2}(0^-)$\\
2. & $K^\ast_0(700)^0$ -- $845\pm 17$ & $859,50$ & $+1,71$ & $(57/2,57/2)$ & $\frac{1}{2}(0^+)$\\
3. & $K^\ast_0(1430)^0$ -- $1425\pm 50$ & $1437,18$ & $+0,85$ & $(37,37)$ & $\frac{1}{2}(0^+)$\\
4. & $K(1460)^0$ -- $1482,4\pm 3,58$ & $1475,77$ & $-0,45$ & $(75/2,75/2)$ & $\frac{1}{2}(0^-)$\\
5. & $K(1830)^0$ -- $1874\pm 43$ & $1889,68$ & $+0,83$ & $(85/2,85/2)$ & $\frac{1}{2}(0^-)$\\
6. & $K^\ast_0(1950)^0$ -- $1944\pm 18$ & $1933,88$ & $-0,52$ & $(43,43)$ & $\frac{1}{2}(0^+)$\\
\hline
   &                 &          &         &             & $D^0=c\bar{u}$,\\
   &&&&&$\bar{D}^0=\bar{c}u$\\
7. & $D^0$ -- $1864,84\pm 0,05$ & $1845,99$ & $-1,01$ & $(85/2,85/2)$ & $\frac{1}{2}(0^-)$\\
8. & $D^\ast_0(2300)^0$ -- $2343\pm 10$ & $2354,69$ & $+0,49$ & $(95/2,95/2)$ & $\frac{1}{2}(0^+)$\\
9. & $D_0(2550)^0$ -- $2549\pm 19$ & $2555$ & $+0,23$ & $(99/2,99/2)$ & $\frac{1}{2}(0^-)$\\
\hline
   &                 &          &         &             & $X_0=\bar{c}d\bar{s}u$\\
10.& $X_0(2900)$ -- $2866\pm 7$ & $2870,79$ & $+0,17$ & $(105/2,105/2)$ & $?(0^+)$,\\
&&&&& exotic state\\
\hline
   &                 &          &         &             & $B^0=d\bar{b}$,\\
   &&&&&$\bar{B}^0=\bar{d}b$\\
11. & $B^0$ -- $5279,66\pm 0,12$ & $5298,05$ & $+0,35$ & $(143/2,143/2)$ & $\frac{1}{2}(0^-)$\\
\hline
   &                 &          &         &             & $B^0_s=s\bar{b}$,\\
   &&&&&$\bar{B}^0_s=\bar{s}b$\\
12. & $B^0_s$ -- $5366,92\pm 0,10$ & $5371,89$ & $+0,09$ & $(72,72)$ & $\frac{1}{2}(0^-)$\\
\hline
\end{tabular}
}
\end{center}
\begin{center}
{\textbf{Table 26. Neutral mesons, spin 1, $\K=\BH$}\\
(Subspace $\bsH^{(0,0,1)}_{\rm phys}(\BH)$, vectors $\left|\K,b,\ell,l-\dot{l}\right\rangle=\left|\BH,0,0,1\right\rangle$).}
\vspace{0.1cm}
{\renewcommand{\arraystretch}{1.0}
\begin{tabular}{|c||l|l|c|l|l|}\hline
 & State and mass & $m_\omega$ & Error \% & $(l,\dot{l})$ & $q\bar{q}$, $I^G(J^{PC})$\\ \hline\hline
   &                 &          &         &             & $K^0=d\bar{s}$,\\
   &&&&&$\bar{K}^0=\bar{d}s$\\
1. & $K^\ast(892)^0$ -- $891,67\pm 0,26$ & $889,14$ & $-0,28$ & $(59/2,57/2)$ & $\frac{1}{2}(1^-)$\\
2. & $K_1(1270)^0$ -- $1253\pm 7$  & $1251,69$ & $-0,10$ & $(35,34)$ & $\frac{1}{2}(1^+)$\\
3. & $K_1(1400)^0$ -- $1403\pm 7$  & $1398,86$ & $-0,29$ & $(37,36)$ & $\frac{1}{2}(1^+)$\\
4. & $K^\ast(1410)^0$ -- $1414\pm 15$  & $1398,86$ & $-1,07$ & $(37,36)$ & $\frac{1}{2}(1^-)$\\
5. & $K_1(1650)^0$ -- $1650\pm 50$  & $1634,94$ & $-0,91$ & $(40,39)$ & $\frac{1}{2}(1^+)$\\
6. & $K^\ast(1680)^0$ -- $1718\pm 18$  & $1717,73$ & $-0,01$ & $(41,40)$ & $\frac{1}{2}(1^-)$\\
\hline
   &                 &          &         &             & $D^0=c\bar{u}$,\\
   &&&&&$\bar{D}^0=\bar{c}u$\\
7. & $D^\ast(2007)^0$ -- $2006,85\pm 0,05$ & $2023,56$ & $+0,83$ & $(89/2,87/2)$ & $\frac{1}{2}(1^-)$\\
8. & $D_1(2420)^0$ -- $2422,1\pm 0,6$ & $2403,74$ & $-0,76$ & $(97/2,95/2)$ & $\frac{1}{2}(1^+)$\\
9. & $D_1(2430)^0$ -- $2412\pm 9$ & $2403,74$ & $-0,34$ & $(97/2,95/2)$ & $\frac{1}{2}(1^+)$\\
10. & $D^\ast(2600)^0$ -- $2627\pm 10$ & $2606,10$ & $-0,79$ & $(101/2,99/2)$ & $\frac{1}{2}(1^-)$\\
11. & $D^\ast(2760)^0$ -- $2781\pm 22$ & $2763,23$ & $-0,64$ & $(52,51)$ & $\frac{1}{2}(1^-)$\\
\hline
\end{tabular}
}
\end{center}
\begin{center}
{\renewcommand{\arraystretch}{1.0}
\begin{tabular}{|c||l|l|c|l|l|}\hline
 & State and mass & $m_\omega$ & Error \% & $(l,\dot{l})$ & $q\bar{q}$, $I^G(J^{PC})$\\ \hline\hline
   &                 &          &         &             & $X_1=\bar{c}d\bar{s}u$\\
12.& $X_1(2900)$ -- $2904\pm 5$ & $2924,96$ & $+0,72$ & $(107/2,105/2)$ & $?(1^-)$, \\
&&&&& exotic state\\
\hline
   &                 &          &         &             & $B^0=d\bar{b}$,\\
   &&&&&$\bar{B}^0=\bar{d}b$\\
13. & $B^\ast{}^0$ -- $5324,71\pm 0,21$ & $5297,79$ & $-0,50$ & $(72,71)$ & $\frac{1}{2}(1^-)$\\
14. & $B_1(5721)^0$ -- $5725,9^{+2,5}_{-2,7}$ & $5748,49$ & $+0,39$ & $(75,74)$ & $\frac{1}{2}(1^+)$\\
\hline
   &                 &          &         &             & $B^0_s=s\bar{b}$,\\
   &&&&&$\bar{B}^0_s=\bar{s}b$\\
15. & $B^\ast_s{}^0$ -- $5415,4^{+1,8}_{-1,5}$ & $5445,98$ & $+0,56$ & $(73,72)$ & $0(1^-)$\\
13. & $B_{s1}(5830)^0$ -- $5828,70\pm 0,20$  & $5825,40$ & $-0,05$ & $(151/2,149/2)$ & $0(1^+)$\\
\hline
\end{tabular}
}
\end{center}
\begin{center}
{\textbf{Table 27. Neutral mesons, spin 2, $\K=\BH$}\\
(Subspace $\bsH^{(0,0,2)}_{\rm phys}(\BH)$, vectors $\left|\K,b,\ell,l-\dot{l}\right\rangle=\left|\BH,0,0,2\right\rangle$).}
\vspace{0.1cm}
{\renewcommand{\arraystretch}{1.0}
\begin{tabular}{|c||l|l|c|l|l|}\hline
 & State and mass & $m_\omega$ & Error \% & $(l,\dot{l})$ & $q\bar{q}$, $I^G(J^{PC})$\\ \hline\hline
   &                 &          &         &             & $K^0=d\bar{s}$,\\
   &&&&&$\bar{K}^0=\bar{d}s$\\
1. & $K^\ast_2(1430)^0$ -- $1427,3\pm 1,5$ & $1436,16$ & $+0,62$ & $(38,36)$ & $\frac{1}{2}(2^+)$\\
2. & $K_2(1580)^0\approx 1580$ & $1593,55$ & $+0,85$ & $(40,38)$ & $\frac{1}{2}(2^-)$\\
3. & $K_2(1770)^0$ -- $1773\pm 8$ & $1759,17$ & $-0,78$ & $(42,40)$ & $\frac{1}{2}(2^-)$\\
4. & $K_2(1820)^0$ -- $1819\pm 12$ & $1801,78$ & $-0,94$ & $(85/2,81/2)$ & $\frac{1}{2}(2^-)$\\
5. & $K^\ast_2(1980)^0$ $1994^{+60}_{-50}$ & $1977,57$ & $-0,82$ & $(89/2,85/2)$ & $\frac{1}{2}(2^+)$\\
6. & $K_2(2250)^0$ -- $2247\pm 17$ & $2256,57$ & $+0,42$ & $(95/2,91/2)$ & $\frac{1}{2}(2^-)$\\
\hline
   &                 &          &         &             & $D^0=c\bar{u}$,\\
   &&&&&$\bar{D}^0=\bar{c}u$\\
7. & $D^\ast_2(2460)^0$ -- $2461,1^{+0,7}_{-0,8}$ & $2452,80$ & $-0,34$ & $(99/2,95/2)$ & $\frac{1}{2}(2^+)$\\
8. & $D_2(2740)^0$ -- $2747\pm 6$ & $2762,46$ & $+0,56$ & $(105/2,101/2)$ & $\frac{1}{2}(2^-)$\\
\hline
   &                 &          &         &             & $B^0=d\bar{b}$,\\
   &&&&&$\bar{B}^0=\bar{d}b$\\
9. & $B^\ast_2(5747)^0$ -- $5737,2\pm 0,7$ & $5747,73$ & $+0,18$ & $(151/2,147/2)$ & $\frac{1}{2}(2^+)$\\
\hline
   &                 &          &         &             & $B^0_s=s\bar{b}$,\\
   &&&&&$\bar{B}^0_s=\bar{s}b$\\
10. & $B^\ast_{s2}(5840)^0$ -- $5839,86\pm 0,12$ & $5824,63$ & $-0,26$ & $(153/2,149/2)$ & $0(2^+)$\\
\hline
\end{tabular}
}
\end{center}
\begin{center}
{\textbf{Table 28. Neutral mesons, spin 3, $\K=\BH$}\\
(Subspace $\bsH^{(0,0,3)}_{\rm phys}(\BH)$, vectors $\left|\K,b,\ell,l-\dot{l}\right\rangle=\left|\BH,0,0,3\right\rangle$).}
\vspace{0.1cm}
{\renewcommand{\arraystretch}{1.0}
\begin{tabular}{|c||l|l|c|l|l|}\hline
 & State and mass & $m_\omega$ & Error \% & $(l,\dot{l})$ & $q\bar{q}$, $I^G(J^{PC})$\\ \hline\hline
   &                 &          &         &             & $K^0=d\bar{s}$,\\
   &&&&&$\bar{K}^0=\bar{d}s$\\
1. & $K^\ast_3(1780)^0$ -- $1778\pm 8$ & $1757,84$ & $-1,13$ & $(85/2,79/2)$ & $\frac{1}{2}(3^-)$\\
2. & $K_3(2320)^0$ -- $2324\pm 24$ & $2303,09$ & $-0,88$ & $(97/2,91/2)$ & $\frac{1}{2}(3^+)$\\
\hline
   &                 &          &         &             & $D^0=c\bar{u}$,\\
   &&&&&$\bar{D}^0=\bar{c}u$\\
3. & $D^\ast_3(2750)^0$ -- $2763,1\pm 3,2$ & $2761,19$ & $-0,07$ & $(53,50)$ & $\frac{1}{2}(3^-)$\\
\hline
\end{tabular}
}
\end{center}
\newpage
\begin{center}
{\textbf{Table 29. Neutral mesons, spin 4, $\K=\BH$}\\
(Subspace $\bsH^{(0,0,4)}_{\rm phys}(\BH)$, vectors $\left|\K,b,\ell,l-\dot{l}\right\rangle=\left|\BH,0,0,4\right\rangle$).}
\vspace{0.1cm}
{\renewcommand{\arraystretch}{1.0}
\begin{tabular}{|c||l|l|c|l|l|}\hline
 & State and mass & $m_\omega$ & Error \% & $(l,\dot{l})$ & $q\bar{q}$, $I^G(J^{PC})$\\ \hline\hline
   &                 &          &         &             & $K^0=d\bar{s},\bar{K}^0=\bar{d}s$\\
1. & $K^\ast_4(2045)^0$ -- $2048^{+8}_{-9}$ & $2065,46$ & $+0,85$ & $(93/2,85/2)$ & $\frac{1}{2}(4^+)$\\
2. & $K_4(2500)^0$ -- $2247\pm 17$ & $2253,51$ & $+0,29$ & $(97/2,89/2)$ & $\frac{1}{2}(4^-)$\\
\hline
\end{tabular}
}
\end{center}
\begin{center}
{\textbf{Table 30. Neutral mesons, spin 5, $\K=\BH$}\\
(Subspace $\bsH^{(0,0,5)}_{\rm phys}(\BH)$, vectors $\left|\K,b,\ell,l-\dot{l}\right\rangle=\left|\BH,0,0,5\right\rangle$).}
\vspace{0.1cm}
{\renewcommand{\arraystretch}{1.0}
\begin{tabular}{|c||l|l|c|l|l|}\hline
 & State and mass & $m_\omega$ & Error \% & $(l,\dot{l})$ & $q\bar{q}$, $I^G(J^{PC})$\\ \hline\hline
   &                 &          &         &             & $K^0=d\bar{s},\bar{K}^0=\bar{d}s$\\
1. & $K^\ast_5(2380)^0$ -- $2382\pm 24$ & $2397,61$ & $+0,65$ & $(101/2,91/2)$ & $\frac{1}{2}(5^-)$\\
\hline
\end{tabular}
}
\end{center}

\subsubsection{$\R$-subspaces/truly neutral mesons\label{sec5_2_3}}

Unlike the baryon sector, the meson sector contains truly neutral states that make up the coherent subspaces $\bsH^{(s)}_{\rm phys}(\R)$. In this case, the operation of charge conjugation is reduced to an identical transformation, which leads to the absence of anti-states. In addition, subspaces $\bsH^{(0,0,s)}_{\rm phys}(\R)$ contain a large number of exotic states, the description of which goes beyond the quark model (see Appendix B).
%\newpage
\begin{center}
{\textbf{Table 31. Truly neutral mesons, spin 0, $\K=\R$}\\
(Subspace $\bsH^{(0,0,0)}_{\rm phys}(\R)$, vectors $\left|\K,b,\ell,l-\dot{l}\right\rangle=\left|\R,0,0,0\right\rangle$).}
\vspace{0.1cm}
{\renewcommand{\arraystretch}{1.0}
\begin{tabular}{|c||l|l|c|l|l|}\hline
 & State and mass & $m_\omega$ & Error \% & $(l,\dot{l})$ & $q\bar{q}$, $I^G(J^{PC})$\\ \hline\hline
   &                 &          &         &             & $\pi^0,a^0=$\\
   &&&&&$(u\bar{u}-d\bar{d})/\sqrt{2}$\\
1. & $\pi^0$ -- $134,9768$ & $135,16$ & $+0,13$ & $(11,11)$ & $1^-(0^{-})$\\
2. & $a_0(980)^0$ -- $980\pm 20$ & $982,14$ & $+0,29$ & $(61/2,61/2)$ & $1^-(0^{++})$\\
3. & $\pi(1300)^0$ -- $1300\pm 100$ & $1287,97$ & $-0,92$ & $(35,35)$ & $1^-(0^{-+})$\\
4. & $a_0(1450)^0$ -- $1474\pm 19$ & $1475,77$ & $+0,13$ & $(75/2,75/2)$ & $1^-(0^{++})$\\
5. & $a_0(1700)^0$ -- $1704\pm 5$  & $1717,98$ & $+0,82$ & $(81/2,81/2)$ & $1^-(0^{++})$\\
6. & $\pi(1800)^0$ -- $1800^{+9}_{-11}$ & $1802,81$ & $-0,39$ & $(83/2,83/2)$ & $1^-(0^{-+})$\\
7. & $a_0(1950)^0$ -- $1931\pm 26$ & $1933,88$ & $+0,15$ & $(43,43)$ & $1^-(0^{++})$\\
8. & $a_0(2020)^0$ -- $2025\pm 30$ & $2023,81$ & $-0,06$ & $(44,44)$ & $1^-(0^{++})$\\
9. & $\pi(2070)^0$ -- $2070\pm 35$ & $2069,55$ & $-0,02$ & $(89/2,89/2)$ & $1^-(0^{-+})$\\
10. & $\pi(2360)^0$ -- $2060\pm 60$ & $2354,69$ & $-0,22$ & $(95/2,95/2)$ & $1^-(0^{-+})$\\
\hline
   &                 &          &         &             & $\eta,\eta^\prime,f=$\\
   &                 &          &         &             &$c_1(u\bar{u}+d\bar{d})$\\
   &&&&&$+c_2(s\bar{s})$\\
11. & $f_0(500)^0$ -- $513\pm 32$  & $517,39$ & $+0,85$ & $(22,22)$ & $0^+(0^{++})$\\
12. & $\eta^0$ -- $547,862$ & $540,64$ & $-1,32$ & $(45/2,45/2)$ & $0^+(0^{-+})$\\
13. & $\eta^\prime(958)^0$ -- $957,78\pm 0,06$ & $950,71$ & $-0,74$ & $(30,30)$ & $0^+(0^{-+})$\\
14. & $f_0(980)^0$ -- $990\pm 20$  & $982,14$ & $-0,79$ & $(61/2,61/2)$ & $0^+(0^{++})$\\
15. & $\eta(1295)^0$ -- $1294\pm 4$ & $1287,97$ & $-0,46$ & $(35,35)$ & $0^+(0^{-+})$\\
16. & $f_0(1370)^0\approx 1370$  & $1361,56$ & $-0,61$ & $(36,36)$ & $0^+(0^{++})$\\
17. & $\eta(1405)^0$ -- $1408,8\pm 2$ & $1399,12$ & $-0,68$ & $(73/2,73/2)$ & $0^+(0^{-+})$\\
18. & $\eta(1475)^0$ -- $1475\pm 4$ & $1475,77$ & $+0,05$ & $(75/2,75/2)$ & $0^+(0^{-+})$\\
\hline
\end{tabular}
}
\end{center}
\begin{center}
{\renewcommand{\arraystretch}{1.0}
\begin{tabular}{|c||l|l|c|l|l|}\hline
 & State and mass & $m_\omega$ & Error \% & $(l,\dot{l})$ & $q\bar{q}$, $I^G(J^{PC})$\\ \hline\hline
19. & $f_0(1500)^0$ -- $1506\pm 6$  & $1514,86$ & $+0,59$ & $(38,38)$ & $0^+(0^{++})$\\
20. & $f_0(1710)^0$ -- $1704\pm 12$  & $1717,98$ & $+0,82$ & $(81/2,81/2)$ & $0^+(0^{++})$\\
21. & $\eta(1760)^0$ -- $1751\pm 15$ & $1760,14$ & $+0,52$ & $(41,41)$ & $0^+(0^{-+})$\\
22. & $\eta(2010)^0$ -- $2010^{+35}_{-60}$ & $2023,81$ & $+0,68$ & $(44,44)$ & $0^+(0^{-+})$\\
23. & $f_0(2020)^0$ -- $1992\pm 16$ & $1978,59$ & $-0,67$ & $(87/2,87/2)$ & $0^+(0^{++})$\\
24. & $f_0(2060)^0\approx 2050$  & $2069,55$ & $+0,95$ & $(89/2,89/2)$ & $0^+(0^{++})$\\
25. & $\eta(2100)^0$ -- $2050^{+30}_{-54}$ & $2069,55$ & $+0,95$ & $(89/2,89/2)$ & $0^+(0^{-+})$\\
26. & $f_0(2100)^0$ -- $2095^{+17}_{-19}$ & $2115,79$ & $+0,99$ & $(45,45)$ & $0^+(0^{++})$\\
27. & $\eta(2190)^0$ -- $2190\pm 50$ & $2162,55$ & $-1,25$ & $(91/2,91/2)$ & $0^+(0^{-+})$\\
28. & $f_0(2200)^0$ -- $2187\pm 14$ & $2162,55$ & $-1,12$ & $(91/2,91/2)$ & $0^+(0^{++})$\\
29. & $\eta(2225)^0$ -- $2221^{+13}_{-10}$ & $2209,82$ & $-0,50$ & $(46,46)$ & $0^+(0^{-+})$\\
30. & $\eta(2320)^0$ -- $2320\pm 15$ & $2305,89$ & $-0,61$ & $(47,47)$ & $0^+(0^{-+})$\\
31. & $f_0(2330)^0$ -- $2340\pm 20$  & $2354,69$ & $+0,63$ & $(95/2,95/2)$ & $0^+(0^{++})$\\
\hline
   &                 &          &         &             & $\eta_c,\chi_{c0}=c\bar{c}$\\
32. & $\eta_c(1S)$ -- $2983,9\pm 0,4$ & $2980,15$ & $-0,12$ & $(107/2,107/2)$ & $0^+(0^{-+})$\\
33. & $\chi_{c0}(1P)$ -- $3414,71\pm 0,3$ & $3438,01$ & $+0,68$ & $(115/2,115/2)$ & $0^+(0^{++})$\\
34. & $\eta_c(2S)$ -- $3637,5\pm 1,1$ & $3618,13$ & $-0,53$ & $(59,59)$ & $0^+(0^{-+})$\\
35. & $\chi_{c0}(3860)$ -- $3862^{+50}_{-35}$ & $3865,46$ & $+0,09$ & $(61,61)$ & $0^+(0^{++})$\\
36. & $\chi_{c0}(3915)$ -- $3921,7\pm 1,8$ & $3928,57$ & $+0,17$ & $(123/2,123/2)$ & $0^+(0^{++})$\\
37. & $R_{c0}(4240)$ -- $4239^{+50}_{-21}$ & $4251,77$ & $+0,30$ & $(64,64)$ & $1^+(0^{--})$, \\
&&&&& exotic state\\
38. & $\chi_{c0}(4500)$ -- $4474\pm 4$ & $4451,83$ & $-0,49$ & $(131/2,131/2)$ & $0^+(0^{++})$, \\ &&&&& exotic state\\
39. & $\chi_{c0}(4700)$ -- $4694^{+16}_{-5}$ & $4725,73$ & $+0,67$ & $(135/2,135/2)$ & $0^+(0^{++})$,\\
&&&&& exotic state\\
\hline
   &                 &          &         &             & $\eta_b,\chi_{b0}=b\bar{b}$\\
40. & $\eta_b(1S)$ -- $9398,7\pm 2,0$ & $9418,75$ & $+0,21$ & $(191/2,191/2)$ & $0^+(0^{-+})$\\
41. & $\chi_{b0}(1P)$ -- $9859,44\pm 0,42$ & $9815,29$ & $-0,44$ & $(195/2,195/2)$ & $0^+(0^{++})$\\
42. & $\eta_b(2S)$ -- $9999\pm 4$ & $10016,62$ & $+0,17$ & $(197/2,197/2)$ & $0^+(0^{-+})$\\
43. & $\chi_{b0}(2P)$ -- $10232,5\pm 0,4$ & $10220$ & $-0,12$ & $(199/2,199/2)$ & $0^+(0^{++})$\\
\hline
44. & $H^0$ -- $125250\pm 17$ & $125195$ & $-0,04$ & $(699/2,699/2)$ & \\
\hline
\end{tabular}
}
\end{center}
\begin{center}
{\textbf{Table 32. Truly neutral mesons, spin 1, $\K=\R$}\\
(Subspace $\bsH^{(0,0,1)}_{\rm phys}(\R)$, vectors $\left|\K,b,\ell,l-\dot{l}\right\rangle=\left|\R,0,0,1\right\rangle$).}
\vspace{0.1cm}
{\renewcommand{\arraystretch}{1.0}
\begin{tabular}{|c||l|l|c|l|l|}\hline
 & State and mass & $m_\omega$ & Error \% & $(l,\dot{l})$ & $q\bar{q}$, $I^G(J^{PC})$\\ \hline\hline
   &                 &          &         &             & $\pi^0,a^0,\rho^0,b^0=$\\
   &                 &          &         &             & $(u\bar{u}-d\bar{d})/\sqrt{2}$\\
1. & $\rho(770)^0$ -- $775,26\pm 0,23$ & $772,63$ & $-0,34$ & $(55/2,53/2)$ & $1^+(1^{--})$\\
2. & $b_1(1235)^0$ -- $1229,5\pm 3,2$ & $1216,18$ & $-1,08$ & $(69/2,67/2)$ & $1^+(1^{+-})$\\
3. & $a_1(1260)^0$ -- $1230\pm 40$ & $1251,69$ & $+1,76$ & $(35,34)$ & $1^-(1^{++})$\\
4. & $\pi_1(1400)^0$ -- $1354\pm 25$ & $1361,30$ & $+0,54$ & $(73/2,71/2)$ & $1^-(1^{-+})$\\
5. & $\rho(1450)^0$ -- $1465\pm 25$ & $1475,51$ & $+0,72$ & $(38,37)$ & $1^+(1^{--})$\\
6. & $\rho(1570)^0$ --$1570\pm 70$ & $1554,21$ & $-1,00$ & $(39,38)$ & $1^+(1^{--})$\\
7. & $\pi_1(1600)^0$ -- $1661^{+15}_{-11}$ & $1676,08$ & $+0,91$ & $(81/2,79/2)$ & $1^-(1^{-+})$\\
8. & $a_1(1640)^0$ -- $1655\pm 16$ & $1634,94$ & $-1,21$ & $(40,39)$ & $1^-(1^{++})$\\
\hline
\end{tabular}
}
\end{center}
\begin{center}
{\renewcommand{\arraystretch}{1.0}
\begin{tabular}{|c||l|l|c|l|l|}\hline
 & State and mass & $m_\omega$ & Error \% & $(l,\dot{l})$ & $q\bar{q}$, $I^G(J^{PC})$\\ \hline\hline
9. & $\rho(1700)^0$ -- $1720\pm 20$ & $1717,73$ & $-0,13$ & $(41,40)$ & $1^+(1^{--})$\\
10. & $\rho(1900)^0$ -- $1909\pm 17$ & $1889,42$ & $-1,02$ & $(43,42)$ & $1^+(1^{--})$\\
11. & $a_1(1930)^0$ -- $1930^{+30}_{-70}$ & $1933,62$ & $+0,19$ & $(87/2,85/2)$ & $1^-(1^{++})$\\
12. & $b_1(1960)^0$ -- $1960\pm 35$ & $1978,33$ & $+0,93$ & $(44,43)$ & $1^+(1^{+-})$\\
13. & $\rho(2000)^0$ -- $2000\pm 30$ & $1978,33$ & $-1,08$ & $(44,43)$ & $1^+(1^{--})$\\
14. & $\pi_1(2015)^0$ -- $2014\pm 20$ & $2023,56$ & $+0,47$ & $(89/2,87/2)$ & $1^-(1^{-+})$\\
15. & $a_1(2095)^0$ -- $2096\pm 17$ & $2069,29$ & $-1,27$ & $(45,44)$ & $1^-(1^{++})$\\
16. & $\rho(2150)^0$ -- $2111\pm 43$ & $2115,54$ & $+0,21$ & $(91/2,89/2)$ & $1^+(1^{--})$\\
17. & $b_1(2240)^0$ -- $2240\pm 35$ & $2257,34$ & $+0,77$ & $(47,46)$ & $1^+(1^{+-})$\\
18. & $\rho(2270)^0$ -- $2265\pm 40$ & $2257,34$ & $-0,34$ & $(47,46)$ & $1^+(1^{--})$\\
19. & $a_1(2270)^0$ -- $2270^{+50}_{-40}$ & $2257,34$ & $-0,56$ & $(47,46)$ & $1^-(1^{++})$\\
\hline
   &                 &          &         &             & $\omega,\phi,h,f=$\\
   &                 &          &         &             &$c_1(u\bar{u}+d\bar{d})$\\
   &                 &          &         &  & $+c_2(s\bar{s})$\\
20. & $\omega(782)$ -- $782,66\pm 0,13$ & $772,63$ & $-1,28$ & $(55/2,53/2)$ & $0^-(1^{--})$\\
21. & $\phi(1020)$ -- $1019,461\pm 0,016$ & $1013,82$ & $-0,55$ & $(63/2,61/2)$ & $0^-(1^{--})$\\
22. & $h_1(1170)$ -- $1166\pm 6$ & $1181,18$ & $+1,30$ & $(34,33)$ & $0^-(1^{+-})$\\
23. & $f_1(1285)$ -- $1281,9\pm 0,5$ & $1287,72$ & $+0,45$ & $(71/2,69/2)$ & $0^+(1^{++})$\\
24. & $h_1(1415)$ -- $1416\pm 8$ & $1398,86$ & $-1,21$ & $(37,36)$ & $0^-(1^{+-})$\\
25. & $f_1(1420)$ -- $1426,3\pm 0,9$ & $1436,93$ & $+0,74$ & $(75/2,73/2)$ & $0^+(1^{++})$\\
26. & $\omega(1420)$ -- $1410\pm 60$ & $1398,86$ & $-0,79$ & $(37,36)$ & $0^-(1^{--})$\\
27. & $f_1(1510)$ -- $1518\pm 5$ & $1514,60$ & $-0,22$ & $(77/2,75/2)$ & $0^+(1^{++})$\\
28. & $h_1(1595)$ -- $1594^{+12}_{-60}$ & 1594,32 & $+0,02$ & $(79/2,77/2)$ & $0^-(1^{+-})$\\
29. & $\omega(1650)$ -- $1670\pm 30$ & $1676,08$ & $+0,36$ & $(81/2,79/2)$ & $0^-(1^{--})$\\
30. & $\phi(1680)$ -- $1680\pm 20$  & $1676,08$ & $-0,23$ & $(81/2,79/2)$ & $0^-(1^{--})$\\
31. & $\omega(1960)$ -- $1960\pm 25$  & $1978,33$ & $+0,93$ & $(44,43)$ & $0^-(1^{--})$\\
32. & $h_1(1965)$ -- $1965\pm 45$  & $1978,33$ & $+0,68$ & $(44,43)$ & $0^-(1^{+-})$\\
33. & $f_1(1970)$ -- $1971\pm 15$  & $1978,33$ & $+0,37$ & $(44,43)$ & $0^+(1^{++})$\\
34. & $\phi(2170)$ -- $2162\pm 7$ & $2162,29$ & $+0,01$ & $(46,45)$ & $0^-(1^{--})$\\
35. & $\omega(2205)$ -- $2205\pm 30$  & $2209,56$ & $+0,21$ & $(93/2,91/2)$ & $0^-(1^{--})$\\
\hline
   &                 &          &         &             & $J/\psi,h=c\bar{c}$\\
36. & $h_1(2215)$ -- $2215\pm 40$  & $2209,56$ & $-0,24$ & $(93/2,91/2)$ & $0^-(1^{+-})$\\
37. & $\omega(2290)$ -- $2290\pm 20$  & $2305,63$ & $+0,68$ & $(95/2,93/2)$ & $0^-(1^{--})$\\
38. & $f_1(2310)$ -- $2310\pm 60$  & $2303,63$ & $-0,19$ & $(95/2,93/2)$ & $0^+(1^{++})$\\
39. & $\omega(2330)$ -- $2330\pm 30$  & $2354,43$ & $+1,04$ & $(48,47)$ & $0^-(1^{--})$\\
40. & $J/\psi(1S)$ -- $3096,9\pm 0,006$ & $3091,29$ & $-0,18$ & $(55,54)$ & $0^-(1^{--})$\\
41. & $\chi_{c1}(1P)$ -- $3510,67\pm 0,05$ & $3497,28$ & $-0,38$ & $(117/2,115/2)$ & $0^+(1^{++})$\\
42. & $h_{c}(1P)$ -- $3525,38\pm 0,11$ & $3557,32$ & $+0,90$ & $(59,58)$ & $0^-(1^{+-})$\\
43. & $\psi(2S)$ -- $3686,10\pm 0,06$ & $3678,94$ & $-0,19$ & $(60,59)$ & $0^-(1^{--})$\\
44. & $\psi(3770)$ -- $3773,7\pm 0,4$ & $3802,61$ & $+0,76$ & $(61,60)$ & $0^-(1^{--})$\\
45. & $\chi_{c1}(3872)$ -- $3871,65\pm 0,06$ & $3865,20$ & $-0,16$ & $(123/2,121/2)$ & $0^+(1^{++})$,\\
&&&&& exotic state\\
46. & $Z_{c}(3900)$ -- $3887,1\pm 2,6$ & $3865,20$ & $-0,56$ & $(123/2,121/2)$ & $1^+(1^{+-})$,\\ &&&&& exotic state\\
\hline
\end{tabular}
}
\end{center}
\begin{center}
{\renewcommand{\arraystretch}{1.0}
\begin{tabular}{|c||l|l|c|l|l|}\hline
 & State and mass & $m_\omega$ & Error \% & $(l,\dot{l})$ & $q\bar{q}$, $I^G(J^{PC})$\\ \hline\hline
47. & $\psi(4040)$ -- $4039\pm 1$ & $4056,06$ & $+0,42$ & $(63,62)$ & $0^-(1^{--})$\\
48. & $\chi_{c1}(4140)$ -- $4146,5\pm 3,0$ & $4120,70$ & $-0,62$ & $(127/2,125/2)$ & $0^+(1^{++})$, \\
&&&&& exotic state\\
49. & $\psi(4160)$ -- $4191\pm 5$ & $4185,86$ & $-0,12$ & $(64,63)$ & $0^-(1^{--})$\\
50. & $Z_{c}(4200)$ -- $4196^{+35}_{-32}$ & $4185,86$ & $-0,24$ & $(64,63)$ & $1^+(1^{+-})$,\\ &&&&& exotic state\\
51. & $\psi(4230)$ -- $4222,7\pm 2,6$ & $4251,52$ & $+0,68$ & $(129/2,127/2)$ & $0^-(1^{--})$\\
52. & $\chi_{c1}(4274)$ -- $4286^{+8}_{-9}$ & $4317,69$ & $+0,74$ & $(65,64)$ & $0^+(1^{++})$,\\ &&&&& exotic state\\
53. & $\psi(4360)$ -- $4372\pm 9$ & $4384,38$ & $+0,28$ & $(131/2,129/2)$ & $0^-(1^{--})$,\\ &&&&& exotic state\\
54. & $\psi(4415)$ -- $4421\pm 4$ & $4451,57$ & $+0,69$ & $(66,65)$ & $0^-(1^{--})$\\
55. & $Z_{c}(4430)$ -- $4478^{+15}_{-18}$ & $4451,57$ & $-0,59$ & $(66,65)$ & $1^+(1^{+-})$,\\ &&&&& exotic state\\
56. & $\psi(4660)$ -- $4630\pm 6$ & $4656,23$ & $+0,56$ & $(135/2,133/2)$ & $0^-(1^{--})$,\\
&&&&& exotic state\\
57. & $\chi_{c1}(4685)$ -- $4684^{+15}_{-17}$ & $4656,23$ & $-0,59$ & $(135/2,133/2)$ & $0^+(1^{++})$,\\
&&&&& exotic state\\
\hline
   &                 &          &         &             & $\Upsilon,\chi_{b1}$,\\
   &&&&&$h_b=b\bar{b}$\\
58. & $\Upsilon(1S)$ -- $9460,30\pm 0,26$ & $9516,86$ & $+0,59$ & $(193/2,191/2)$ & $0^-(1^{--})$\\
59. & $\chi_{b1}(1P)$ -- $9892,78\pm 0,26$ & $9915,44$ & $+0,23$ & $(197/2,195/2)$ & $0^+(1^{++})$\\
60. & $h_b(1P)$ -- $9899,3\pm 0,8$ & $9915,44$ & $+0,16$ & $(197/2,195/2)$ & $0^-(1^{+-})$\\
61. & $\Upsilon(2S)$ -- $10023,26\pm 0,31$ & $10016,37$ & $-0,07$ & $(99,98)$ & $0^-(1^{--})$\\
62. & $\chi_{b1}(2P)$ -- $10255,46\pm 0,22$ & $10219,74$ & $-0,35$ & $(100,99)$ & $0^+(1^{++})$\\
63. & $h_b(2P)$ -- $10259,8\pm 1,2$ & $10219,74$ & $-0,39$ & $(100,99)$ & $0^-(1^{+-})$\\
64. & $\Upsilon(3S)$ -- $10355,2\pm 0,5$ & $10322,20$ & $-0,32$ & $(201/2,199/2)$ & $0^-(1^{--})$\\
65. & $\chi_{b1}(3P)$ -- $10513,4\pm 0,7$ & $10528,64$ & $+0,14$ & $(203/2,201/2)$ & $0^+(1^{++})$\\
66. & $\Upsilon(4S)$ -- $10579,4\pm 1,2$ & $10528,64$ & $-0,48$ & $(203/2,201/2)$ & $0^-(1^{--})$\\
67. & $Z_{b}(10610)$ -- $10607,2\pm 2,0$ & $10632,63$ & $+0,24$ & $(102,101)$ & $1^+(1^{+-})$,\\ &&&&& exotic state\\
68. & $Z_{b}(10650)$ -- $10652,2\pm 1,5$ & $10632,63$ & $-0,18$ & $(102,101)$ & $1^+(1^{+-})$,\\ &&&&& exotic state\\
69. & $\Upsilon(10753)$ -- $10753\pm 6$ & $10737,13$ & $-0,15$ & $(205/2,203/2)$ & $?^?(1^{--})$\\
70. & $\Upsilon(10860)$ -- $10885,2^{+2,6}_{-1,6}$ & $10842,14$ & $-0,39$ & $(103,102)$ & $0^-(1^{--})$\\
71. & $\Upsilon(11020)$ -- $11000\pm 4$ & $11053,69$ & $+0,49$ & $(104,103)$ & $0^-(1^{--})$\\
\hline
72. & $Z^0$ -- $91187,6\pm 0,0021$ & $91062,24$ & $-0,14$ & $(597/2,595/2)$ & \\
\hline
\end{tabular}
}
\end{center}
\newpage
\begin{center}
%\newpage
{\textbf{Table 33. Truly neutral mesons, spin 2, $\K=\R$}\\
(Subspace $\bsH^{(0,0,2)}_{\rm phys}(\R)$, vectors $\left|\K,b,\ell,l-\dot{l}\right\rangle=\left|\R,0,0,2\right\rangle$).}
\vspace{0.1cm}
{\renewcommand{\arraystretch}{1.0}
\begin{tabular}{|c||l|l|c|l|l|}\hline
 & State and mass & $m_\omega$ & Error \% & $(l,\dot{l})$ & $q\bar{q}$, $I^G(J^{PC})$\\ \hline\hline
   &                 &          &         &             & $\pi^0,a^0,\rho^0=$\\
   &                 &          &         &             & $(u\bar{u}-d\bar{d})/\sqrt{2}$\\
1. & $a_2(1320)^0$ -- $1318,2\pm 0,6$ & $1323,49$ & $+0,40$ & $(73/2,69/2)$ & $1^-(2^{++})$\\
2. & $\pi_2(1670)^0$ -- $1670,6^{+2,9}_{-1,2}$ & $1675,31$ & $+0,28$ & $(41,39)$ & $1^-(2^{-+})$\\
3. & $a_2(1700)^0$ -- $1698\pm 40$ & $1716,96$ & $+1,11$ & $(83/2,79/2)$ & $1^-(2^{++})$\\
4. & $\pi_2(1880)^0$ -- $1874^{+26}_{-5}$ & $1888,65$ & $+0,78$ & $(87/2,53/2)$ & $1^-(2^{-+})$\\
5. & $\rho_2(1940)^0$ -- $1940\pm 40$ & $1932,86$ & $-0,37$ & $(44,42)$ & $1^+(2^{--})$\\
6. & $a_2(1950)^0$ -- $1950^{+30}_{-70}$ & $1932,86$ & $-0,88$ & $(44,42)$ & $1^-(2^{++})$\\
7. & $a_2(1990)^0$ -- $2050\pm 10$ & $2068,53$ & $+0,90$ & $(91/2,87/2)$ & $1^-(2^{++})$\\
8. & $\pi_2(2005)^0$ -- $1953^{+17}_{-27}$ & $1977,57$ & $+0,74$ & $(89/2,85/2)$ & $1^-(2^{-+})$\\
9. & $a_2(2030)^0$ -- $2030\pm 20$ & $2022,79$ & $-0,35$ & $(45,43)$ & $1^-(2^{++})$\\
10. & $\pi_2(2100)^0$ -- $2090\pm 29$ & $2114,77$ & $+1,18$ & $(46,44)$ & $1^-(2^{-+})$\\
11. & $a_2(2175)^0$ -- $2175\pm 40$ & $2161,53$ & $-0,62$ & $(93/2,89/2)$ & $1^-(2^{++})$\\
12. & $\rho_2(2225)^0$ -- $2225\pm 35$ & $2208,80$ & $-0,73$ & $(47,45)$ & $1^+(2^{--})$\\
13. & $a_2(2255)^0$ -- $2255\pm 20$ & $2256,57$ & $+0,07$ & $(95/2,91/2)$ & $1^-(2^{++})$\\
14. & $\pi_2(2285)^0$ -- $2285\pm 20$ & $2304,86$ & $+0,87$ & $(48,46)$ & $1^-(2^{-+})$\\
\hline
   &                 &          &         &             & $f,\eta,\omega=$\\
   &                 &          &         &             &$c_1(u\bar{u}+d\bar{d})$\\
   &&&&&$+c_2(s\bar{s})$\\
15. & $f_2(1270)$ -- $1275,5\pm 0,8$ & $1286,95$ & $+0,89$ & $(36,34)$ & $0^+(2^{++})$\\
16. & $f_2(1430)\approx 1430$  & $1436,16$ & $+0,43$ & $(38,36)$ & $0^+(2^{++})$\\
17. & $f^\prime_2(1525)$ -- $1517,4\pm 2,5$ & $1513,84$ & $+0,23$ & $(39,37)$ & $0^+(2^{++})$\\
18. & $f_2(1565)$ -- $1542\pm 19$ & $1553,44$ & $+0,74$ & $(79/2,75/2)$ & $0^+(2^{++})$\\
19. & $f_2(1640)$ -- $1639\pm 6$  & $1634,18$ & $-0,29$ & $(81/2,77/2)$ & $0^+(2^{++})$\\
20. & $\eta_2(1645)$ -- $1617\pm 5$ & $1634,18$ & $+1,06$ & $(81/2,77/2)$ & $0^+(2^{-+})$\\
21. & $f_2(1750)$ -- $1755\pm 10$ & $1759,17$ & $+0,24$ & $(42,40)$ & $0^+(2^{++})$\\
22. & $f_2(1810)$ -- $1815\pm 12$ & $1801,78$ & $-0,73$ & $(85/2,81/2)$ & $0^+(2^{++})$\\
23. & $\eta_2(1870)$ -- $1842\pm 8$ & $1844,96$ & $+0,16$ & $(43,41)$ & $0^+(2^{-+})$\\
24. & $f_2(1910)$ -- $1900\pm 9$ & $1888,65$ & $-0,59$ & $(87/2,83/2)$ & $0^+(2^{++})$\\
25. & $f_2(1950)$ -- $1936\pm 12$ & $1932,86$ & $-0,16$ & $(44,42)$ & $0^+(2^{++})$\\
26. & $\omega_2(1975)$ -- $1975\pm 20$ & $1977,57$ & $+0,13$ & $(89/2,85/2)$ & $0^-(2^{--})$\\
27. & $f_2(2000)$ -- $2001\pm 10$  & $2022,79$ & $+1,08$ & $(45,43)$ & $0^+(2^{++})$\\
28. & $f_2(2010)$ -- $2011^{+60}_{-80}$ & $2022,79$ & $+0,58$ & $(45,43)$ & $0^+(2^{++})$\\
29. & $\eta_2(2030)$ -- $2030\pm 5$  & $2022,79$ & $-0,35$ & $(45,43)$ & $0^+(2^{-+})$\\
30. & $f_2(2140)$ -- $2140\pm 12$ & $2161,53$ & $+1,01$ & $(93/2,89/2)$ & $0^+(2^{++})$\\
31. & $f_2(2150)$ -- $2157\pm 12$ & $2161,53$ & $+0,21$ & $(93/2,89/2)$ & $0^+(2^{++})$\\
32. & $\omega_2(2195)$ -- $2195\pm 30$ & $2208,80$ & $+0,63$ & $(47,45)$ & $0^-(2^{--})$\\
33. & $f_2(2240)$ -- $2240\pm 15$  & $2256,57$ & $+0,74$ & $(95/2,91/2)$ & $0^+(2^{++})$\\
34. & $\eta_2(2250)$ -- $2248\pm 20$ & $2256,57$ & $+0,38$ & $(95/2,91/2)$ & $0^+(2^{-+})$\\
35. & $f_2(2295)$ -- $2293\pm 13$  & $2304,86$ & $+0,52$ & $(48,46)$ & $0^+(2^{++})$\\
36. & $f_2(2300)$ -- $2297\pm 28$ & $2304,86$ & $+0,34$ & $(48,46)$ & $0^+(2^{++})$\\
37. & $f_2(2340)$ -- $2345^{+50}_{-40}$ & $2356,66$ & $+0,49$ & $(97/2,93/2)$ & $0^+(2^{++})$\\
\hline
\end{tabular}
}
\end{center}
\begin{center}
{\renewcommand{\arraystretch}{1.0}
\begin{tabular}{|c||l|l|c|l|l|}\hline
 & State and mass & $m_\omega$ & Error \% & $(l,\dot{l})$ & $q\bar{q}$, $I^G(J^{PC})$\\ \hline\hline
   &                 &          &         &             & $\chi,\psi=c\bar{c}$\\
38. & $\chi_{c2}(1P)$ -- $3556,17\pm 0,7$ & $3556,56$ & $+0,01$ & $(119/2,115/2)$ & $0^+(2^{++})$\\
39. & $\psi_2(3823)$ -- $3823,7\pm 0,5$ & $3801,84$ & $-0,57$ & $(123/2,119/2)$ & $0^-(2^{--})$\\
40. & $\chi_{c2}(3930)$ -- $3922,2\pm 1,0$ & $3927,54$ & $+0,13$ & $(125/2,121/2)$ & $0^+(2^{++})$\\
   &                 &          &         &             & $\Upsilon,\chi_{b2}$,\\
   &&&&&$h_b=b\bar{b}$\\
41. & $\chi_{b2}(1P)$ -- $9912,21\pm 0,26$ & $9914,68$ & $+0,02$ & $(99,97)$ & $0^+(2^{++})$\\
42. & $\Upsilon_2(1D)$ -- $10163,7\pm 1,4$ & $10117,03$ & $-0,46$ & $(100,98)$ & $0^-(2^{--})$\\
43. & $\chi_{b2}(2P)$ -- $10268,65\pm 0,22$ & $10218,98$ & $-0,48$ & $(201/2,197/2)$ & $0^+(2^{++})$\\
44. & $\chi_{b2}(3P)$ -- $10513,4\pm 0,7$ & $10527,88$ & $+0,14$ & $(102,100)$ & $0^+(2^{++})$\\
\hline
\end{tabular}
}
\end{center}
\begin{center}
{\textbf{Table 34. Truly neutral mesons, spin 3, $\K=\R$}\\
(Subspace $\bsH^{(0,0,3)}_{\rm phys}(\R)$, vectors $\left|\K,b,\ell,l-\dot{l}\right\rangle=\left|\R,0,0,3\right\rangle$).}
\vspace{0.1cm}
{\renewcommand{\arraystretch}{1.0}
\begin{tabular}{|c||l|l|c|l|l|}\hline
 & State and mass & $m_\omega$ & Error \% & $(l,\dot{l})$ & $q\bar{q}$, $I^G(J^{PC})$\\ \hline\hline
   &                 &          &         &             & $a^0,b^0,\rho^0=$\\
   &                 &          &         &             & $(u\bar{u}-d\bar{d})/\sqrt{2}$\\
1. & $\rho_3(1690)^0$ -- $1688,8\pm 2,1$ & $1674,04$ & $-0,87$ & $(83/2,77/2)$ & $1^+(3^{--})$\\
2. & $a_3(1875)^0$ -- $1874\pm 43$ & $1887,38$ & $+0,71$ & $(44,41)$ & $1^-(3^{++})$\\
3. & $\rho_3(1990)^0$ -- $1982\pm 14$ & $1976,29$ & $-0,29$ & $(45,42)$ & $1^+(3^{--})$\\
4. & $b_3(2030)^0$ -- $2032\pm 12$ & $2021,52$ & $-0,51$ & $(91/2,85/2)$ & $1^+(3^{+-})$\\
5. & $a_3(2030)^0$ -- $2031 \pm 12$ & $2021,52$ & $-0,47$ & $(91/2,85/2)$ & $1^-(3^{++})$\\
6. & $b_3(2245)^0$ -- $2245\pm 50$ & $2255,29$ & $+0,46$ & $(48,45)$ & $1^+(3^{+-})$\\
7. & $\rho_3(2250)^0$ -- $2248\pm 17$ & $2255,29$ & $+0,32$ & $(48,45)$ & $1^+(3^{--})$\\
8. & $a_3(2275)^0$ -- $2275\pm 35$ & $2303,59$ & $+1,25$ & $(97/2,91/2)$ & $1^-(3^{++})$\\
\hline
   &                 &          &         &             & $f,\phi,\omega,h=$\\
   &                 &          &         &             &$c_1(u\bar{u}+d\bar{d})$\\
   &&&&&$+c_2(s\bar{s})$\\
9. & $\omega_3(1670)$ -- $1667\pm 4$ & $1674,04$ & $+0,42$ & $(83/2,77/2)$ & $0^-(3^{--})$\\
10. & $\phi_3(1850)$ -- $1854\pm 7$ & $1843,69$ & $-0,55$ & $(87/2,81/2)$ & $0^-(3^{--})$\\
11. & $\omega_3(1945)$ -- $1950^{+30}_{-70}$  & $1931,58$ & $-0,94$ & $(89/2,83/2)$ & $0^-(3^{--})$\\
12. & $h_3(2025)$ -- $2025\pm 20$  & $2021,52$ & $-0,17$ & $(91/2,85/2)$ & $0^-(3^{+-})$\\
13. & $f_3(2050)$ -- $2048\pm 8$  & $2067,25$ & $+0,94$ & $(46,43)$ & $0^+(3^{++})$\\
14. & $\omega_3(2255)$ -- $2255\pm 15$  & $2255,29$ & $+0,01$ & $(48,45)$ & $0^-(3^{--})$\\
15. & $h_3(2275)$ -- $2275\pm 25$  & $2255,29$ & $-0,36$ & $(48,45)$ & $0^-(3^{+-})$\\
16. & $\omega_3(2285)$ -- $2278\pm 28$  & $2303,59$ & $+1,12$ & $(97/2,91/2)$ & $0^-(3^{--})$\\
17. & $f_3(2300)$ -- $2334\pm 35$  & $2352,39$ & $+0,78$ & $(49,46)$ & $0^+(3^{++})$\\
\hline
   &                 &          &         &             & $\psi=c\bar{c}$\\
18. & $\psi_3(3842)$ -- $3842,71\pm 0,20$ & $3863,13$ & $+0,53$ & $(125/2,119/2)$ & $0^-(3^{--})$\\
\hline
\end{tabular}
}
\end{center}
\newpage
\begin{center}
{\textbf{Table 34. Truly neutral mesons, spin 4, $\K=\R$}\\
(Subspace $\bsH^{(0,0,4)}_{\rm phys}(\R)$, vectors $\left|\K,b,\ell,l-\dot{l}\right\rangle=\left|\R,0,0,4\right\rangle$).}
\vspace{0.1cm}
{\renewcommand{\arraystretch}{1.0}
\begin{tabular}{|c||l|l|c|l|l|}\hline
 & State and mass & $m_\omega$ & Error \% & $(l,\dot{l})$ & $q\bar{q}$, $I^G(J^{PC})$\\ \hline\hline
   &                 &          &         &             & $a^0,\rho^0,\pi^0=$\\
   &                 &          &         &             & $(u\bar{u}-d\bar{d})/\sqrt{2}$\\
1. & $a_4(1970)^0$ -- $1967\pm 16$ & $1974,50$ & $+0,38$ & $(91/2,83/2)$ & $1^-(4^{++})$\\
2. & $\rho_4(2230)^0$ -- $2230\pm 25$ & $2253,51$ & $+1,05$ & $(97/2,89/2)$ & $1^+(4^{--})$\\
3. & $\pi_4(2250)^0$ -- $2250\pm 15$ & $2253,51$ & $+0,15$ & $(97/2,89/2)$ & $1^-(4^{-+})$\\
4. & $a_4(2255)^0$ -- $2237\pm 5$ & $2253,51$ & $+0,74$ & $(97/2,89/2)$ & $1^-(4^{++})$\\
\hline
   &                 &          &         &             & $f,\omega,\eta=$\\
   &                 &          &         &             &$c_1(u\bar{u}+d\bar{d})+c_2(s\bar{s})$\\
5. & $f_4(2050)$ -- $2018\pm 11$ & $2019,73$ & $+0,08$ & $(46,42)$ & $0^+(4^{++})$\\
6. & $\omega_4(2250)$ -- $2250\pm 30$  & $2253,51$ & $+0,15$ & $(97/2,89/2)$ & $0^-(4^{--})$\\
7. & $f_4(2300)\approx 2314$  & $2301,80$ & $-0,53$ & $(49,45)$ & $0^+(4^{++})$\\
8. & $\eta_4(2330)$ -- $2328\pm 38$ & $2350,60$ & $+0,97$ & $(99/2,91/2)$ & $0^+(4^{-+})$\\
\hline
\end{tabular}
}
\end{center}
%\newpage
\begin{center}
{\textbf{Table 36. Truly neutral mesons, spin 5, $\K=\R$}\\
(Subspace $\bsH^{(0,0,5)}_{\rm phys}(\R)$, vectors $\left|\K,b,\ell,l-\dot{l}\right\rangle=\left|\R,0,0,5\right\rangle$).}
\vspace{0.1cm}
{\renewcommand{\arraystretch}{1.0}
\begin{tabular}{|c||l|l|c|l|l|}\hline
 & State and mass & $m_\omega$ & Error \% & $(l,\dot{l})$ & $q\bar{q}$, $I^G(J^{PC})$\\ \hline\hline
   &                 &          &         &             & $\rho^0=$\\
   &                 &          &         &             & $(u\bar{u}-d\bar{d})/\sqrt{2}$\\
1. & $\rho_5(2350)^0$ -- $2330\pm 35$ & $2348,30$ & $+0,78$ & $(50,45)$ & $1^+(5^{--})$\\
\hline
   &                 &          &         &             & $\omega=$\\
   &                 &          &         &             &$c_1(u\bar{u}+d\bar{d})+c_2(s\bar{s})$\\
2. & $\omega_5(2250)$ -- $2250\pm 70$  & $2251,21$ & $+0,05$ & $(49,44)$ & $0^-(5^{--})$\\
\hline
\end{tabular}
}
\end{center}
\begin{center}
{\textbf{Table 37. Truly neutral mesons, spin 6, $\K=\R$}\\
(Subspace $\bsH^{(0,0,6)}_{\rm phys}(\R)$, vectors $\left|\K,b,\ell,l-\dot{l}\right\rangle=\left|\R,0,0,6\right\rangle$).}
\vspace{0.1cm}
{\renewcommand{\arraystretch}{1.0}
\begin{tabular}{|c||l|l|c|l|l|}\hline
 & State and mass & $m_\omega$ & Error \% & $(l,\dot{l})$ & $q\bar{q}$, $I^G(J^{PC})$\\ \hline\hline
   &                 &          &         &             & $a^0=$\\
   &                 &          &         &             & $(u\bar{u}-d\bar{d})/\sqrt{2}$\\
1. & $a_6(2450)^0$ -- $2450\pm 130$ & $2444,62$ & $-0,22$ & $(103/2,91/2)$ & $1^-(6^{++})$\\
\hline
   &                 &          &         &             & $f=$\\
   &                 &          &         &             &$c_1(u\bar{u}+d\bar{d})$\\
   &&&&&$+c_2(s\bar{s})$\\
2. & $f_6(2510)$ -- $2465\pm 50$ & $2444,62$ & $-0,83$ & $(103/2,91/2)$ & $0^+(6^{++})$\\
3. & $f_6(3100)$ -- $3100\pm 100$  & $3082,35$ & $-0,57$ & $(115/2,103/2)$ & $0^+(6^{++})$\\
\hline
\end{tabular}
}
\end{center}

%%%%%%%%%%%%%%%%%%%%%%%%%%%%%%%%%%%%%%%%%%
\section{Conclusion\label{sec6}}

It is well known that search for the primary constituents of matter is the general motivation of theoretical physics. This search, however, still adheres to the main thesis of atomism and reductionism that matter is an aggregate consisting of fundamental (i.e., indivisible) subunits. Such position is opposed to the opinion of Heisenberg who repeatedly stressed that the “consists of” principle does not work at the microlevel, at least because an elementary particle may decay in several different ways.

The quark model is actually a modern incarnation of the Democritus hypothesis, where a two-quark $q\bar{q}$ and a three-quark $qqq$ compositions are prescribed to mesonic and baryon parts of the matter spectrum. Such $q\bar{q}$- and $qqq$-schemes do not account for a number of experimentally detected states (exotic hadrons, tetraquarks, and pentaquarks); moreover, there are many other serious problems facing the quark model, such as only a low accuracy of the baryon spectrum predicted by the $\SU(6)\otimes\GO(3)$-symmetry model, insufficient understanding of the confinement mechanism, the proton spin crisis (for more details see \cite{Var21}). In addition, leptons fall beyond the scope of the quark model, which fact obviously precludes quarks from being universal subunits of matter. Being integrated into the Standard Model (SM), the quark model implies the following list of 17 “fundamental particles”: fermions ($u$, $d$, $s$, $c$, $b$, $t$, $e$, $\mu$, $\tau$, $\nu_e$, $\nu_\mu$, $\nu_\tau$) and bosons ($\gamma$, $g$, $ Z$, $W$, $H$). This list contains seven fictitious particles ($u$, $d$, $s$, $c$, $b$, $t$, $g$) that have never been observed in free state. Five particles ($\mu$, $\tau$, $Z$, $W$, $H$) are certainly not elementary since they decay by several different channels. Three types of neutrinos ($\nu_e$, $\nu_\mu$, $\nu_\tau$) transform into each other via neutrino oscillations, i.e. the lepton number for neutrinos is not conserved. Within the SM framework, this motley set of 17 entities is interrelated by the $G_{SM}=\GU(1)\times\SU(2)\times\SU(3)$ dynamic symmetry and 18 parameters. It is easily seen that the SM is very far from the Democritus ideal of a universal primary element of matter. However, after we exclude the fictitious and decaying particles from the list, each of the remaining three $\nu$, $e$, and $\gamma$ can claim to be a primary element.

Concerning the spinor representations of the Lorentz group, note that the \textit{fundamental representation} of this group over the field of complex numbers $\F=\C$ acts in a two-dimensional spin space whose vector is a two-component spinor. And further, any finite-dimensional irreducible spinor representation of the Lorentz group can be factorized as a tensor product of two-dimensional fundamental representations. In turn, the spin space of the fundamental representation is a minimal left ideal of the quaternion algebra (biquaternion algebras $\C_2\simeq\C\otimes\BH$ in the case of field $\F=\C$ and quaternion algebras $\cl_{0,2}\simeq\BH$, $\cl_{1,1}\simeq\R(2)$, $\cl_{0,2}\simeq\R(2)$ in the case of field $\F=\R$, with the division rings $\K\simeq\BH$ for $\cl_{0,2}$ and $\K\simeq\R$ for $\cl_{1,1}$ and $\cl_{2,0} $). The elements of minimal left ideals of four-dimensional quaternion algebras are two-component spinors.

It is remarkable that information bit is structurally identical to a two-component spinor. On the other hand, two-component spinor describes the lowest state corresponding to neutrino. Thus, the spinor concept should be a keypoint in the search for the primary element of matter. A mathematical object of spinor had appeared due to Cartan \cite{Car13}. Van der Waerden \cite{Waer} noted that the name “spinor” was coined by Ehrenfest after the well-known paper by Goudsmit and Uhlenbeck \cite{UG} on the rotating electron, where they tried to explain the electron spin mechanical using the model of a spinning top. It is known that since the work of Uhlenbeck and Goudsmit \cite{UG} all attempts to describe the electron spin classically (i.e., mechanically) have failed. The first theory that had given a correct mathematical formulation of spin “two-valuedness that denies classical description” was proposed by Pauli in 1927 \cite{Pauli}. Avoiding visual kinematic models, Pauli introduced a doubled Hilbert space $\bsH_2\otimes\bsH_\infty$ of wave functions whose vectors are two-component spinors. That was the first appearance of two-component spinors and the doubling in physics. The next doubling (\textit{bispinors} and the space $\bsH_4\otimes\bsH_\infty$) was proposed by Dirac in 1928 \cite{Dir28}. One more doubling leads to \textit{hypervistors} in the $\K$-Hilbert space $\bsH_8\otimes\bsH_\infty$, see \cite{Var19a,VPB22}. Hypertwistors are vectors of the fundamental representation of the Rumer-Fet group $\SO(2,4)\otimes\SU(2)\otimes\SU(2)^\prime$, which has a deep relation to the periodic system of elements \cite{Fet}. Spinors, bispinors, and twistors are special cases of hypertwistors. In \cite{Heisen} Heisenberg noted that the process of bifurcation, or dividing in two parts, expands the space of nature in a very natural way, thus creating the possibility of new symmetries. This doubling of symmetry is the most essential thing in the definition of spin. In fact, the spin is the doubling (bifurcation) of symmetry.

As noted above, the concept of spin stemmed from an attempt to describe the “classically nondescribable two-valuedness” of an electron \textit{viewed as a particle}. However, a particle is a classical concept. Van der Waerden asked why didn't Pauli endow electron with a momentum $m_s=\pm 1/2$ and a magnetic momentum $2m_s$ \cite[p.~243]{Waer}. But whether a classical particle can have a nonclassical property? Obviously, no. And this is an answer to the question of Van der Waerden. An innovative interpretation of spin was given by Yu. Rumer and A. Fet: “Until now, we believed that \textit{one and the same particle}, e.g., an electron can exist in two states: with spin +1/2 and $-1/2$. However, an electron without a definite spin value have been never observed and is only an abstract concept. It is quite natural then that there exist \textit{two} elementary particles: an electron with spin +1/2 and an electron with spin $-1/2$, while “just an electron” does not occur in nature" \cite[p.~161-162]{RF70}. The objects like “just an electron”, “just a proton”, etc., are commonly assumed to live in spacetime individually, independently of each other, and carrying all their inherent quantum properties at once. Such an interpretation is incompatible with the results of the recent experiments to test Bell's inequalities by Friedman-Clauser, Aspe, Greenberg-Horn-Zeilinger, and others. The classical concept of particle is obviously invalid at the microlevel, as well as the mechanical models associated with this concept. Getting rid of the visual image (and the mental template) of a “particle", one can perceive this concept just as an embodiment of the idea of \textit{discreteness}, impossibility to divide a value into infinitely small parts, the denial of \textit{continuum}. The quantum and continuum are eternal thesis and antithesis going back to Democritus and Empedocles. In contemporary physics, however, the quantum concept is based on the Planck's fundamental discreteness of energy, in contrast to the purely speculative process of division to infinity.

In case we identify the minimal structural component $\left|\mathfrak{q}\right\rangle$ (energy quantum) with neutrino $\left|\nu\right\rangle$, $\left|\mathfrak{q}\right\rangle\equiv\left|\nu\right\rangle$, we will come to a “neutrino theory of everything” with neutrino being the primary element of matter. That would be another incarnation of reductionism. But the states, be they neutrinos or any other entities, are secondary with respect to the substance and do not define it (see the first theorem of Spinoza, Section 4). This is the main principle of holism. From the holistic viewpoint, the primary substance, which we call energy or matter, has an infinite spectrum of states, the lowest of which corresponds to neutrino.

The data presented in this paper show that the mass distribution of states fits the grid of the weight diagram (known as extended Weyl diagram) with an average accuracy of 0.41\%. According to the mass formula (\ref{Mass2}), each node $(l,\dot{l})$ of the weight diagram is associated with the quantum numbers $l$, $\dot{l}$, which are the eigenvalues of the Casimir operators of the Lorentz group. Thus, each node of the weight diagram is a state of the spectrum of matter (a "particle") and at the same time a representation of the Lorentz group in complete compliance with the canonical correspondence of the GNS construction and Wigner's interpretation.

In conclusion, this paper was aimed to emphasize the fundamental role of two-component spinors, which is to describe the minimum structural component of matter, i.e., a quantum of energy. We assert that the quantum character of the matter spectrum is deeply connected with the factorization of cyclic vectors of $\K$-Hilbert space (which is separable in this case) in the form of a tensor product of the fundamental states $\left|\mathfrak{q}\right\rangle$ by means of the GNS construction and algebraic quantization. Owing to the $\K$-structure, the state $\left|\mathfrak{q}\right\rangle$ splits into two states $\left|\mathfrak{q}_a\right\rangle$ and $\left|\mathfrak{ q}_s\right\rangle$. This is true for both the Fermi and Bose states.

\textbf{Acknowledgments}: I am deeply grateful to Anna Sidorova-Biryukova for her comprehensive assistance in preparing the acticle.

\section*{Appendix A: Cartan Subalgebra and Weyl Generators\label{app1}}
\setcounter{equation}{0}
\setcounter{section}{0}
\setcounter{subsection}{0}
\renewcommand{\thesubsection}{A.\arabic{subsection}}
\renewcommand{\theequation}{A.\arabic{equation}}

Let $\fg$ be a Lie algebra of dimension $n$ formed by generators $\bsX_i$ ($i=1\rightarrow n$) that satisfy permutation conditions
\[
\ld\bsX_i,\bsX_j\rd=\sum^n_{k=1}f_{ijk}\bsX_k\equiv f_{ijk}\bsX_k,
\]
where $f_{ijk}$ are structural constants. $n$ generators $\bsX_i$ form the \textit{basis} $\mathcal{B}$ of the Lie algebra $\fg$.
\begin{defn} (Maximal Abelian subalgebra). An Abelian subalgebra $\fK$ of a Lie algebra $\fg$ is called maximal when there are no additional elements of the algebra $\fg$ that commute with all elements of the subalgebra $\fK$.
\end{defn}
The subalgebra $\fK$ is better known as the \textit{Cartan subalgebra} of the algebra $\fg$, and the number $m$ of elements of the Cartan subalgebra is called the \textit{rank} of the Lie algebra $\fg$.
\begin{defn} (Cartan subalgebra of the Lie algebra). Let $\fg$ be an $n$-dimensional Lie algebra, then the set of all mutually commuting basis elements $\lf\bsX_i=\bsH_i\rf$ ($i=1\rightarrow m$) of the algebra $\fg$ forms the basis of the maximal Abelian subalgebra $\fK\subset\fg$ of the algebra $\fg$.
\end{defn}
\begin{defn} (The rank of the Lie algebra). The dimension $m$ of the Cartan subalgebra $\fK\subset\fg$ determines the \textit{rank} of the Lie algebra $\fg$.
\end{defn}
The elements $\bsH_i$ of the Cartan subalgebra $\fK$ are called \textit{Cartan generators} or \textit{Cartan elements} ($\bsH_i$ also called \textit{Abelian generators} or \textit{diagonal operators}). Cartan generators satisfy permutation conditions
\begin{equation}\label{Per0}
\ld\bsH_i,\bsH_j\rd=0,\quad\forall i,j=1,\ldots,m.
\end{equation}
This means that all $\bsH_i$ are simultaneously diagonalizable. Denoting their eigenvalues by $h_i$, we get
\[
\bsH_i\left|h_1,h_2,\ldots,h_m\right\rangle=h_i\left|h_1,h_2,\ldots,h_m\right\rangle,\quad\forall i=1\rightarrow m.
\]
The eigenvalues $h_i$ are called \textit{weights}. $h_i$ can be considered as components of an $m$-dimensional vector $\boldsymbol{h}$, which is called a \textit{weight vector}. The weights of the Cartan subalgebra are used as quantum numbers to denote the states of a given multiplet.

Further, from the remaining generators $\bsX_i$ of the algebra $\fg$ ($i=1\rightarrow n-m$), which are not elements of the Cartan subalgebra $\fK$, we form linear combinations, which, in turn, form a linearly independent set of \textit{rising} and \textit{lowering} operators (\textit{ladder operators} $\bsE_\alpha$ -- \textit{Weyl generators} or \textit{Weyl elements}). Along with Cartan generators $\bsH_i$, they form the \textit{Cartan-Weyl basis} $\lf\bsH_i,\bsE_\alpha\rf$ of the algebra $\fg$.

Thus, the Lie algebra $\fg$ can be decomposed into a direct sum consisting of the Cartan subalgebra $\fK$ ($m$ generators $\bsH_i$) and $n-m$ one-dimensional subalgebras $\fW_\alpha$ formed by the Weyl generators $\bsE_\alpha$:
\[
\fg=\fK\bigoplus^{n-m}_{\alpha=1}\fW_\alpha=\fK\oplus\fW_1\oplus\fW_2\oplus\ldots\oplus\fW_{n-m}.
\]
There are permutation relations
\begin{equation}\label{Per1}
\ld\bsH_i,\bsE_\alpha\rd=\alpha_i\bsE_\alpha,\quad\forall i=1\rightarrow m,\;\alpha=1\rightarrow n-m.
\end{equation}
The relations (\ref{Per1}) can be written as follows:
\[
\begin{bmatrix}
\ld\bsH_1,\bsE_\alpha\rd\\
\ld\bsH_2,\bsE_\alpha\rd\\
\vdots\\
\ld\bsH_m,\bsE_\alpha\rd
\end{bmatrix}=
\begin{bmatrix}
\alpha_1\\
\alpha_2\\
\vdots\\
\alpha_m
\end{bmatrix}\bsE_\alpha=\boldsymbol{\alpha}\bsE_\alpha.
\]
The various $\alpha_i$ are called \textit{roots} of the generator $\bsE_\alpha$. The set of roots $\alpha_i$ can be considered as components of the vector $\boldsymbol{\alpha}$ called the \textit{root vector}, which belongs to the $m$-dimensional \textit{root space}. The root vector for each Weyl generator $\bsE_\alpha$ can be depicted on a diagram called  a \textit{weight diagram} or \textit{Weyl diagram}, whose dimension is equal to the rank $m$ of the Lie algebra $\fg$. Weight diagrams for rank $m\geq 3$ are difficult to vizualize. An alternative method for depicting weight diagrams for any rank was proposed by Dynkin in 1946 \cite{Dyn46}. It should be noted that by virtue of (\ref{Per0}) all Cartan generators $\bsH_i$ have roots $\alpha_i=0$ and, therefore, are located in the center of the Weyl diagram.

In general, the standard form of commutation relations for generators of the Lie algebra $\fg$ is written as follows:
\begin{equation}\label{Per2}
\begin{array}{lclcl}
\ld\bsH_i,\bsH_j\rd&=&0,&&\forall i,j=1,\ldots,m;\\
\ld\bsH_i,\bsE_\alpha\rd&=&\alpha_i\bsE_\alpha,&&\forall i=1\rightarrow m,\;\alpha=1\rightarrow n-m;\\
\ld\bsE_\alpha,\bsE_{-\alpha}\rd&=&\alpha^i\bsH_i;&&\\
\ld\bsE_\alpha,\bsE_\beta\rd&=&N^\gamma_{\alpha\beta}\bsE_\gamma,&&\beta\neq-\alpha.
\end{array}
\end{equation}
The quantities $N^\gamma_{\alpha\beta}$ can also be expressed in terms of root vectors, so we know the algebra $\fg$ if its roots are known. These roots have the property
\[
\sum_\alpha\alpha_i\alpha_j=\delta_{ij},
\]
where $\alpha$ can take only $n-m$ values:
\begin{equation}\label{Roots}
\alpha=\pm 1,\,\pm 2,\,\ldots,\,\pm\frac{1}{2}(n-m).
\end{equation}

Further, the \textit{Casimir invariants} (operators) $\bsC_\mu$ commute with all generators $\bsX_i$ of the algebra $\fg$:
\[
\ld\bsC_\mu,\bsX_i\rd=0,\quad\forall\mu=1\rightarrow m,\; i=1\rightarrow n.
\]
The number of Casimir operators for a given Lie algebra $\fg$ is determined by the rank of this algebra.
\begin{thm} (Racah Theorem) For each $n$-dimensional Lie algebra $\fg$ of rank $m$, there are a total of $m$ Casimir operators $\bsC_\mu$ ($\mu=1\rightarrow m$) that commute with the generators $\bsX_i$ ($i=1\rightarrow n$) of the algebra $\fg$.
\end{thm}

As a consequence, all $\bsC_\mu$ also commute with Cartan generators $\bsH_i$:
\[
\ld\bsC_\mu,\bsH_i\rd=0,\quad\forall\mu,i=1\rightarrow m.
\]
Thus, it is possible to find a complete set of states that are simultaneously the eigenstates of all $\bsC_\mu$ and $\bsH_i$. Let's define a ket-vector of the following form:
\[
\left|c_1,c_2,\ldots,c_m;h_1,h_2,\ldots,h_m\right\rangle=\left|c_\mu;h_i\right\rangle,
\]
where $c_\mu$, $h_i$ are eigenvalues of the operators $\bsC_\mu$, $\bsH_i$. Then
\[
\bsC_\mu\left|c_\mu;h_i\right\rangle=c_\mu\left|c_\mu;h_i\right\rangle,\quad
\bsH_i\left|c_\mu;h_i\right\rangle=h_i\left|c_\mu;h_i\right\rangle
\]
for all $\mu,i=1\rightarrow m$. We conclude that the Casimir operators $\bsC_\mu$ and the generators $\bsH_i$ of the Cartan subalgebra $\fK$ allow us to label each state of the multiplet, while the ladder operators $\bsE_\alpha$ allow us to move between states inside the multiplet, as shown on the Weyl diagram. Thus, when the ladder operator $\bsE_\alpha$ acts on the ket-vector $\left|c_\mu;h_i\right\rangle$, it shifts the eigenvalues of the operators $\bsH_i$ by the value $\alpha_i$ in accordance with
\[
\bsE_\alpha\left|c_\mu;h_i\right\rangle\approx\left|c_\mu;h_i+\alpha_i\right\rangle.
\]

\section*{Appendix B: Quark Model\label{app2}}
\setcounter{equation}{0}
\setcounter{section}{0}
\setcounter{subsection}{0}
\renewcommand{\thesubsection}{B.\arabic{subsection}}
\renewcommand{\theequation}{B.\arabic{equation}}

The quark hypothesis was first introduced by Gell-Mann and Zweig \cite{Gel61} in the 60s in an attempt to build a systematics of the hadron spectra known at that time. According to the quark model, all existing hadrons (mesons and baryons) ``consist of'' some fundamental ``subunits'' (quarks). To date, the quark model, endowed with a very specific terminology (flavor, strangeness, charm, beauty, etc.), considers six types (flavors) of quarks, combined in three generations. The quantum numbers of quarks are given in Table A.1. (see \cite{Ams19}).
\begin{center}
{\textbf{Table B.1.} Quantum numbers of quarks}
\vspace{0.1cm}
{\renewcommand{\arraystretch}{1.1}
\begin{tabular}{|l|cccccc|}\hline
\phantom{$\sI_z$ -- $z$-component of isospin} & $d$ & $u$ & $s$ & $c$ & $b$ & $t$\\ \hline\hline
$\sQ$ -- electric charge & $-\frac{1}{3}$ & $+\frac{2}{3}$ & $-\frac{1}{3}$ & $+\frac{2}{3}$ & $-\frac{1}{3}$ & $+\frac{2}{3}$\\
$\sI$\phantom{${}_z$} -- isospin & \phantom{$-$}$\frac{1}{2}$ & \phantom{$-$}$\frac{1}{2}$ & \phantom{$-$}$0$ & \phantom{$-$}$0$ & \phantom{$-$}$0$ & \phantom{$-$}$0$\\
$\sI_z$ -- $z$-component of isospin & $-\frac{1}{2}$ & $+\frac{1}{2}$ & \phantom{$-$}$0$ & \phantom{$-$}$0$ & \phantom{$-$}$0$ & \phantom{$-$}$0$\\
$\sS$ -- strangeness & \phantom{$-$}$0$ & \phantom{$-$}$0$ & $-1$ & \phantom{$-$}$0$ & \phantom{$-$}$0$ & \phantom{$-$}$0$\\
$\sC$ -- charm & \phantom{$-$}$0$ & \phantom{$-$}$0$ & \phantom{$-$}$0$ & $+1$ & \phantom{$-$}$0$ & \phantom{$-$}$0$\\
$\sB$ -- beauty & \phantom{$-$}$0$ & \phantom{$-$}$0$ & \phantom{$-$}$0$ & \phantom{$-$}$0$ & $-1$ & \phantom{$-$}$0$\\
$\sT$ -- truth & \phantom{$-$}$0$ & \phantom{$-$}$0$ & \phantom{$-$}$0$ & \phantom{$-$}$0$ & \phantom{$-$}$0$ & $+1$\\
\hline
\end{tabular}
}
\end{center}
The quantum numbers of quarks are related to the electric charge $\sQ$ by means of the generalized Gell-Mann--Nishijima formula
\[
\sQ=\sI_z+\frac{B+\sS+\sC+\sB+\sT}{2},
\]
where $B$ is the baryon number. Quarks are considered to be point-like particles of spin 1/2 with positive parity. It should be noted that the spin and parity of quarks are not objects of the $\SU(3)$-theory (quark model), but are introduced into it from the outside ``by hands'' (as, indeed, the quarks themselves). A consequence is the artificial manner to determine the spin and parity of elementary particles based on the quark model. Antiquarks have negative parity. The baryon number of quarks is 1/3, the antiquarks have a negative baryon number ($-1/3$). By convention, the flavor of a quark has the same sign as its charge $\sQ$. Antiquarks have opposite signs of flavor \cite{PDG}. The hypercharge is determined by the formula
\[
\sY=B+\sS-\frac{\sC-\sB+\sT}{3}.
\]
Therefore, the hypercharge of $u$- and $d$-quarks is 1/3, $-2/3$ for the $s$-quark and 0 for the other quarks.

According to $\SU(3)$-systematics, mesons are $q\bar{q}$-systems of quark-antiquark pairs ($M=q\bar{q}$), baryons consist of three quarks ($B=qqq$), antibaryons consist of three antiquarks ($\bar{B}=\bar{q}\bar{q}\bar{q}$).

Mesons have a baryon number $B=0$. In the quark model, these states are represented by a $q\bar{q}$-system (the flavors of the quark $q$ and the antiquark $\bar{q}$ can be different). The total angular momentum of the  $q\bar{q}$-system (see Figure A.1) is equal to
\begin{equation}\label{Angular}
\mathbf{J}=\mathbf{L}+\mathbf{S},
\end{equation}
where
\[
|l-s|\leq J\leq l+s.
\]
\begin{figure}[t]
\unitlength=1mm
\begin{center}
\begin{picture}(20,13)
\put(10,0){\vector(0,1){17}}
\put(9.15,7){$\bullet$}
\put(2,7.15){\circle{4}}
\put(1,7){$\scriptstyle q$}
\put(18,7.5){\circle{4}}
\put(17,7){$\scriptstyle\bar{q}$}
\put(4,7.5){\line(1,0){12}}
\put(20,7.5){\vector(1,0){8}}
\put(0,7.5){\vector(-1,0){8}}
\put(1,12){$\uparrow$}
\put(17,12){$\downarrow$}
\put(1,1){$\uparrow$}
\put(17,1){$\uparrow$}
\put(-7,5){$n$}
\put(8,15){$l$}
\put(20,11){$s=0$}
\put(20,2){$s=1$}
%\put(10,10){\oval(3,2)}
\put(-55,-5){\begin{minipage}{35pc}{\small {\bf Figure B.1.} Orbital angular momentum $l$ and radial vibrations of the $q\bar{q}$-meson.}\end{minipage}}
\end{picture}
\end{center}
\end{figure}
In this case, $s=s_q+s_{\bar{q}}$ is the total spin of quarks, taking the value $s=0$ (the spins of quarks are antiparallel) or $s=1$ (the spins of quarks are parallel), $l$ is the orbital angular momentum of the $q\bar{q}$-system, taking the values $l=0,1,2,\ldots$. Radial excitations (vibrations) of the $q\bar{q}$-system are denoted by the quantum number $n\geq 1$ (the principal quantum number). Thus, the quantum numbers of the $q\bar{q}$-meson have the form
\[
n^{2s+1}l_J\quad\text{or}\quad I^G(J^{PC}),
\]
where
\begin{equation}\label{PCG}
P=(-1)^l,\quad C=(-1)^{l+s},\quad G=(-1)^{I+l+s}
\end{equation}
respectively, are parity, charge parity and $G$-parity, $I$ -- isospin.

The spectrum of excited (radial and orbital) meson states is similar to the spectrum of a hydrogen atom. So, for the hydrogen atom, the lower vibrational states are designated $1s$, $2p$, $3d$, $\ldots$, in the quark model, the designations $1S$, $1P$, $1D$, etc. are used. Mesons are classified by $J^{PC}$-multiplets. States with $l=0$ are pseudoscalars ($0^{-+}$) and vectors ($1^{--}$). Orbital excitations with $l=1$ are vectors ($1^{++}$) and ($1^{+-}$), as well as tensors ($2^{++}$), etc.

However, not all experimentally detected mesons fit into the standard $q\bar{q}$ scheme. The so-called \textit{exotic mesons}, whose characteristics go beyond the quark model, are divided into three types \cite{LanM}.\\
1) \textit{Exotics of the first kind}. These are states with obviously exotic values of such basic quantum numbers as electric charge $|\sQ|\geq 2$, or strangeness $|\sS|\geq 2$, or isotopic spin $\sI>1$. Such states cannot have usual $q\bar{q}$-type quark structure and must necessarily be exotic multi-quark states (for example, tetraquarks $qq\bar{q}\bar{q}$).\\
2) \textit{Exotics of the second kind}. These are states having exotic combinations of such quantum numbers as spin $J$, parity $P$ and charge parity $C$, which mesons with the usual quark $q\bar{q}$-structure cannot have. Thus, for neutral $q\bar{q}$-mesons with the total spin of quarks $s$ and their orbital moment $l$, parity and charge parity are determined by the relations (\ref{PCG}). It follows that such mesons can only have combinations of quantum numbers $C=P=(-1)^J$ or $(-1)^{J+1}$, as well as  $C=(-1)^J$, $P=(-1)^{J+1}$. There cannot be $q\bar{q}$ states with $C=(-1)^{J+1}$ and $P=(-1)^J$ or $J=0$ and $C=-1$ (if $J=0$, then $s=l=0,1$, $C=+1$). Thus, exotic sets of quantum meson numbers are combinations $J^{PC}=0^{+-}$, $0^{--}$, $1^{-+}$, $2^{+-}$, $3^{-+}$, etc. For example, vector mesons $\pi_1(1400)$ and $\pi_1(1600)$ with an exotic $J^{PC}=1^{-+}$ combination were found (for more details, see \cite{PDG}). All kinds of exotic mesons can have such $J^{PC}$ values (both multi-quark states, as well as hybrids and glueballs).\\
3) \textit{Exotics of the third kind}. These are states with hidden exotics (cryptoexotic mesons). Such states have no external exotic features, and their structure can be determined indirectly by some specific features in their characteristics (abnormally small widths, abnormal decay channels, special mechanisms of formation, etc.). Exotic mesons of all kinds can also belong to the exotics of the third kind.

Further, baryons are fermions with a baryon number $B=1$. In the quark model, these states are represented by a $qqq$-system (anti-baryons with a $\bar{q}\bar{q}\bar{q}$-structure have a baryon number $B=-1$). The total angular momentum of the $q_1q_2q_3$-system (see Figure A.2) is also determined by the formula (\ref{Angular}). In this case, $s=s_{q_1}+s_{q_2}+s_{q_3}$ is the total spin of quarks, taking the value $s=1/2$ (the spin of one quark is antiparallel to the spins of the other two) or $s=3/2$ (the spins of all quarks are parallel).
\begin{figure}[t]
\unitlength=0.95mm
\begin{center}
\begin{picture}(20,20)
\put(0,0){\circle{5}}
\put(20,0){\circle{5}}
\put(10,17){\circle{5}}
\put(-1.4,-0.35){$q_1$}
\put(18.6,-0.35){$q_2$}
\put(8.6,16.35){$q_3$}
\put(2.5,0){\line(1,0){15}}
\put(10,0){\line(0,1){14.5}}
\put(9,-4){$l_\rho$}
\put(11,6){$l_\lambda$}
\put(-35,-10){\begin{minipage}{35pc}{\small {\bf Figure B.2.} Angular moments $l_\rho$ and $l_\lambda$ for the $qqq$-baryon.}\end{minipage}}
\end{picture}
\end{center}
\end{figure}
The total orbital moment for a $q_1q_2q_3$ system is
\begin{equation}\label{Orbital}
\mathbf{L}=\mathbf{l}_\rho+\mathbf{l}_\lambda,
\end{equation}
where
\[
\boldsymbol{\rho}=\frac{1}{\sqrt{2}}(\mathbf{r}_1-\mathbf{r}_2),\quad
\boldsymbol{\lambda}=\frac{1}{\sqrt{6}}(\mathbf{r}_1+\mathbf{r}_2-2\mathbf{r}_3)
\]
are the Jacobi parameters of a three-body system. Here $\mathbf{r}_1$, $\mathbf{r}_2$, $\mathbf{r}_3$ are the radius vectors of quarks $q_1$, $q_2$, $q_3$, the parameter $\boldsymbol{\rho}$ describes the relative motion of quarks $q_1$ and $q_2$, and $\boldsymbol{\lambda}$ describes the motion of the quark $q_3$ with respect to the ``diquark'' system $q_1,q_2$. The parity of the $qqq$-baryon is determined by the relation
\begin{equation}\label{Parity}
P=(-1)^{l_\rho+l_\lambda}.
\end{equation}
Charge parity $C$ and $G$-parity for baryons are not defined. This is due to the fact that the charge operator
$\hat{Q}$ does not commute with the charge conjugation operator $\hat{C}$. The converse is true only for truly neutral states or for neutral systems of type $q\bar{q}$. Truly neutral (Majorana) fermions (baryons) have not yet been found in nature.

Thus, all $qqq$-baryons in the quark model are classified according to combinations of quantum numbers $J^P$. However (as in the case with mesons), not all experimentally detected baryons fit into $qqq$ scheme. \textit{Exotic baryons} \cite{LanB} whose characteristics go beyond the quark model are states with $|\sQ|>2$, or $\sI>3/2$, or $\sS>0$. Such states cannot have the usual $qqq$-type quark structure and must necessarily be exotic multi-quark states (for example, pentaquarks $qqqq\bar{q}$ or dibaryons $qqqqqq$). In 2015, LHCb collaboration published the first experimental evidence of the existence of charmed pentaquark states $c\bar{c}uud$ with a mass of about 4,4 GeV (states $P_c(4312)^+$, $P_c(4440)^+$ and $P_c(4457)^+$, see \cite{Liu19}).

\end{document}